\documentclass{article}

\usepackage{graphicx}
\usepackage{multirow}
\usepackage{amssymb} 
\usepackage{amsthm}
\usepackage{amsmath}
\usepackage{fullpage}
\usepackage[longnamesfirst]{natbib}
\usepackage{enumerate}
\usepackage{amsfonts, bm, amsmath, color, natbib, rotating, amsthm, subfigure, lscape, multicol, epstopdf, verbatim, hyperref}  
\usepackage{authblk}
\usepackage{xr}
\usepackage{grffile}
\bibpunct{(}{)}{;}{a}{,}{,}

\usepackage{setspace}
\theoremstyle{plain} \newtheorem{theorem}{Theorem} \newtheorem{proposition}{Proposition}  \newtheorem{corollary}{Corollary}

\theoremstyle{definition}    
\newcommand{\utwi}[1]{\mbox{\boldmath $ #1$}}
\theoremstyle{remark}

\begin{document}

\newif\ifblinded

\title{Functional Autoregression for Sparsely Sampled Data}

\ifblinded
\author{}
\else

\author{Daniel R. Kowal, David S. Matteson, and David Ruppert\thanks{Kowal is PhD Candidate, Department of Statistical Science, Cornell University, 301 Malott Hall, Ithaca, NY 14853 (E-mail: \href{mailto:drk92@cornell.edu}{drk92@cornell.edu}). Matteson is Assistant Professor, Department of Statistical Science and ILR School, Cornell University, 1196 Comstock Hall, Ithaca, NY 14853 (E-mail: \href{mailto:matteson@cornell.edu}{matteson@cornell.edu}; Webpage: \url{http://www.stat.cornell.edu/\~matteson/}). Ruppert is Andrew Schultz, Jr. Professor of Engineering,  Department of Statistical Science and School of Operations Research and Information Engineering, Cornell University, 1170 Comstock Hall, Ithaca, NY 14853 (E-mail: \href{mailto:dr24@cornell.edu}{dr24@cornell.edu}; Webpage: \url{http://people.orie.cornell.edu/\~davidr/}). Financial support from NSF grant AST-1312903 (Kowal and Ruppert) and a Xerox PARC Faculty Research Award, and NSF grant DMS-1455172 (Matteson)   is gratefully acknowledged. We thank Professor Robert A. Jarrow for suggesting the yield curve application and for his helpful discussions. The authors thank the editors and two referees for very helpful comments.}}

\fi

\maketitle

\large

\vspace{-12mm}

\begin{abstract}
We develop a hierarchical Gaussian process model for forecasting and inference of functional time series data. Unlike existing methods, our approach is especially suited for sparsely or irregularly sampled curves and for curves sampled with non-negligible measurement error. The latent process is dynamically modeled as a functional autoregression (FAR) with Gaussian process innovations. We propose a fully nonparametric dynamic functional factor model for the dynamic innovation process, with broader applicability and improved computational efficiency over standard Gaussian process models. We prove finite-sample forecasting and interpolation optimality properties of the proposed model, which remain valid with the Gaussian assumption relaxed. An efficient Gibbs sampling algorithm is developed for estimation, inference, and forecasting, with extensions for FAR($p$) models with model averaging over the lag $p$. Extensive simulations demonstrate substantial improvements in forecasting performance and recovery of the autoregressive surface over competing methods, especially under sparse designs.  {\color{blue}  We apply the proposed methods to forecast nominal and real yield curves using daily U.S.\ data. Real yields are observed more sparsely than nominal yields, yet the proposed methods are highly competitive in both settings.}
\end{abstract}

\clearpage

\section{Introduction}
We develop a hierarchical Gaussian process model for forecasting and inference of functional time series data.  A{ \it functional time series} is a time-ordered sequence of random functions, $Y_1,\ldots,Y_T$, on some compact index set  $\mathcal{T} \subset \mathbb{R}^D$, typically with $D=1$. Unlike existing methods, our approach is especially suited for sparsely or irregularly sampled curves,  {\color{blue}  in which the functions $Y_t(\tau)$ are observed at a small number of possibly unequally-spaced points $\tau \in \mathcal{T}$},  and for curves sampled with non-negligible measurement error, which occur frequently in financial applications. Applications of functional time series are abundant, including:
daily or weekly interest rate curves as a function of time to maturity, such as daily Eurodollar futures contracts \citep{kargin2008curve} and weekly yield curves \citep{FDFM,kowal2014bayesian}; yearly sea surface temperature as a function of time-of-year \citep{besse2000autoregressive}; 
yearly mortality and fertility rates as a function of age \citep{hyndman2007robust}; daily pollution curves as a function of time-of-day \citep{damon1z2002inclusion,aue2015prediction}; 
and a vast collection of spatio-temporal applications in which a time-dependent variable is measured as a function of spatial location (e.g., \citealp{cressie2011statistics}). 
The primary goal of functional time series analysis is usually forecasting $\{Y_t\}$, but we are also interested in performing inference and 
obtaining an interpretable representation of the time evolution of $\{Y_t\}$.

The most prevalent model for functional time series data is the {\it functional autoregressive model of order 1}, written FAR(1):
\begin{equation} \label{far}
Y_t - \mu= \Psi(Y_{t-1} - \mu)+ \epsilon_t,
\end{equation}
where $Y_t \in L^2(\mathcal{T})$, $\Psi$ is a bounded linear operator on $L^2(\mathcal{T})$, $\epsilon_t \in L^2(\mathcal{T})$ is a sequence of independent mean zero random innovation functions with $\mathbb{E}||\epsilon_t||^2 < \infty$, and $\mu$ is the  mean of $\{Y_t\}$ under stationarity. The FAR(1) model,  developed by \cite{bosq2000linear}, is an extension of two highly successful models:  the functional linear model for function-on-function regression and the vector autoregressive model for multivariate time series, and has been successfully applied in a variety of applications. {\color{blue} 
Importantly, the FAR(1) model 
provides a mechanism for modeling the evolution of $\{Y_t\}$ jointly over the entirety of the domain $\mathcal{T}$.}
More generally, \eqref{far} can be extended for multiple lags to the FAR($p$) model:
$
Y_t - \mu= \sum_{\ell=1}^p \Psi_\ell (Y_{t-\ell} - \mu)+ \epsilon_t.$

Existing approaches for estimating the FAR($p$) model typically use an eigendecomposition of the empirical (contemporaneous and lagged) covariance operators \citep{damon1z2002inclusion,damon2005estimation,horvath2012inference,kokoszka2012dependent} or kernel-based procedures for modeling the conditional expectation \citep{besse2000autoregressive}. 
A related approach is to estimate a multivariate time series model for the functional principal component (FPC) scores of the observed data \citep{aue2015prediction}. Extensions of the FAR(1) model for nonstationary functional time series are available, such as the time-dependent FAR kernels proposed in \cite{chen2015adaptive}. 

In general, existing methods for FAR($p$)  are designed for functional data observed on dense grids without measurement error, and typically require pre-smoothing discretized functional observations. However, such procedures may exhibit erratic behavior for sparse designs and are inappropriate in such settings. More generally, under an FAR($p$) model that includes measurement error and discretization of the functional observations, we prove that the two most common approaches for functional data analysis|estimators that are linear in the FPC scores or the pre-smoothed observations|produce predictions that are inadmissible (in a decision theory sense). Indeed, the presence of measurement error fundamentally alters the behavior of the observable process: if an FAR process is observed with measurement error, then the observable process is no longer an FAR process, but rather a functional autoregressive moving average process (see Proposition 1). Even under dense designs, existing methods produce poor estimates of the FAR operator $\Psi$ \citep{didericksen2012empirical}, which inhibits interpretability of the time evolution of $\{Y_t\}$, and do not provide finite-sample inference. We propose new methodology that simultaneously addresses all of these challenges.  

We propose a general two-level hierarchy for modeling functional time series: an \emph{observation equation} addresses measurement error and discretization of the functional data, while an \emph{evolution equation} defines a process model for the underlying functional time series. The latent process is dynamically modeled as an FAR($p$). We parsimoniously specify the FAR model with mean zero Gaussian process innovations, which are fully specified by covariance functions without parameterizing sample paths. The dynamic innovation process is further specified by a dynamic functional factor model. In contrast with standard approaches for Gaussian processes, this avoids selecting and estimating a parametric covariance function, and allows greater computational stability and efficiency, and broader applicability. Interpolating curves at unsampled locations and forecasting future curves are primary objectives in functional time series modeling; the proposed model produces optimal (best linear) predictions under both sparse and dense designs in the presence of measurement error, even with the Gaussian assumption relaxed. We propose an efficient Gibbs sampling algorithm for estimation, inference, and forecasting. Extensive simulations demonstrate substantial improvements in forecasting performance and recovery of the autoregressive surface over competing methods, especially under sparse designs.

 {\color{blue}   We apply our methodology to model and forecast {\it nominal} and {\it real} yield curves using daily U.S.\ data. For a given currency and level of risk of a debt, the nominal yield curve, $Y_t^N(\tau)$, describes the interest rate at time $t$ as a function of the length of the borrowing period, or time to maturity, $\tau$. Similarly, the real yield curve, $Y_t^R(\tau)$, corresponds to an interest rate that is adjusted for inflation. Both $Y_t^N$ and $Y_t^R$ may be modeled as functional time series. However, real yields are sparsely observed for each time $t$, and only at longer maturities, which is problematic for existing functional time series models. The proposed methods provide a natural hierarchical framework for modeling both nominal yield curves and real yield curves, and in both cases produce highly competitive forecasts.}

Bayesian methods for functional time series are limited, with the exception of \cite{laurini2014dynamic} and \cite{kowal2014bayesian}.
The primary contributions of this article are the following: 
(i) development of a hierarchical framework for FAR($p$) (Section \ref{hierSection}), which produces optimal (best linear) predictions under both sparse and dense designs in the presence of measurement error; 
(ii) a dynamic functional factor model for the innovation covariance, which is nonparametric, computationally convenient, and offers useful generalizations to non-Gaussian distributions (Section \ref{gaussSection});
(iii) a procedure for model averaging over the lag, $p$, within a hierarchical FAR($p$) model (Section \ref{psiEst});
(iv) comparisons of the proposed methods to existing methods for FAR($p$) using theoretical results (Section \ref{finiteOpt}), an extensive simulation study (Section \ref{sims}), and a real data application (Section \ref{yields}); 
(v) {\color{blue} a comparative forecasting study of daily U.S.\! nominal and real yield curve data (Section \ref{yields});}
and (vi) an efficient Gibbs sampling algorithm, which uses common full conditional distributions and existing \texttt{R} software (Appendix). Details of our Gibbs sampling algorithm and additional theoretical and simulation results are in the web supplement.


\section{Hierarchical Gaussian Processes for FAR}\label{hierSection}
Let $Y_1,\ldots, Y_T$ be  a time-ordered sequence of random functions in $L^2(\mathcal{T})$, where $\mathcal{T} \subset \mathbb{R}^D$ is a compact index set. We focus on $D=1$ with $\mathcal{T} = [0,1]$, but the methods can be developed more generally. For interpretability and computational convenience, we restrict our attention to the integral operators defined by 
$
\Psi_\ell(Y)(\tau) = \int \psi_\ell(\tau, u) Y(u)\, du,
$ so the FAR($p$) model  is
\begin{equation} \label{farpInt}
Y_t(\tau) - \mu(\tau)= \sum_{\ell=1}^p \int  \psi_\ell(\tau, u)\left\{Y_{t-\ell}(u) - \mu(u)\right\} \, du+ \epsilon_t(\tau) \quad \forall \tau \in \mathcal{T}.
\end{equation}
Using integral operators, the FAR($p$) model resembles the functional linear model, in which   $(Y_t - \mu)$ is regressed on $(Y_{t-1}-\mu), \ldots, (Y_{t-p}-\mu)$. The functional linear model is widely popular in functional data analysis, and has been extensively studied  (e.g., \citealp{cardot1999functional, ramsay2006functional}).   

In practice, model  \eqref{farpInt} is incomplete: the functional observations $\{Y_t\}$ are not observed directly, but rather via discrete samples of each curve, and typically with measurement error.  
Suppose that we observe $y_{i,t} \in \mathbb{R}$ sampled with noise $\nu_{i,t}$ from $Y_t \in L^2(\mathcal{T})$:
\begin{equation} \label{obs}
y_{i,t} = Y_t(\tau_{i,t}) + \nu_{i,t}
\end{equation}
for $i=1,\ldots, m_t$, where  $\tau_{1,t},\ldots,\tau_{m_t,t}$ are the observation points of $Y_t$ and  $\nu_{i,t}$ is a mean zero measurement error with finite variance. Typically for functional data, $m_t$ will be large and $\mathcal{T}_t = \{\tau_{1,t},\ldots, \tau_{m_t,t}\}$ will be dense in $\mathcal{T}$. However, for our procedures, we allow $m_t$ to be small for some (or all) $t$, with observation points $\mathcal{T}_o \equiv \cup_t \mathcal{T}_t$ dense or sparse in $\mathcal{T}$. 
 Combining \eqref{obs} with \eqref{farpInt} for $p=1$ and defining  $\mu_t \equiv Y_t - \mu$,
 we obtain the two-level hierarchical model
\begin{equation}\label{hierfar}
\begin{cases}
{y}_{i,t} = \mu(\tau_{i,t}) + \mu_t(\tau_{i,t}) + \nu_{i,t}, & i=1,\ldots, m_t,\\
\mu_t(\tau)  = \int \psi(\tau, u) \mu_{t-1}(u) \, du + \epsilon_t(\tau), & \forall \tau \in \mathcal{T} 
\end{cases}
\end{equation}
for $t=2,\ldots, T$, where we assume that $\{\nu_{i,t}\}$ and $\{\epsilon_t\}$ are mutually independent sequences. 

The measurement error is a nontrivial component of model \eqref{hierfar}, which we demonstrate in the following proposition: 
\begin{proposition}\label{farma}
Let $Y_t -\mu = \sum_{\ell=1}^p \Psi_\ell(Y_{t-\ell}-\mu) + \epsilon_t$, and suppose that we observe $y_t = Y_t + \nu_t$, where $\{\epsilon_t\}$ and $\{\nu_t\}$ are independent white noise processes. Then the observable process $\{y_t\}$ follows a \emph{functional autoregressive moving average} (FARMA) process of order $(p,p)$. 
\end{proposition}
We define a FARMA process and prove Proposition \ref{farma} in Section B.1 of the web supplement. 
The implication of Proposition \ref{farma} is that, if the true model for $Y_t$ is FAR($p$), yet $Y_t$ is observed with error, then the FAR($p$) model for the observables is inappropriate. As a result, estimation of $\Psi_\ell$ will be  inefficient and forecasting will  deteriorate, due to both increased estimation error of $\Psi_\ell$ and model misspecification. By comparison, the hierarchical model decomposes the observed data into a functional (autoregressive) process and measurement error, and in doing so circumvents the model misspecification issues implied by Proposition \nolinebreak \ref{farma}.





We model the random functions $\mu$, $\psi$, and $\{\epsilon_t\}$ as {\it Gaussian processes}: $\mu \sim \mathcal{GP}(0, K_\mu)$, $\psi  \sim \mathcal{GP}(0, K_\psi)$, and $\epsilon_t \stackrel{indep}{\sim} \mathcal{GP}(0, K_\epsilon)$, where the notation $\mathcal{GP}(m, K)$ denotes a Gaussian process with mean function $m$ and covariance function $K$. Gaussian processes have a long history in machine learning \citep{rasmussen2006gauss} and spatial statistics \citep{cressie2011statistics},  and have seen increased application in  functional data analysis, especially for hierarchical modeling \citep{behseta2005hierarchical,kaufman2010bayesian,shi2011gaussian,earls2014bayesian}. The conditional distribution of $\mu_t = Y_t - \mu$ is $[\mu_t | \mu_{t-1}, \psi, K_\epsilon] \sim  \mathcal{GP} (\int \psi(\cdot, u) \mu_{t-1}(u) \, du, K_\epsilon)$, which models the evolution of $\mu_t$ and serves as the prior distribution for the observation level of \eqref{hierfar}.   Notably, the model only requires {\it conditionally} Gaussian processes, and therefore may accommodate more general distributional assumptions, such as scale-mixtures of Gaussian distributions and stochastic volatility. Moreover, the posterior expectations derived from the hierarchical Gaussian process model are best linear predictors, and therefore are optimal among linear predictors for interpolation and forecasting  of $Y_t$, even for non-Gaussian distributions (see Section \ref{finiteOpt}). We assume $\nu_{i,t} \stackrel{iid}{\sim} N(0, \sigma_\nu^2)$ for the measurement errors; priors for  $\sigma_\nu^2$ and the parameters associated with $K_\mu$, $K_\epsilon$, and $K_\psi$ will be discussed later.



\subsection{Dynamic Linear Models for FAR($p$)}
For practical implementation of model \eqref{hierfar}, we must select a finite set of evaluation points,  $\mathcal{T}_e \equiv \{\tau_1,\ldots,\tau_M\} \subset \mathcal{T}$, at which we wish to estimate, forecast, or perform inference on the random functions, in particular $\mu_t = Y_t - \mu$. Naturally, we assume that $\mathcal{T}_t \subseteq \mathcal{T}_e$ for all $t$, but this assumption may be relaxed. Notably, $\mathcal{T}_e$   provides a convenient structure for forecasting and inference of $y_{i,t}$ and $Y_t(\tau_{i,t})$ at the observations points $\tau_{i,t} \in \mathcal{T}_t$, as well as interpolation of $Y_t$ at  any unobserved points, $\tau^* \in \mathcal{T}_e \setminus \mathcal{T}_o$. By definition, for any Gaussian process $x \sim \mathcal{GP}(m, K)$ defined on $\mathcal{T}$, we have $\utwi{x} \sim N(\utwi{m}, \utwi{K})$, where $\utwi{x} = (x(\tau_1), \ldots,x(\tau_M))'$, $\utwi{m} = (m(\tau_1), \ldots,m(\tau_M))'$, and $\utwi{K} =\{K(\tau_i, \tau_k)\}_{i,k = 1}^M$. This result is particularly useful for constructing an estimation procedure and deriving the optimality results of Section \ref{finiteOpt}.

By selecting $M$ large and $\mathcal{T}_e$ dense in $\mathcal{T}$, we can accurately approximate the integral in \eqref{hierfar} using quadrature methods:
\begin{equation}\label{quad}
\int \psi(\tau, u) \mu_{t-1}(u) \, du \approx \left(\psi(\tau, \tau_1), \ldots, \psi(\tau, \tau_M)\right) \utwi{Q} \utwi{\mu}_{t-1},
\end{equation}
where  $\utwi{Q}$ is a known quadrature weight matrix and  $\utwi{\mu}_{t-1} = (\mu_{t-1}(\tau_1),\ldots,\mu_{t-1}(\tau_M))'$. The approximation in \eqref{quad} is important for computational tractability in estimation of both $\mu_t$ and $\psi$. {\color{blue}  Practical implementations of  functional data methods require discretization or finite approximations; the quadrature approximation in \eqref{quad} is a natural approach, and does not impose restrictive assumptions on the functional forms of $\psi$ and $\mu_{t-1}$. In addition, our simulation analysis suggests that the quadrature approximation does not noticeably inhibit estimation or forecasting, especially relative to existing FAR methods.} In practice, the trapezoidal rule for computing $\utwi{Q}$ works well, and {\color{blue}  for simulated data $M=20$ is sufficiently large}. {\color{blue}  We include a sensitivity analysis in the web supplement to assess the effects of $M$ on the approximation error in \eqref{quad}, which supports this choice of $M$}.

 Assuming $\mathcal{T}_o \subseteq \mathcal{T}_e$, let $\utwi{Z}_t$ be the $m_t \times M$ incidence matrix that identifies the observations points   observed at time $t$, i.e., $(\tau_{1,t},\ldots, \tau_{m_t ,t})' = \utwi{Z}_t (\tau_1,\ldots,\tau_M)'$.  We can write the hierarchical model \eqref{hierfar} as a {\it dynamic linear model} (DLM;   \citealp{westDLM}) in $\utwi{\mu}_t$: 
\begin{equation}\label{dlmEval}
\begin{cases}
\utwi{y}_{t} =    \utwi{Z}_t\utwi{\mu} +  \utwi{Z}_t\utwi{\mu}_t + \utwi{\nu}_t,  & [\utwi{\nu}_t| \sigma_\nu^2]  \stackrel{indep}{\sim} N\left(\utwi{0}, \sigma_\nu^2 \utwi{I}_{m_t}\right) \mbox{ for } t=1,\ldots,T,\\
\utwi{\mu}_t = \utwi{\Psi}\utwi{Q}\utwi{\mu}_{t-1} + \utwi{\epsilon}_t, & [\utwi{\epsilon}_t| \utwi{K}_\epsilon] \stackrel{indep}{\sim} N\left(\mathbf{0}, \utwi{K}_\epsilon\right)\mbox{ for } t=2,\ldots,T, \\
\utwi{\mu}_1 \sim N(\utwi{0}, \utwi{K}_\epsilon),
\end{cases}
\end{equation}
where $\utwi{y}_t = (y_{1,t},\ldots, y_{m_t,t})'$, $\utwi{\mu} = (\mu(\tau_{1}), \ldots, \mu(\tau_{M}))'$,  $\utwi{\Psi} = \{\psi(\tau_i, \tau_k)\}_{i,k = 1}^M$, and $\utwi{K}_\epsilon =\{K_\epsilon(\tau_i, \tau_k)\}_{i,k = 1}^M$. Model \eqref{dlmEval} can be extended for multiple lags to the FAR($p$) model by replacing the second level with $\utwi{\mu}_t = \sum_{\ell=1}^p \utwi{\Psi}_\ell\utwi{Q} \utwi{\mu}_{t-\ell} + \utwi{\epsilon}_t$ for $\utwi{\Psi}_\ell = \{\psi_\ell(\tau_i, \tau_k)\}_{i,k = 1}^M$. The DLM formulation of the FAR($p$) is  useful for MCMC sampling,  since efficient samplers exist for the vector-valued state variables, $\{\utwi{\mu}_t\}$ (e.g., \citealp{durbin2002simple}). The proposed Gibbs sampling algorithm for model \eqref{dlmEval} (see Section A of the web supplement) is a moderate extension of traditional DLM samplers, and iteratively samples the state vectors $\{\utwi{\mu}_t\}$, the measurement error variance $\sigma_\nu^2$,  the innovation covariance $\utwi{K}_\epsilon$, and the unknown evolution matrix $\utwi{\Psi}$.  {\color{blue}  The DLM also facilitates non-Bayesian parameter estimation and forecasting, such as an EM algorithm for the latent state variables $\{\utwi{\mu}_t\}$ with the parameters $\{\sigma_\nu^2, \utwi{K}_\epsilon, \utwi{\Psi}\}$ (e.g., \citealp{cressie2011statistics}).}

{\color{blue} 
The connection between the hierarchical FAR model \eqref{hierfar} and the DLM \eqref{dlmEval} is further illuminated by considering the autocovariance properties of the respective models. 
Recalling $\mu_t(\tau) = Y_t(\tau) - \mu(\tau)$, let $C_\ell(\tau_1, \tau_2) = \mathbb{E}\left[\mu_t(\tau_1) \mu_{t-\ell}(\tau_2)\right]$ be the lag-$\ell$ autocovariance function of $\{Y_t\}$, which is time-invariant under stationarity of $\{Y_t\}$. Under model \eqref{hierfar} and assuming stationarity of $\{Y_t\}$, the lag-1 autocovariance function is equivalently $C_1(\tau_1, \tau_2)  =  \mathbb{E}\left[\mu_t(\tau_1) \mu_{t-1}(\tau_2)\right]  =\mathbb{E}\left[ \left\{\int \psi(\tau_1, u) \mu_{t-1}(u) \, du + \epsilon_t(\tau_1)\right\} \mu_{t-1}(\tau_2)\right]  = \int \psi(\tau_1, u) C_0(u, \tau_2) \, du$. For $\ell \ge 1$, we have the more general recursion $C_\ell(\tau_1, \tau_2) = \int \psi(\tau_1, u) C_{\ell-1}(u, \tau_2) \, du$, from which it is clear that each $C_\ell$ is completely determined by the pair $(\psi, C_0)$.  Now let $\bm C_\ell = \mathbb{E}\left[\bm \mu_t \bm \mu_{t-\ell}'\right]$ be the lag-$\ell$ autocovariance matrix for the vector-valued time series $\{\bm \mu_t\}$ in \eqref{dlmEval}. Under stationarity of $\{\bm \mu_t\}$, the lag-1 autocovariance matrix of $\bm \mu_t$ is $\bm C_1 =  \mathbb{E}\left[\bm \mu_t \bm \mu_{t-1}'\right]  = \mathbb{E}\left[ \left\{ \utwi{\Psi}\utwi{Q}\bm \mu_{t-1} + \bm \epsilon_t\right\} \bm \mu_{t-1}'\right]  = \bm{\Psi}\bm{Q} \bm C_0$. Notably, the relationship $\bm{C}_1 = \bm{\Psi}\bm{Q} \bm C_0$ is an approximation to the continuous version, $C_1(\tau_1, \tau_2)  =\int \psi(\tau_1, u) C_0(u, \tau_2) \, du$, using the same quadrature approximation as in \eqref{quad}. More generally, the matrix  recursion  $\bm C_\ell = \bm{\Psi}\bm{Q} \bm C_{\ell-1}$ is a quadrature-based approximation to the continuous recursion, $C_\ell(\tau_1, \tau_2) = \int \psi(\tau_1, u) C_{\ell-1}(u, \tau_2) \, du$  for $\ell \ge 1$. Therefore, the evolution matrix $\utwi{\Psi}\utwi{Q}$ in the DLM \eqref{dlmEval} induces a discrete approximation to the autocovariance structure in the hierarchical FAR model \eqref{hierfar}.
}

The evolution equation of \eqref{dlmEval} resembles a VAR(1) on $\utwi{\mu}_{t} =  (\mu_{t}(\tau_1),\ldots,\mu_{t}(\tau_M))'$, but differs from a standard  VAR on $\utwi{y}_t$ for a few critical reasons. 
First, fitting a VAR to $\utwi{y}_t$ is only well-defined if both the dimension $m_t$ and  the observation points $\mathcal{T}_t$ are fixed over time. If this does not hold, then imputation is necessary. Our procedure imputes automatically and optimally using the conditional mean function and the conditional covariance function of the corresponding Gaussian process. 
Second,  the components of $\utwi{y}_t$ are likely highly correlated due to the functional nature of the observations. Strong collinearity in VARs can cause overfitting and adversely affect forecasting and inference. In our model, the kernel function $\psi$ is regularized using a smoothness prior (see Section \ref{psiEst}), which mitigates the adverse effects of collinearity on  estimation of  $\psi$.  {\color{blue}  The smoothness prior on $\psi$ is a nonstandard regularization technique for VARs, but is appropriate in this setting.} Finally, the quadrature matrix, $\utwi{Q}$, is absorbed into the VAR coefficient matrix $\utwi{\Psi}\utwi{Q}$, and reweights the vector $\utwi{\mu}_{t-1}$ using information from the evaluation points $\mathcal{T}_e$. This reweighting incorporates not only  the vector values $\utwi{\mu}_t$, but also  
the information that the components of $\utwi{\mu}_t$ correspond to  ordered elements of $\mathcal{T}_e$, which need not be equally spaced. The simulations of Section \ref{sims} demonstrate the substantial improvements in forecasting of our procedure relative to a VAR on $\utwi{y}_t$. 

\section{A Dynamic Functional Factor Model for the Innovation Process}\label{gaussSection} 
The standard approach for Gaussian process models is to select a parametric covariance function that only depends  on a few parameters, and then estimate those parameters using either fully Bayesian methods or empirical Bayes \citep{rasmussen2006gauss}. 
The choice of the covariance function determines the properties of the sample trajectories, such as smoothness and periodicity, but notably does {\it not} imply a parametric form for the sample trajectories. 
Indeed, the FAR($1$) model \eqref{dlmEval} may be estimated using these standard approaches; we provide one implementation in Section \ref{sims}.


However, there are substantial computational limitations that accompany standard parametric covariance functions. Even when the covariance function is known up to some parameters $\utwi{\rho}$, in general we cannot directly sample from the  full conditional posterior distribution  for $\utwi{\rho}$. As a result,  posterior sampling for   $\utwi{\rho}$ can be inefficient. Gaussian processes also require computation of the $M\times M$ innovation covariance matrix $\utwi{K}_\epsilon$, which must be inverted|both for evaluating the conditional likelihood of $\utwi{\rho}$ and for  sampling $\{\utwi{\mu}_t\}$ and $\psi$. Most common choices for parametric covariance functions do not offer any simplifying structure for computing this inverse, which may be computationally inefficient and unstable. In addition, extensions for time-dependent covariance functions or non-Gaussian distributions are not readily available, and further increase the difficulties with posterior sampling.



We propose a  low-rank, fully nonparametric approach for modeling the innovation covariance function. Using the {\it functional dynamic linear model} (FDLM) of \cite{kowal2014bayesian}, we estimate the unknown covariance function using a functional factor model, which does not require specification of a parametric form for the covariance function. This method avoids the need for inversion of the full $M \times M$ covariance matrix, and is more computationally stable and efficient. The integration of the FDLM into \eqref{dlmEval} retains the fully Bayesian hierarchical structure, and permits joint inference for all parameters via an efficient MCMC sampling algorithm. 
A {\it functional} factor model is most appropriate because $\epsilon_t$ is a Gaussian process with covariance function $K_\epsilon$, so $K_\epsilon$ must be  well-defined on $\mathcal{T} \times \mathcal{T}$. 
Notably, the FDLM offers convenient generalizations for stochastic volatility models \citep{kim1998stochastic} and more robust models using scale-mixtures of Gaussian distributions  \citep{fernandez2000bayesian}.

The FDLM decomposes the innovations  $\epsilon_t$  into   {\it factor loading curves} (FLCs), $\phi_j \in L^2(\mathcal{T})$, and time-dependent {\it factors}, $e_{j,t} \in \mathbb{R}$, for $j=1,\ldots,J_\epsilon$:
\begin{equation}\label{fdlm}
\epsilon_t(\tau) = \sum_{j=1}^{J_\epsilon} e_{j,t} \phi_j(\tau) + \eta_t(\tau)
\quad  \color{blue}  \forall \tau \in \mathcal{T},\color{black} \rm 
\end{equation}
where $J_\epsilon$ is the number of factors  and $\{\eta_t\}$ is the mean zero approximation error with $\eta_t \stackrel{iid}{\sim} \mathcal{GP}(0, K_\eta)$, where  $K_\eta(\tau, u) = \sigma_\eta^2 \mathbf{1}(\tau=u)$ and $\mathbf{1}(\cdot)$ is the indicator function. 
We model each FLC $\phi_j$ as a smooth function admitting the basis expansion $\phi_j(\tau) = \utwi{b}_\phi'(\tau) \utwi{\xi}_{j}$, where $\utwi{b}_\phi$ is a $J_\phi$-dimensional vector of known basis functions and $\utwi{\xi}_j$ is an unknown vector of coefficients. For superior MCMC performance, we prefer the  low-rank thin plate spline basis for $\utwi{b}_\phi$ (e.g., \citealp{crainiceanu2005bayesian}) with knot locations selected using the quantiles of the observation points, $\mathcal{T}_o$. 
We place a smoothness prior on each $\utwi{\xi}_j$, which is expressed via a conditionally conjugate Gaussian  distribution and is convenient for efficient posterior sampling (see the Appendix). The smoothness assumption typically produces more interpretable FLCs $\{\phi_j\}$ and can improve estimation for unobserved points  $\tau^* \not \in  \mathcal{T}_o$.  For the factors $\utwi{e}_t = (e_{1,t},\ldots,e_{J_\epsilon,t})'$, we assume  $[\utwi{e}_{t} | \utwi{\Sigma}_e] \stackrel{indep}{\sim}N(\utwi{0}, \utwi{\Sigma}_e)$, with $\utwi{\Sigma}_e = \mbox{diag}\left(\{\sigma_j^2\}_{j=1}^{J_\epsilon}\right)$
for simplicity. By comparison, the factors in \cite{kowal2014bayesian} are time-dependent; we assume independence to obtain a special case of the FDLM in which the implied innovation process $\{\epsilon_t\}$ is an independent sequence, which also improves computational efficiency of the FDLM sampling algorithm. Importantly, we obtain a nonparametric, low-rank approximation to the innovation covariance, $K_\epsilon$, with useful computational simplifications.

For identifiability, we order the factors according to variability of $\epsilon_t$ explained, $\sigma_1^2 > \sigma_2^2 > \cdots > \sigma_{J_\epsilon}^2 > 0$, and require orthonormality of the FLCs.  It is computationally convenient to enforce the discrete orthonormality constraint $\utwi{\Phi}'\utwi{\Phi} = \utwi{I}_{J_\epsilon}$, where $\utwi{\Phi} = \utwi{B}_\phi \utwi{\Xi}$  is the $M \times J_\epsilon$ matrix of FLCs evaluated at $\mathcal{T}_e$, $\utwi{B}_\phi = (\utwi{b}_\phi(\tau_1), \ldots, \utwi{b}_\phi(\tau_M))'$ is the $M\times J_\phi$ matrix of basis functions evaluated at $\mathcal{T}_e$,  and $\utwi{\Xi} = (\utwi{\xi}_1, \ldots, \utwi{\xi}_{J_\epsilon})$ is the $J_\phi \times J_\epsilon$ matrix of unknown FLC basis coefficients. The implied covariance matrix for $\utwi{\epsilon}_t = (\epsilon_t(\tau_1), \ldots, \epsilon_t(\tau_M))'$ under \eqref{fdlm} is  
$\utwi{K}_\epsilon = \utwi{\Phi} \utwi{\Sigma}_e\utwi{\Phi}' + \sigma_\eta^2 \utwi{I}_M$, conditional on $\{\phi_j, \sigma_j^2\}$ and $\sigma_\eta^2$. Importantly, the discretized orthonormality constraint offers a substantial simplification for computing the inverse of $\utwi{K}_\epsilon$ using the Woodbury identity:
\begin{equation}\label{covMatInv}
\utwi{K}_\epsilon^{-1}  =  \sigma_\eta^{-2} \utwi{I}_M - \sigma_\eta^{-2}\utwi{\Phi}\utwi{\tilde{\Sigma}}_e\utwi{\Phi}',
 \end{equation}
 where $\utwi{\tilde{\Sigma}}_e = \sigma_\eta^{-2}\left(\utwi{\Sigma}_e^{-1} +\sigma_\eta^{-2} \utwi{\Phi}'\utwi{\Phi}\right)^{-1} = \mbox{diag}\left(\{\sigma_j^2/(\sigma_\eta^2+\sigma_j^2)\}_{j=1}^{J_\epsilon}\right)$.
 As a result, $\utwi{K}_\epsilon^{-1}$ may be computed without any matrix inversions. By comparison, parametric  covariance functions  not only fail to offer   computational simplifications for $\utwi{K}_\epsilon^{-1}$, but also require {\it additional} computations of $\utwi{K}_\epsilon^{-1}$ in the estimation of the covariance function parameters, $\utwi{\rho}$.  The FDLM sampling algorithm for the factors $\{e_{j,t}\}$, the FLCs $\{\phi_j\}$, and the variances $\{\sigma_j^2\}$ and $\sigma_\eta^2$ is computationally inexpensive and MCMC efficient.  Note that the  approximation error  is a nontrivial addition to model \eqref{fdlm}:  $\eta_t$ is necessary  for nondegeneracy of $\utwi{K}_\epsilon$, which is invertible only when $\sigma_\eta^2 >0$. And while $\sigma_\eta^2 > 0$ implies that the innovations $\epsilon_t$, and therefore $\mu_t$, are not smooth, we find that in practice, the sample paths of $\epsilon_t$ and $\mu_t$ do appear smooth for sufficiently small  $\sigma_\eta^2$. {\color{blue}  Generalizations to non-nugget approximation error variance functions $K_\eta(\tau, u) = \sigma_\eta^2(\tau) \mathbf{1}(\tau=u)$ for $\sigma_\eta^2 \colon \mathcal{T}\rightarrow \mathbb{R}^+$ are available, but may introduce additional model complexity and computational costs.}
 
 An important application of the FDLM simplification in \eqref{covMatInv} is given in Theorem \ref{fdlmKriging}, in which we derive a computationally convenient form for estimating the out-of-sample posterior distribution $[\mu_t(\tau^*) | \{\utwi{y}_r\}_{r=1}^s]$ for $\tau^* \not\in\mathcal{T}_e$, which includes as special cases the forecasting distribution ($s<t$), the filtering distribution ($s=t$), and the smoothing distribution ($s > t$). 

\section{Modeling the FAR Kernel}\label{psiEst}
An accurate predictor of $\psi$ is important not only for forecasting and inference, but also for interpreting the time evolution of $\{Y_t\}$. The likelihood for $\psi$ is specified by the evolution equation in model \eqref{dlmEval}, which may be extended for multiple lags. We select a Gaussian process prior for $\psi$, which encourages smoothness of the surface and produces more interpretable results. Using the basis approximation $\psi_\ell(\tau, u) = \utwi{b}_0'(\tau, u)\utwi{\theta}_{\psi_\ell}$, we place a Gaussian prior on $\utwi{\theta}_{\psi_\ell}$, which induces a Gaussian process prior for $\psi_\ell$.  A tensor product 
basis $\utwi{b}_0'(\tau, u) = (\utwi{b}_\psi'(u) \otimes \utwi{b}_\psi'(\tau))$ for $\bm{b}_\psi$ a $J_\psi$-dimensional vector of B-spline basis functions is computationally efficient in our setting, especially for large $M$. The details are presented in the Appendix. {\color{blue} Since  $J_\psi < M$, the evolution matrix $\utwi{\Psi}\utwi{Q}$ in \eqref{dlmEval} has $J_\psi^2 < M^2$ unknown parameters, so the evolution equation in the DLM \eqref{dlmEval} has fewer parameters than a standard VAR(1) on $\bm{\mu}_t$.}
Notably, the posterior distribution for $\psi_\ell$ depends on $\utwi{K}_\epsilon^{-1}$, which is computationally unstable for many common parametric covariance functions. By comparison, the nonparametric FDLM estimate of $\utwi{K}_\epsilon^{-1}$ in \eqref{covMatInv} is computationally stable, which further stabilizes estimates of $\psi_\ell$.

An important choice in the FAR($p$) model is the maximum lag, $p$: a poor choice of $p$ can produce suboptimal forecasts and reduce MCMC efficiency.  A reasonable approach is to compare the DIC or marginal likelihoods for different choices of $p$. However, this  requires recomputing the model for each choice of $p$, which can be computationally intensive. Similarly, \cite{kokoszka2013determining} propose a multistage hypothesis testing procedure based on asymptotic approximations and an FPC decomposition, but would require modification for the hierarchical Bayesian implementation of \eqref{dlmEval}. 

Our approach is to select a maximum lag under consideration, $p_{max}$, and assign each lag $\ell$ a state variable, $s_\ell \in \{0,1\}$, for $\ell=1,\ldots,p_{max}$, to assess whether or not $\psi_\ell$ is included in the model: 
\begin{equation}\label{psiSel}
\mu_t(\tau) = \sum_{\ell=1}^{p_{max}} s_\ell \int \psi_\ell(\tau, u) \mu_{t-\ell}(u) \, du + \epsilon_t(\tau), \end{equation}
which extends \cite{kuo1998variable} and \cite{korobilis2013var} to the FAR($p$) setting.
By averaging over the states $\{s_\ell\}_{\ell=1}^{p_{max}}$, the forecasts of model \eqref{psiSel} are the model-averaged forecasts over the FAR($\ell$) models for $\ell=1,\ldots,p_{max}$. Since we restrict $s_\ell \in\{0,1\}$, rather than strongly shrinking $\psi_\ell$ toward zero, we can substantially improve computational efficiency: at each MCMC iteration, we sample $\{\utwi{\mu}_t\}$ jointly from the FAR($p^*$) extension of the DLM \eqref{dlmEval}, where $p^* = \min\{\ell : s_{\ell+1}=\cdots = s_{p_{max}} = 0\}$ is the largest lag of nonzero autocorrelation. 

The joint distribution of the states  is $[s_1,s_2,\ldots, s_{p_{max}}] = [s_1]  \prod_{\ell=2}^{p_{max}} [s_\ell | s_{\ell-1}, \ldots, s_1]$, where $[s_\ell | s_{\ell-1},\ldots, s_1]$ is the probability that the lag $\ell$ autocorrelation term is included in the model, given whether the autocorrelation terms of the smaller lags $\ell-1,\ldots, 1$ are included in the model. We assume that $s_{\ell} = 0$ implies that $s_k$ is likely also zero for all $k > \ell$, which induces a more parsimonious model. In particular, we use the computationally convenient Markov  assumption $[s_\ell | s_{\ell-1},\ldots, s_1] = [s_\ell | s_{\ell-1} ]$ with a small transition probability for $\mathbb{P}(s_\ell = 1 | s_{\ell-1} = 0 ) = q_{01}$. The reverse transition probability, $\mathbb{P}(s_\ell = 0 | s_{\ell-1} = 1) =q_{10}$, encourages smaller models when it is large. By default, we select $q_{01} = 0.01$, $q_{10} = 0.75$, and complete the joint prior distribution of $\{s_\ell\}_{\ell=1}^{p_{max}}$ with  $\mathbb{P}(s_1 = 1) = 0.9$; for simulated data, the posterior does not appear to be sensitive to these choices. 


\section{Finite-Dimensional Optimality} \label{finiteOpt}
The Gaussian assumptions in model \eqref{dlmEval} provide convenient posterior distributions for MCMC sampling and a useful framework for inference, but are not necessary for model \eqref{farpInt}. 
Suppose we relax the Gaussian assumption to $\epsilon_t \sim \mathcal{SP}(0, K_\epsilon)$, where $\mathcal{SP}(m, K)$ denotes a second-order stochastic process with mean function $m$ and covariance function $K$. Similarly, let $\nu_{i,t}$ be a  mean zero random variable with variance $\sigma_\nu^2$ and let $\mu_1 \equiv \epsilon_1$. Given a finite set of evaluation points, $\mathcal{T}_e \subset \mathcal{T}$, model \eqref{hierfar} implies the distribution-free DLM 
\begin{equation}\label{dlmEvalnon}
\begin{cases}
\utwi{y}_{t} =    \utwi{Z}_t\utwi{\mu} +  \utwi{Z}_t\utwi{\mu}_t + \utwi{\nu}_t,  & \mathbb{E}[\utwi{\nu}_t| \sigma_\nu^2] = 0, \, \mbox{Cov}[\utwi{\nu}_t| \sigma_\nu^2] = \sigma_\nu^2 \utwi{I}_{m_t},\\
\utwi{\mu}_t = \utwi{\Psi}\utwi{Q} \utwi{\mu}_{t-1} + \utwi{\epsilon}_t, 
& \mathbb{E}[\utwi{\epsilon}_t| \utwi{K}_\epsilon]  = 0, \, \mbox{Cov}[\utwi{\epsilon}_t| \utwi{K}_\epsilon]  = \utwi{K}_\epsilon,
\end{cases}
\end{equation}
under the integral approximation \eqref{quad}, where the vectors and matrices are defined as before and $\utwi{\mu}_1 \equiv \utwi{\epsilon}_1$. Since this holds for any finite set of evaluation points $\mathcal{T}_e \subset \mathcal{T}$, we may consider the DLM \eqref{dlmEvalnon} to be a {\it collection} of models indexed by the evaluation points, $\mathcal{T}_e$. The error sequences, $\{\utwi{\nu}_t\}$ and $\{\utwi{\epsilon}_t\}$, are assumed to be uncorrelated, rather than independent. If we additionally assume Gaussianity of $\{\utwi{\nu}_t\}$ and $\{\utwi{\epsilon}_t\}$, then the uncorrelatedness implies independence, and model \eqref{dlmEvalnon} becomes model \eqref{dlmEval}. Extensions for the FAR($p$) models are similar. The results below also hold for time-dependent variances for $\utwi{\nu}_t$ and $\utwi{\epsilon}_t$.



Let $d$ be an estimator of $\delta \in L^2(\mathcal{T})$, and consider the 
squared error loss using the Euclidean norm: $
\mathcal{L}_e({\delta}, {d}) = ( \utwi{\delta} - \utwi{d} )'( \utwi{\delta}  - \utwi{d}),
$
where $\mathcal{L}_e$ is indexed by the set of evaluation points, $\mathcal{T}_e$,  at which $\delta$ and $d$ are evaluated to form the corresponding vectors $\utwi{\delta}$ and $\utwi{d}$. When $\mathcal{T}_e$ is an equally-spaced fine grid on $\mathcal{T}$, the loss function $\mathcal{L}_e$ will approximate the usual loss function for functional data, $\mathcal{L}_{L^2}(\delta, d) = \int (\delta(u) - d(u))^2\, du$, for most reasonable choices of $\delta$ and $d$ (up to a rescaling by $M=|\mathcal{T}_e|$).
In a standard Bayesian analysis, the goal would be to minimize the posterior risk, $\mathbb{E}[\mathcal{L}_e(\delta, d) | \{\utwi{y}_t\}]$, for which the solution is the posterior expectation, $d = \mathbb{E}[\delta | \{\utwi{y}_t\}]$. Indeed, the estimators discussed below minimize the posterior risk under the Gaussian assumptions of model \eqref{dlmEval}. However, by relaxing the distributional assumptions in \eqref{dlmEvalnon} to increase the generality of the model, we no longer have sufficient information to compute posterior distributions or posterior moments. In addition, it is difficult to compare Bayesian and non-Bayesian procedures under the posterior risk, and  most procedures for functional time series modeling are non-Bayesian. Therefore, we consider the overall risk 
$
\mathcal{R}_e(\delta, d) =\mathbb{E}[\mathcal{L}_e(\delta, d)], 
$
which is the expected value of the posterior risk with respect to the sampling distribution. As with the loss function $\mathcal{L}_e$, the risk function $\mathcal{R}_e$  is indexed by the evaluation points, $\mathcal{T}_e$; we seek to minimize $\mathcal{R}_e$  for any choice of $\mathcal{T}_e$.


Let $\mathcal{D}_t = \left\{\utwi{y}_t, \utwi{y}_{t-1},\ldots,\utwi{y}_1\right\} \cup \mathcal{D}_0$ be the information available at time $t$, where $\mathcal{D}_0$ represents the information prior to time $t=1$.  
\begin{theorem} \label{linearOpt}
For \emph{any} finite set of evaluation points $\mathcal{T}_e  \subset  \mathcal{T}$, the \emph{unique best linear predictor} of the conditional random vector $ \utwi{\delta} \sim [\utwi{\delta}|\utwi{Y}, \utwi{\Theta}]$, where $\utwi{\delta}, \utwi{Y} \subseteq \mathcal{D}_T \cup \{\mu_t(\tau): \tau \in \mathcal{T}_e, t=1, \ldots,T\}$ and $\utwi{\Theta} = \{\mu, \sigma_\nu^2, \psi, K_\epsilon\}$, under the risk $\mathcal{R}_e$ and conditional  on model \eqref{hierfar} with the integral approximation \eqref{quad}, is  the conditional expectation $\utwi{\hat{\delta}}(\utwi{Y}|\utwi{\Theta}) \equiv\mathbb{E}[\utwi{\delta}|\utwi{Y}, \utwi{\Theta}]$ as computed under model \eqref{dlmEval}.  
\end{theorem}
The proof of Theorem \ref{linearOpt} is in the Appendix, and extends fundamental results for vector-valued DLMs. The best linear predictors of Theorem \ref{linearOpt} equivalently minimize the risk $\mathcal{R}(\delta, d) = \sup_{\mathcal{T}_e} \mathcal{R}_e(\delta, d)$ among all linear estimators, where the $\sup$ is taken over all finite $\mathcal{T}_e \subset  \mathcal{T}$. The most useful examples of $[\utwi{\delta}|\utwi{Y}, \utwi{\Theta}]$ in Theorem \ref{linearOpt} are the forecasting distributions $[\utwi{y}_{t+h}| \mathcal{D}_t, \utwi{\Theta}]$ and $[\utwi{\mu}_{t+h}|\mathcal{D}_t, \utwi{\Theta}]$ for $h>0$, the smoothing distributions $[\utwi{\mu}_t | \mathcal{D}_T, \utwi{\Theta}]$, and the filtering distributions $[\utwi{\mu}_t|\mathcal{D}_t, \utwi{\Theta}]$, for $t=1,\ldots, T$. 
  Theorem \ref{linearOpt} depends on the observation points $\mathcal{T}_o$ only via the assumption that $\utwi{Z}_t$ is known. In general, we assume  $\mathcal{T}_o \subseteq \mathcal{T}_e$, so $\utwi{Z}_t$ is an incidence matrix and therefore known.  Theorem \ref{linearOpt} does {\it not} require $\mathcal{T}_o$ to become arbitrarily dense in $\mathcal{T}$, and is valid for both sparse and dense designs. For implementation, we compute the relevant expectations within the Gibbs sampling algorithm (see Section A of the web supplement), and then average over the Gibbs sample of $\utwi\Theta$. 
     {\color{blue}  Alternatively, an EM algorithm could be used to estimate the relevant expectations \citep{cressie2011statistics}.} 




There is no intrinsic reason to restrict the estimators to linearity. However, several popular competing methods are linear, and therefore are dominated by the conditional expectations computed from model \eqref{dlmEval} whenever the estimators are distinct. More formally:
\begin{corollary}\label{basisCor}
Consider a basis expansion of the observations $\utwi{y}_t \approx \utwi{B}_t \utwi{\theta}_t$, where $\utwi{B}_t' = (\utwi{b}(\tau_{1,t}), \ldots, \utwi{b}(\tau_{m_t, t})) $, $\utwi{b}$ is a known $J$-dimensional vector of basis functions, and $\utwi{\theta}_t$ is the corresponding $J$-dimensional vector of unknown basis coefficients. If the estimator $\utwi{\hat\theta}_t$ of $\utwi{\theta}_t$ is linear in $\utwi{y}_t$, then estimates or forecasts of the form  $\utwi{H}\utwi{\hat\theta}_t + \utwi{h}$, conditional on the matrix $\utwi{H}$ and the vector $\utwi{h}$, are inadmissible for all $[\utwi{\delta}|\utwi{Y}]$ whenever $\utwi{H}\utwi{\hat\theta}_t + \utwi{h} \ne \utwi{\hat{\delta}}(\utwi{Y}|\utwi{\Theta})$. 
\end{corollary}
The most important application of Corollary \ref{basisCor} is to characterize the inadmissibility of procedures based on FPC scores. In the notation of Corollary \ref{basisCor}, let $\utwi{b}$ be the FPC basis, which we assume is fixed and known. The components of $\utwi{\theta}_t$ correspond to the FPC scores, defined by $\theta_{j,t} = \int \{Y_t(u) - \mu(u)\}b_j(u) \, du = \int \mu_t(u)b_j(u) \, du$.  There are two standard approaches for computing FPC scores: quadrature methods for dense designs absent measurement error, and the PACE procedure of \cite{yao2005functional}, which uses conditional expectations under a Gaussian assumption and applies more generally. 
In both cases, the FPC scores are linear in $\utwi{y}_t$, so Corollary \ref{basisCor} applies. 

Among functional time series methods, the most pertinent procedures are \cite{aue2015prediction} and  \cite{hyndman2007robust}.  \cite{aue2015prediction} provide the more general framework, in which  they compute the best linear predictors for the FPC scores, and then forecast the FPC scores using multivariate time series methods. For time series methods that are {\it linear} in the FPC scores, such forecasts are inadmissible. While \cite{aue2015prediction} undoubtedly provide a simple yet general framework for forecasting  a functional time series, the simulations of Section \ref{sims} confirm the consequence of inadmissibility on forecasting performance.

\begin{corollary}\label{presmooth}
Consider the common functional data pre-processing procedure in which the discrete, noisy observations, $\utwi{y}_t$, are replaced by  estimated functions evaluated on a fine grid, $\utwi{\hat{y}}_t$, and then estimates and forecasts are computed using the functional ``data" $\utwi{\hat{y}}_t$. If $\utwi{\hat{y}}_t$ is linear in $\{\utwi{y}_t\}$,  then any estimator or forecast linear in $\{\utwi{\hat{y}}_t\}$ is inadmissible  for all $[\utwi{\delta}|\utwi{Y}]$ whenever $\utwi{\hat{y}}_t \ne \utwi{\hat{\delta}}(\utwi{Y}|\utwi{\Theta})$.
\end{corollary}
Typically, $\utwi{\hat{y}}_t$ is estimated using splines or kernel smoothers, both of which are linear in $\utwi{y}_t$. 
As an application of Corollary \ref{presmooth}, the simple forecasting method of fitting a VAR to $\utwi{\hat{y}}_t$ evaluated on a grid of points,  conditional on the VAR coefficient matrix, is inadmissible. 


\begin{corollary}\label{interpOpt}
The unique best linear predictor of $[\mu_t(\tau^*) | \mathcal{D}_s]$ for any times $t,s$  and any point $\tau^* \in \mathcal{T}$ is  the corresponding expectation under  model \eqref{dlmEval}. 
\end{corollary}
Model \eqref{dlmEval} achieves the optimality of a  kriging estimator for interpolation of any point $\tau^*\in \mathcal{T}$, simply by adding $\tau^*$ to the evaluation set $\mathcal{T}_e$. 

In practice, we need not include all such $\tau^*$ in $\mathcal{T}_e$: we can estimate the out-of-sample posterior distribution $[\mu_t(\tau^*) | \mathcal{D}_s]$ for $\tau^* \not \in \mathcal{T}_e$ by sampling from the out-of-sample full conditional distribution   $\left[\mu_t(\tau^*) | \{\utwi{\mu}_r\}_{r=1}^T,  \utwi{\Theta}, \mathcal{D}_s\right]$ within the Gibbs sampler, and then averaging over the Gibbs sample of $\{\utwi{\mu}_r\}_{r=1}^T$ and $\utwi{\Theta}$. Let $\utwi{\psi}'(\tau^*) \equiv (\psi(\tau^*, \tau_1),\ldots, \psi(\tau^*, \tau_M))$ and  $\utwi{\phi}'(\tau^*) \equiv (\phi_1(\tau^*),\ldots,\phi_{J_\epsilon}(\tau^*))$.  {\color{blue}  In the special case of model \eqref{hierfar} and using the FDLM \eqref{fdlm}, we have the following computationally efficient alternative for state space imputation:}
\begin{theorem}\label{fdlmKriging}
Suppose $\tau^* \in \mathcal{T}$ such that $\tau^*\not\in\mathcal{T}_e$. Under the FDLM \eqref{fdlm} and conditional  on model \eqref{hierfar} with the integral approximation \eqref{quad},   the out-of-sample full conditional distribution of $\mu_t(\tau^*)$ is $\left[\mu_t(\tau^*) | \{\utwi{\mu}_r\}_{r=1}^T,  \utwi{\Theta}, \mathcal{D}_s\right] 
 \sim N\left(m_t(\tau^*), K_t(\tau^*)\right)$, where   $m_t(\tau^*) =\utwi{\psi}'(\tau^*)\utwi{Q}\utwi{\mu}_{t-1} +   \utwi{\phi}'(\tau^*)\utwi{\tilde\Sigma}_e\utwi{\Phi}'\left(\utwi{\mu}_t - \utwi{\Psi}\utwi{Q}\utwi{\mu}_{t-1}\right)$ and $K_t(\tau^*) =\sigma_\eta^2 +\sigma_\eta^2 \utwi{\phi}'(\tau^*)\utwi{\tilde \Sigma}_e\utwi{\phi}(\tau^*)$. \end{theorem}
The proof of Theorem \ref{fdlmKriging} and extensions for $p>1$ are in Section B.3 of the web supplement. Using Theorem \ref{fdlmKriging}, we can efficiently estimate the out-of-sample posterior distribution $[\mu_t(\tau^*) | \mathcal{D}_s]$ with minimal adjustments to the Gibbs sampling algorithm (see Section A.2 of the web supplement).  {\color{blue}  
Theorem \ref{fdlmKriging} builds upon the approximation in \eqref{quad} and the computational simplifications of the FDLM to produce simple and efficient moment calculations for the full conditional distributions without expanding the dimension of the state vector, $M$.} Note that for implementation purposes, the terms $\utwi{\mu}_t$ and $\utwi{\mu}_{t-1}$ appearing in $m_t(\tau^*)$ are assumed to be sampled from the full conditional distribution $\left[\{\utwi{\mu}_r\}_{r=1}^T | \utwi{\Theta}, \mathcal{D}_s\right]$.



\section{Simulations} \label{sims}


We conducted extensive simulations to evaluate the proposed methods for FAR($p$) relative to several competitive alternatives. We are particularly interested in one-step forecasting and recovery of the FAR kernel $\psi_1$, and in how the associated performance varies with the sample size $T$, the location and number of the observation points $ \tau_{1,t},\ldots,\tau_{m_t,t}$, the kernel $\psi_1$, and the  smoothness of the innovation process $\epsilon_t$. We also assess the performance of the model averaging procedure of Section \ref{psiEst} for $p\in\{1,2\}$, and compare the nonparametric FDLM approach of Section \ref{gaussSection} with a more standard parametric Gaussian process implementation.



\subsection{Sampling Designs}

\begin{figure}[h!]
\begin{center}
\includegraphics[width=0.24\textwidth]{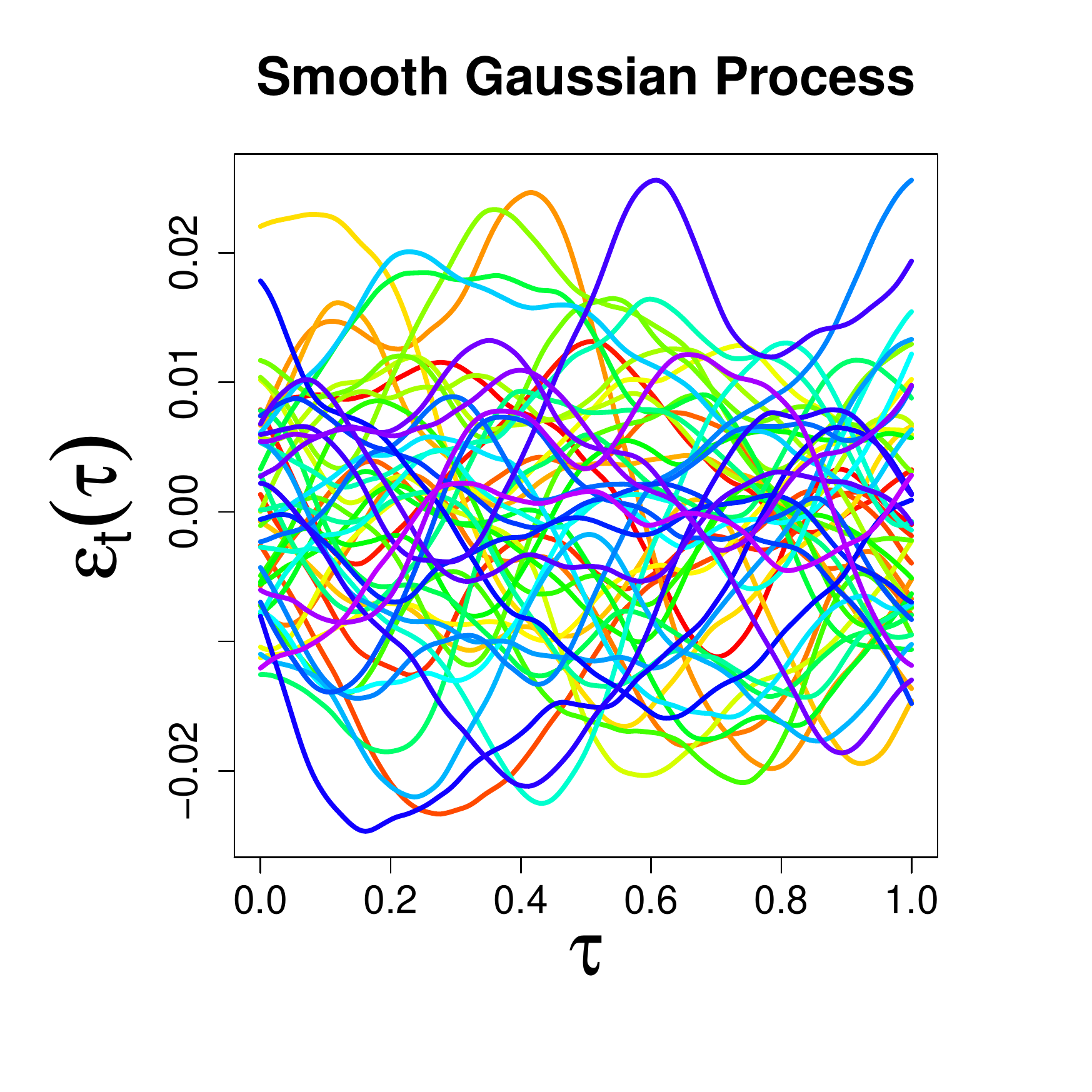}\includegraphics[width=0.24\textwidth]{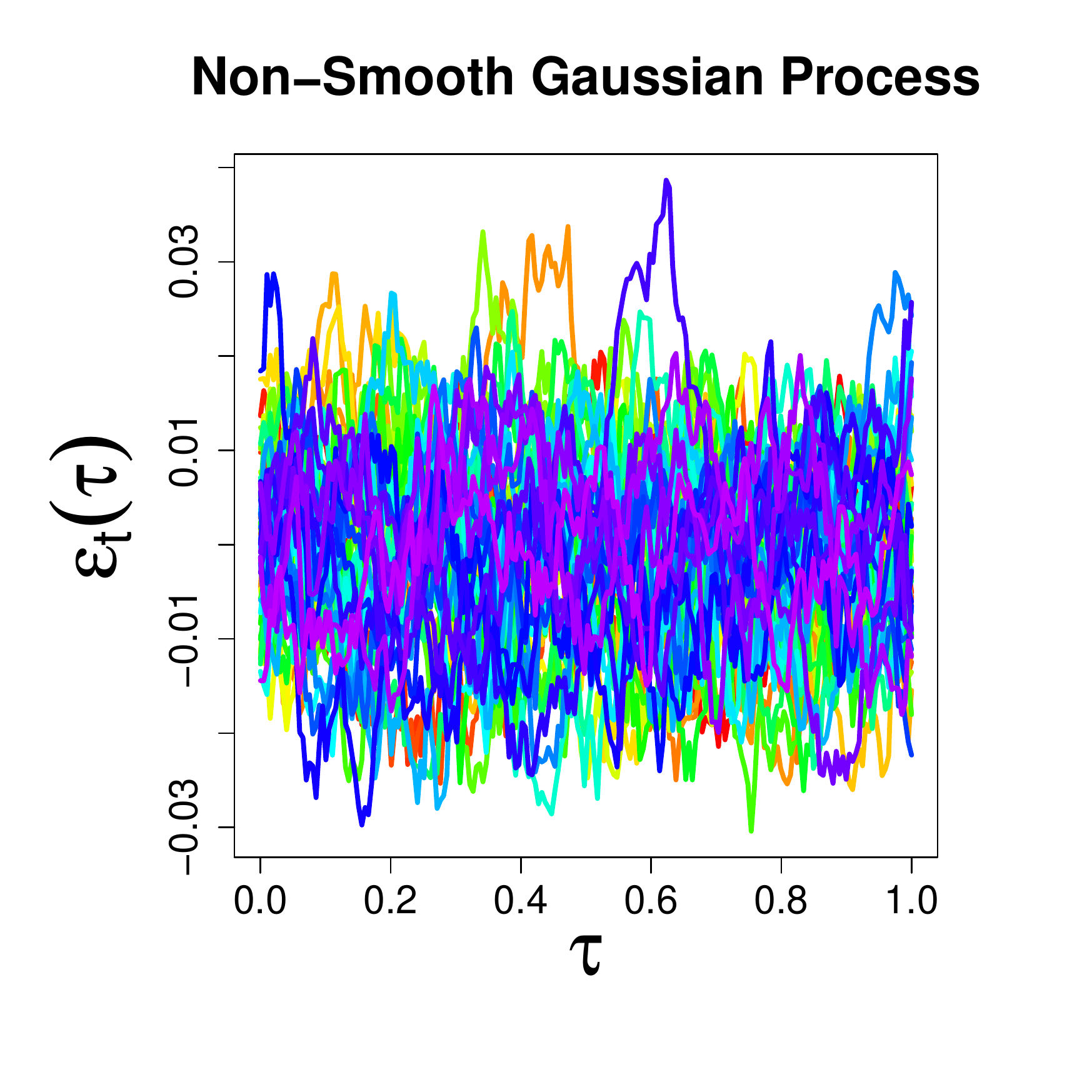}
\includegraphics[width=0.24\textwidth]{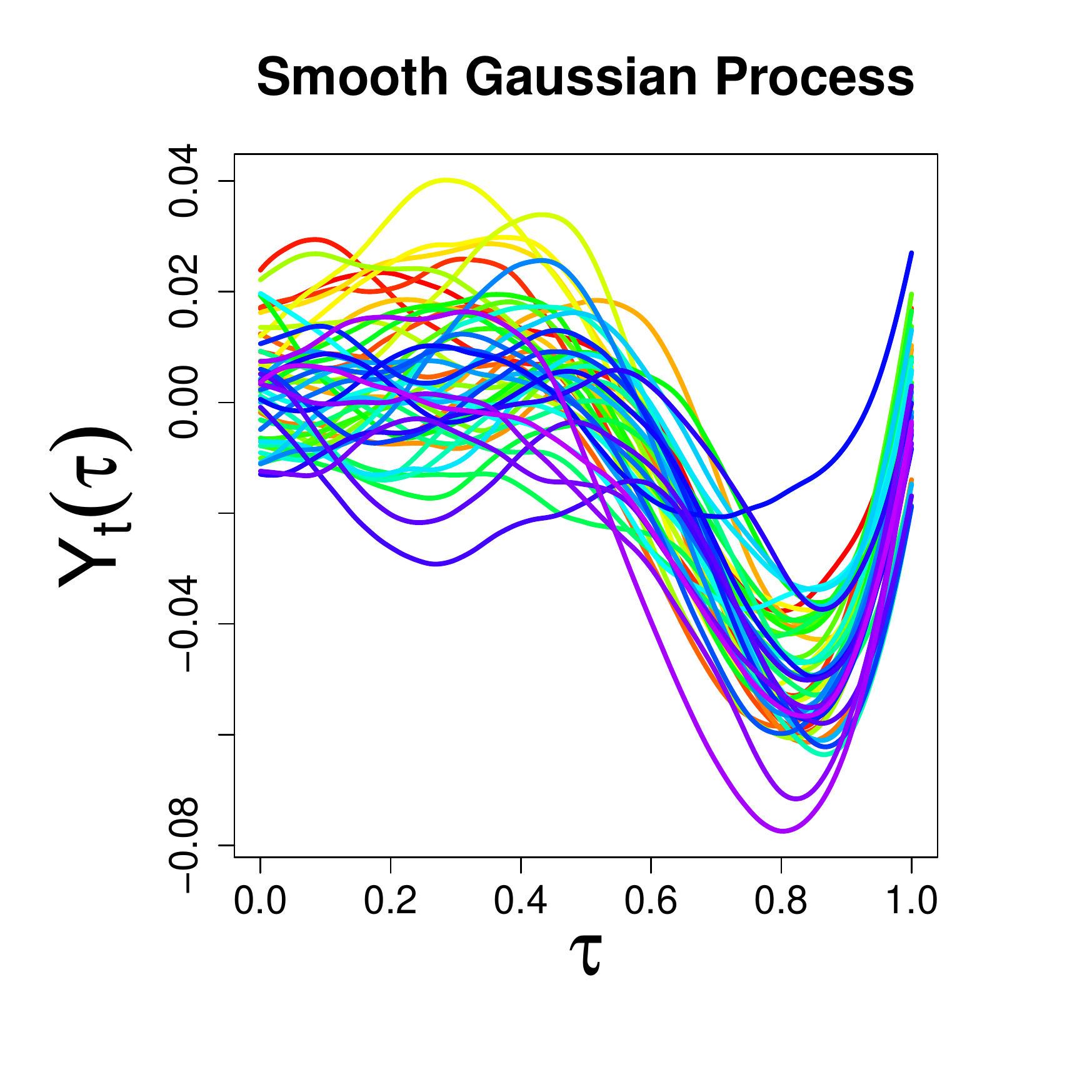}\includegraphics[width=0.24\textwidth]{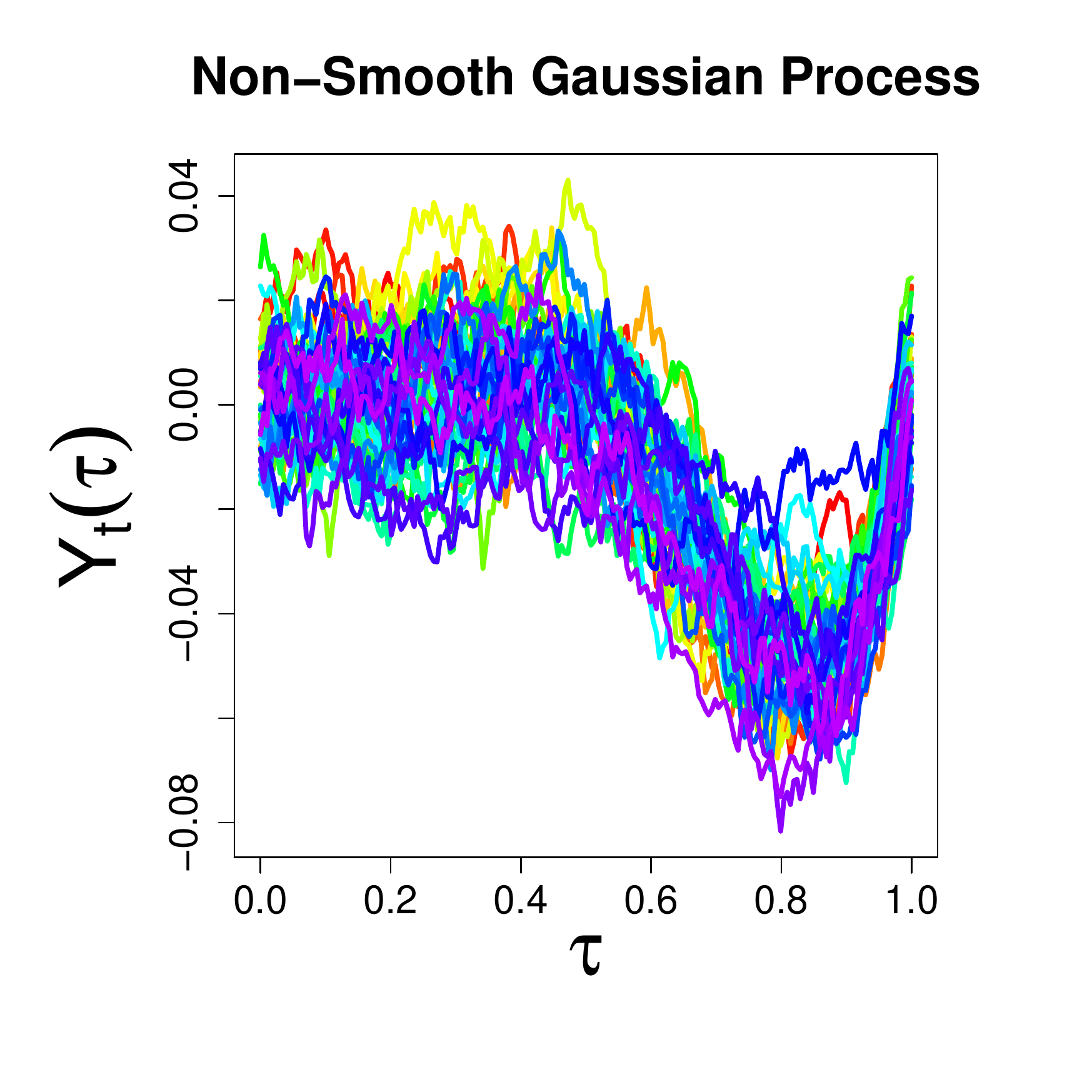}
\caption{Sample paths of $\epsilon_t$ and $Y_t = \mu_t+\mu$ as a function of $\tau$, where $\epsilon_t$ is a Gaussian process with the Mat\'{e}rn correlation function,  $\utwi{\rho} = (\rho_1, 0.1)$, $\sigma = 0.01$, and $Y_t$ is generated using the Bimodal-Gaussian FAR(1) kernel, $t=1,\ldots,T=50$. {\color{blue}  The curves are time-ordered by color (from red/orange to blue/violet).}
{\bf Left to right:} $\epsilon_t(\tau), \rho_1 = 2.5$; $\epsilon_t(\tau), \rho_1 = 0.5$; $Y_t(\tau), \rho_1 = 2.5$; $Y_t(\tau), \rho_1 = 0.5$. 
Note that we do not observe $Y_t$ directly, but rather $y_{i,t} = Y_t(\tau_{i,t}) + \nu_{i,t}$, where $\nu_{i,t} \sim N(0, \sigma_\nu^2)$ is measurement error with $\sigma_\nu =  \sigma/5 =0.002$ and $\mathcal{T}_t=\{\tau_{1,t}, \ldots, \tau_{m_t, t}\}$ are the observation points at time $t$. 
\label{fig:paths}}
\end{center}
\end{figure}

For all simulations, the mean function is $\mu(\tau) = \frac{1}{10}\tau^3 \sin(2\pi\tau)$, which produces the dominate shape in the rightmost panels of Figure \ref{fig:paths}. The measurement errors are  identically distributed for all simulations:  $\nu_{i,t} \stackrel{iid}{\sim}N(0, \sigma_\nu^2)$  with  $\sigma_\nu =0.002$. We vary the sample size from small ($T=50$) to large ($T=350$) for the FAR(1) simulations, and use a moderate sample size ($T=125$) for the FAR(2) simulation. The FAR(1) kernel used for Figure \ref{fig:paths} is the {\it Bimodal-Gaussian} kernel, $\psi(\tau, u) \propto \frac{0.75}{\pi (0.3)(0.4)}\exp\{-(\tau-0.2)^2/(0.3)^2 - (u-0.3)^2/(0.4)^2\}  + \frac{0.45}{\pi (0.3)(0.4)} \exp\{-(\tau-0.7)^2/(0.3)^2 - (u-0.8)^2/(0.4)^2\}$, following \cite{wood2003thin}; {\color{blue} see the web supplement for a plot of the Bimodal-Gaussian kernel.} We also present results for the {\it Linear-$\tau$} kernel, $\psi(\tau,u) \propto \tau$,  and the {\it Linear-$u$} kernel, $\psi(\tau,u)\propto u$. Each kernel is rescaled according to a pre-specified squared norm, $C_{\psi_\ell} = \int\int \psi_\ell^2(\tau,u) \, d \tau \, d u$, with $\sum_{\ell=1}^p C_{\psi_\ell} < 1$ for stationarity. We select $C_{\psi_1} = 0.8$ for the FAR(1) simulations and use $(C_{\psi_1}, C_{\psi_2}) = (0.4, 0.2)$ for the FAR(2) simulation; smaller values of $C_{\psi_\ell}$ produce similar comparative results,  but the forecasting performance deteriorates for all methods. For the innovation process, $\epsilon_t$, we consider both smooth and non-smooth Gaussian processes. We use the covariance function parametrization $K_\epsilon= \sigma^2 R_\rho$,  where $R_\rho$ is the Mat\'{e}rn correlation function $
R_\rho(\tau, u) = \left\{2^{\rho_1-1}\Gamma(\rho_1)\right\}^{-1} \left( ||\tau - u||/\rho_2\right)^{\rho_1} K_{\rho_1}\!\left( ||\tau-u||/\rho_2\right)
$,  $\Gamma(\cdot)$ is the gamma function, $K_{\rho_1}$ is the modified Bessel function of order $\rho_1$, and $\utwi{\rho} = (\rho_1, \rho_2)$ are parameters \citep{matern2013spatial}. We let $\sigma = 0.01$ and $\utwi{\rho} = (\rho_1, 0.1)$, with $\rho_1 = 2.5$ for smooth (twice-differentiable) sample paths and $\rho_1 = 0.5$ for non-smooth (continuous, non-differentiable) sample paths.  

We consider three sampling designs for the observation points: {\it dense}, {\it sparse-random}, and {\it sparse-fixed}. In each case, the set of evaluation points, $\mathcal{T}_e$, is an equally-spaced grid of $M=30$ points on $\mathcal{T}=[0,1]$. 
The {\it dense} design uses $m_t = 25$ equally-spaced observation points on $[0,1]$ for all $t$, for which the results are representative of denser ($m_t \gg 25$) designs and  similar to those of \cite{didericksen2012empirical}; see Section C of the web supplement. The {\it sparse-random} design is generated by first sampling each $m_t$ from a zero-truncated $\mbox{Poisson}\left(5\right)$ distribution, and then sampling $\tau_{1,t}, \ldots, \tau_{m_t,t}$ without replacement from $\mathcal{T}_e$. This is a common design in sparse functional data, in which $m_t$ may be small for some $t$, but $\mathcal{T}_o$ is dense in $\mathcal{T}$. The {\it sparse-fixed} design uses $m_t = 8$ equally-spaced points in $\mathcal{T}$. This is the most challenging design, and one for which multivariate time series methods should be most competitive with functional time series methods.  {\color{blue}  Comparatively, the sparse settings are similar to the dense setting, but with additional missing observations.}

\subsection{Competing Estimators} \label{compete}
Within the proposed framework and using the FDLM of Section \ref{gaussSection} for the innovation covariance function, we compute forecasts for $p=1$ (FDLM-FAR($1$)), and in the FAR(2) simulation, for $p=2$ (FDLM-FAR($2$)) and $p=3$ (FDLM-FAR($3$)). We also compute forecasts using the model averaging procedure with $p_{max} = 4$ (FDLM-FAR($p$)). To assess the performance of the FDLM implementation, we  compute forecasts using model \eqref{dlmEval} with a parametric covariance function for $K_\epsilon = \sigma^2 R_\rho$ (GP-FAR($1$)). We use the Mat\'{e}rn correlation function for $R_\rho$, with $\rho_1 = 2.5$ as in the smooth Gaussian process simulations, and use the priors $\sigma^{-2} \sim \mbox{Gamma}\left(10^{-3}, 10^{-3}\right)$ and $\rho_2 \sim \mbox{Uniform}\left(0, U_{\rho_2} \right)$, where $U_{\rho_2}$ is the maximum value of $\rho_2$ for which the correlation function $R_\rho$ is less than 0.99 for all pairs of evaluation points.  
These models are implemented using the Gibbs sampling algorithm provided in Section A of the web supplement, and estimates are based on 5,000 MCMC simulations after a burn-in of 5,000. For the large sample setting ($T=350$), the mean computation time per 1,000 MCMC simulations was 2.3 minutes for FDLM-FAR($1$) and 4.4 minutes for GP-FAR($1$). {\color{blue} The  computing times are calculated on a 64-bit Windows machine with a 2.40-GHz Intel core i7-4700MQ processor with 8 GB of RAM, and the code is written in \texttt{R}.}

We consider several important competing methods. Let $\utwi{\hat y}_{t+1}$ denote the one-step forecast at time $t$. For baseline comparisons, we use the random-walk (RW) forecast, $\utwi{\hat y}_{t+1} = \utwi{y}_t$, and the mean (Mean) forecast, $\utwi{\hat y}_{t+1} = \utwi{\hat \mu}$, where $\utwi{\hat \mu}$ is a smooth estimate of the mean of $\{\utwi{y}_s\}_{s=1}^{t}$.  {\color{blue}  We estimate $\utwi{\hat \mu}$ using a B-spline basis expansion via the function \texttt{meanfd}() in the \texttt{R} package \texttt{fda} \citep{fdaPackage}}.  
Both estimators are robust against overfitting, and the mean forecast is optimal when $\psi = 0$. We also compute the one-step forecast based on a  VAR(1)  fit to $\{\utwi{y}_s\}_{s=1}^{t}$ (VAR-Y). In the sparse-random design, the observations $\utwi{y}_t$ were used to linear interpolate on $\mathcal{T}_e$ prior to fitting the VAR. In the sparse-fixed design, the VAR was fit to the observation points, and then forecasts for the evaluation points were computed by fitting a spline to the VAR forecasts of the observation points. {\color{blue}  For additional comparisons, we computed forecasts from a simple exponential smoother (SES) applied pointwise to each component of $\bm{y}_t$, i.e., each time series $\{y_{j,t}\}_{t=1}^T$. The SES forecasts are implemented using the $\texttt{ses} $ function in the \texttt{R} package \texttt{forecast} \citep{hyndman2008automatic}, with an identical imputation scheme as VAR-Y.} We also considered two functional data methods. First, we used the Estimated Kernel procedure outlined in \cite{horvath2012inference}, which estimates $\psi_\ell$ in \eqref{farpInt} using FPCs (FAR Classic); we fix $p=1$ for simplicity. This method has well-studied theoretical properties and is a useful baseline for FAR models. Second, we implemented the method of \cite{aue2015prediction}, which we briefly described in Section \ref{finiteOpt}, using a VAR(1) on the FPC scores (VAR-FPC). We compute the FPCs using the \texttt{fda} package {\color{blue}  in \texttt{R}} with B-spline basis functions. To avoid the ill-conditioned estimators discussed in \cite{horvath2012inference}, we regularize via basis truncation, using $8$ equally-spaced interior knots. The number of components is selected to explain at least 95\% of the variability in $\{\utwi{{y}}_t \}$. For the sampling designs considered here, this approach works well. 
Finally, we report the oracle forecast (FAR Oracle) computed using the true one-step forecasts {\color{blue} $ \mathbb{E} \left[\mu_t(\tau) | \{\psi_\ell, \mu_{t-\ell}\}_{\ell=1}^p\right]$} $= \sum_{\ell=1}^p\int \psi_\ell(\tau, u) \mu_{t-\ell}(u) \, du$ within the simulation, {\color{blue}  where $\{\psi_\ell\}_{\ell=1}^p$ are the  FAR kernels from the simulation specification, $\{\mu_t\}$ are the  simulated values of the latent FAR process, and the integral is approximated using the trapezoidal rule with $M=200$ grid points.} The oracle forecast is not actually an estimator, and is unaffected by sparsity or small sample sizes.



We estimate the one-step forecasts $[\utwi{y}_{T+h} | \utwi{y}_{1:(T+h - 1)}]$, $h=1,\ldots, 25,$ for all estimators under consideration, and compare them using the mean squared forecast error
$MSFE_e = \frac{1}{25M} \sum_{h=1}^{25}||\utwi{Y}_{T+h} - \utwi{\hat{Y}}_{T+h}||^2$ where $\utwi{Y}_{T+h} = (Y_{T+h}(\tau_1),\ldots, Y_{T+h}(\tau_M)))'$, which measures the one-step forecasting performance at the {\it evaluation} points, and the mean squared error $MSE_{\psi_1}  = \frac{1}{M^2}\sum_{i=1}^M\sum_{k = 1}^M \{\psi_1(\tau_i, \tau_k) - \hat{\psi}_1(\tau_i, \tau_k)\}^2$, which measures the recovery of the lag-1 kernel $\psi_1$. Because $\mathcal{T}_e$ is relatively dense in $\mathcal{T}$, $MSFE_e$ and $MSE_{\psi_1}$ approximate the integrated squared errors $\int \{Y_{T+h}(u) - \hat{Y}_{T+h}(u)\}^2 \, du$ and $\int \int \{\psi_1(\tau,u) - \hat{\psi}_1(\tau, u)\}^2 \, d \tau \, d u$, respectively. Estimators $\hat{\psi}_1$ are available only for the proposed methods and FAR Classic. For computational convenience in the proposed methods, we update $\{\utwi{\mu}_t\}_{t=1}^{T+h-1}$ using all of the data $\utwi{y}_{1:(T+h - 1)}$, but sample all other parameters only conditional on $\utwi{y}_{1:T}$.  DLM updating algorithms provide recursive one-step forecasts for $\utwi{\mu}_t$, but in general there are no convenient updating algorithms for the other parameters. In practice, this is not a problem, but suggests that our simulation analysis may underestimate the performance of the proposed model.

\subsection{Results} \label{results}
\begin{figure}[h!]
\begin{center}
\includegraphics[width=.47\textwidth]{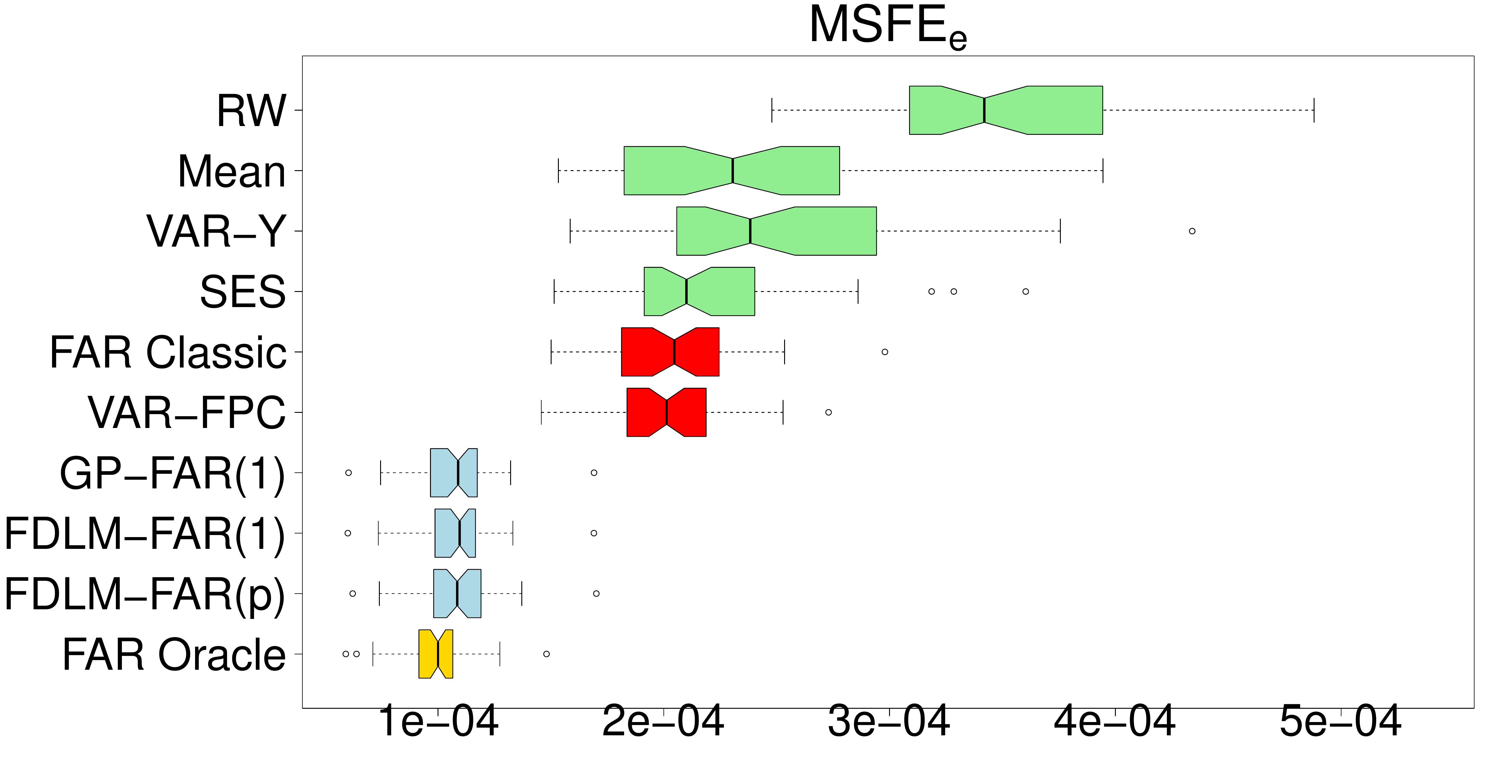}
\includegraphics[width=.47\textwidth]{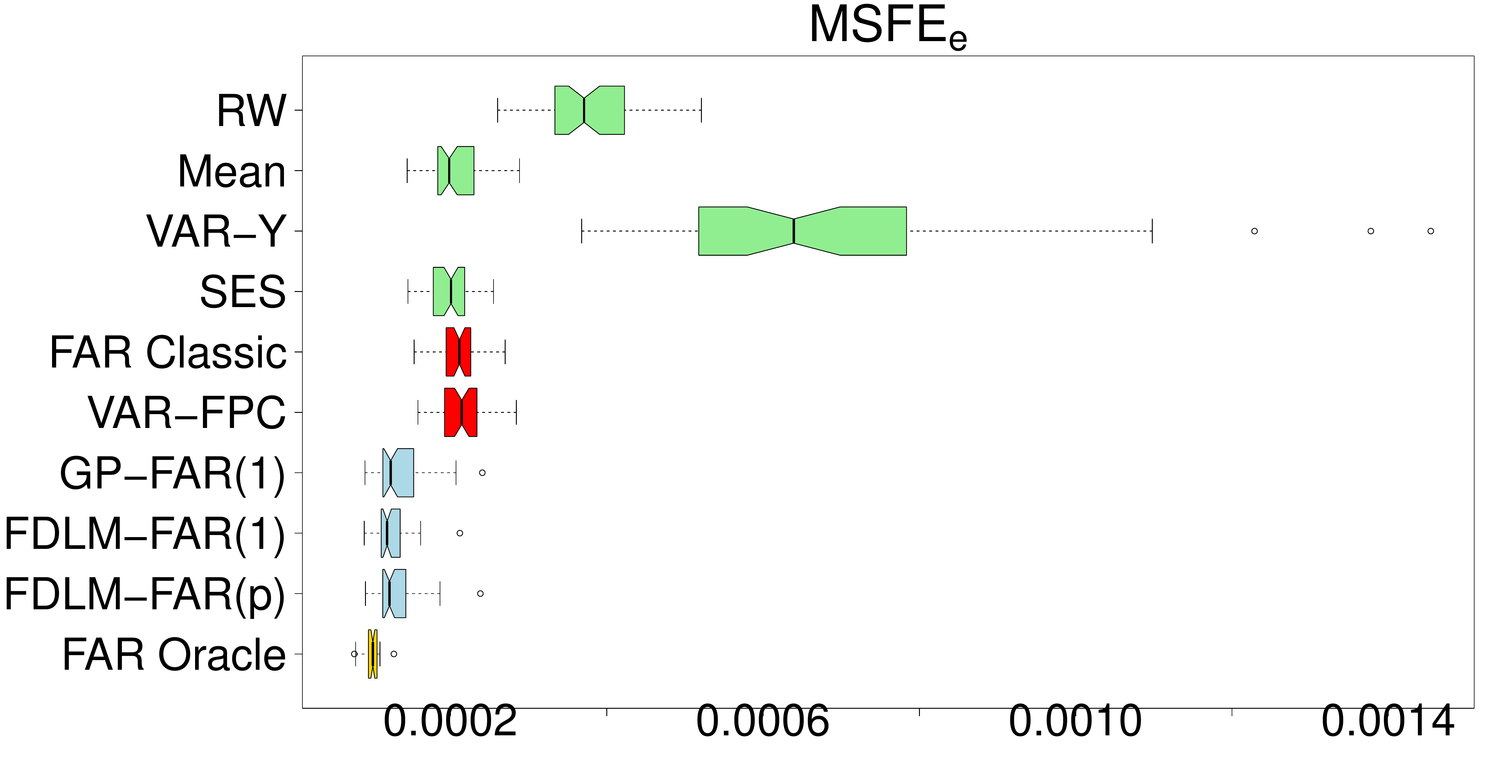}
\includegraphics[width=.47\textwidth]{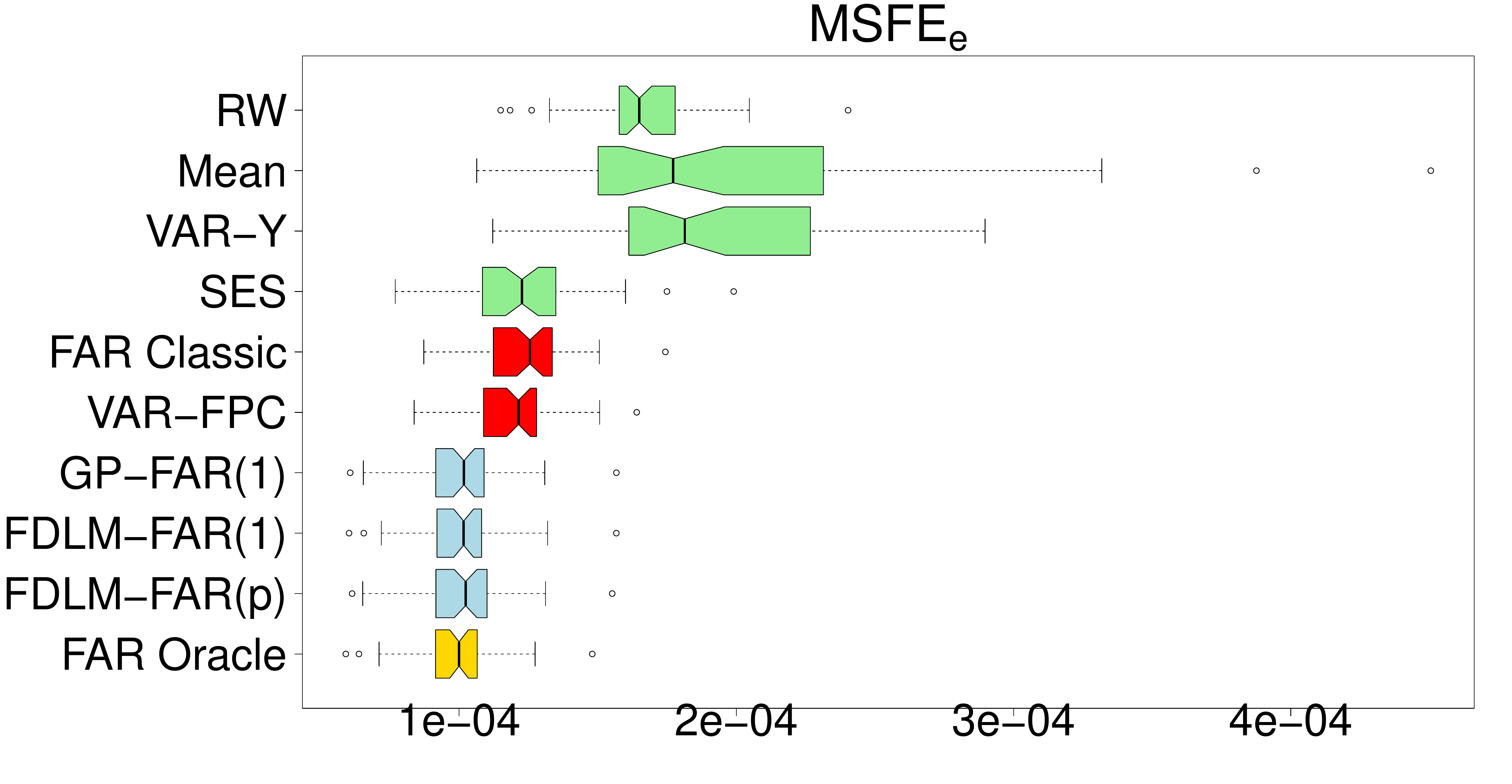}
\includegraphics[width=.47\textwidth]{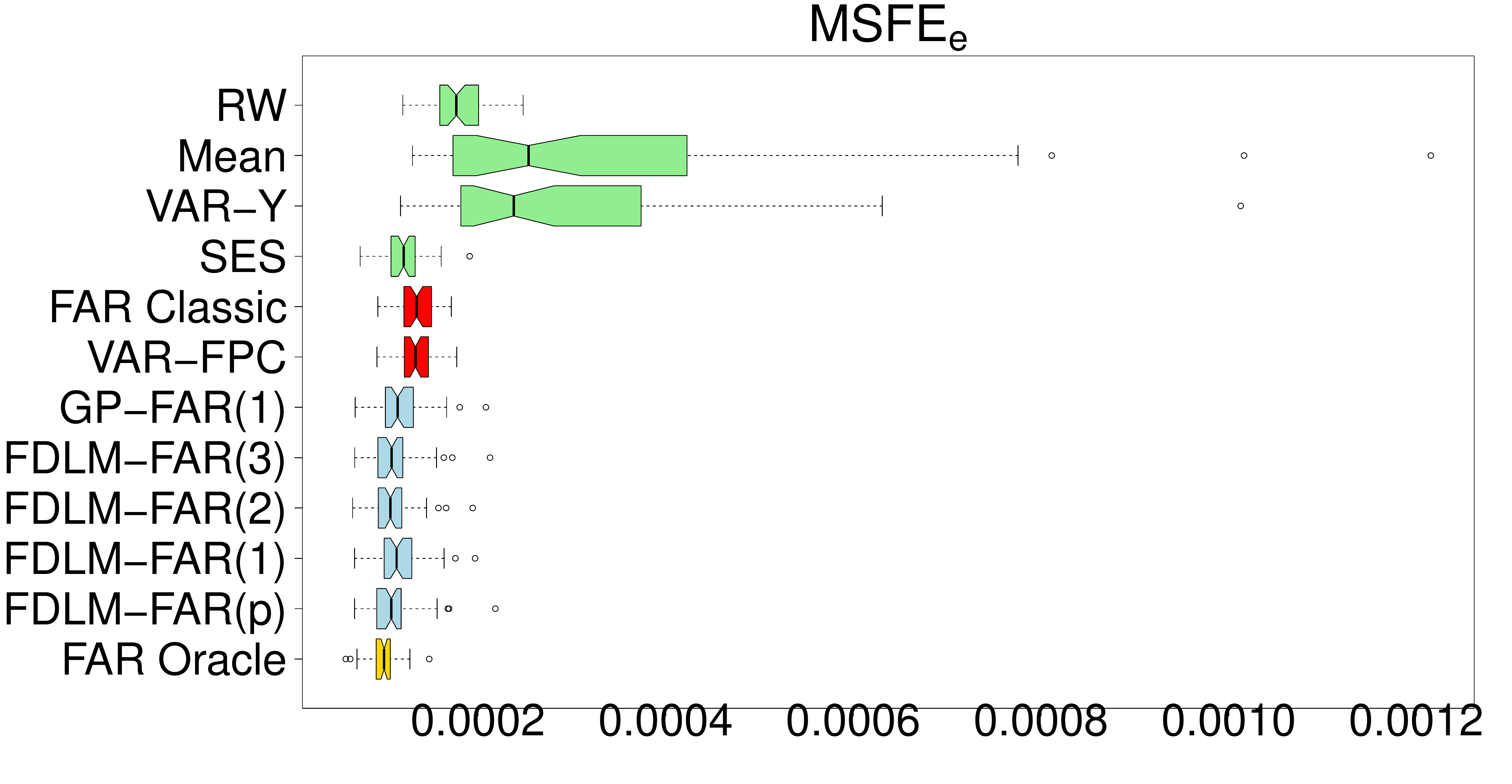}
\caption{$MSFE_e$ under various designs. {\bf Top left:} FAR(1), $T=350$, sparse-random design with the Linear-$u$ kernel and smooth GP innovations. {\bf Top right:}  FAR(1), $T=50$, sparse-random design with the Bimodal-Gaussian kernel and non-smooth GP innovations. {\bf Bottom left:}  FAR(1), $T=350$, sparse-fixed design with the Bimodal-Gaussian kernel and smooth GP innovations. {\bf Bottom right:}  FAR(2), $T=125$, sparse-fixed design with Bimodal-Gaussian and Linear$-\tau$ kernels and smooth GP innovations. The proposed methods provide superior forecasts and nearly achieve the oracle performance, despite the presence of sparsity.  \label{fig:figMSFE}}
\end{center}
\end{figure}

\begin{figure}[h!]
\begin{center}
\includegraphics[width=.47\textwidth]{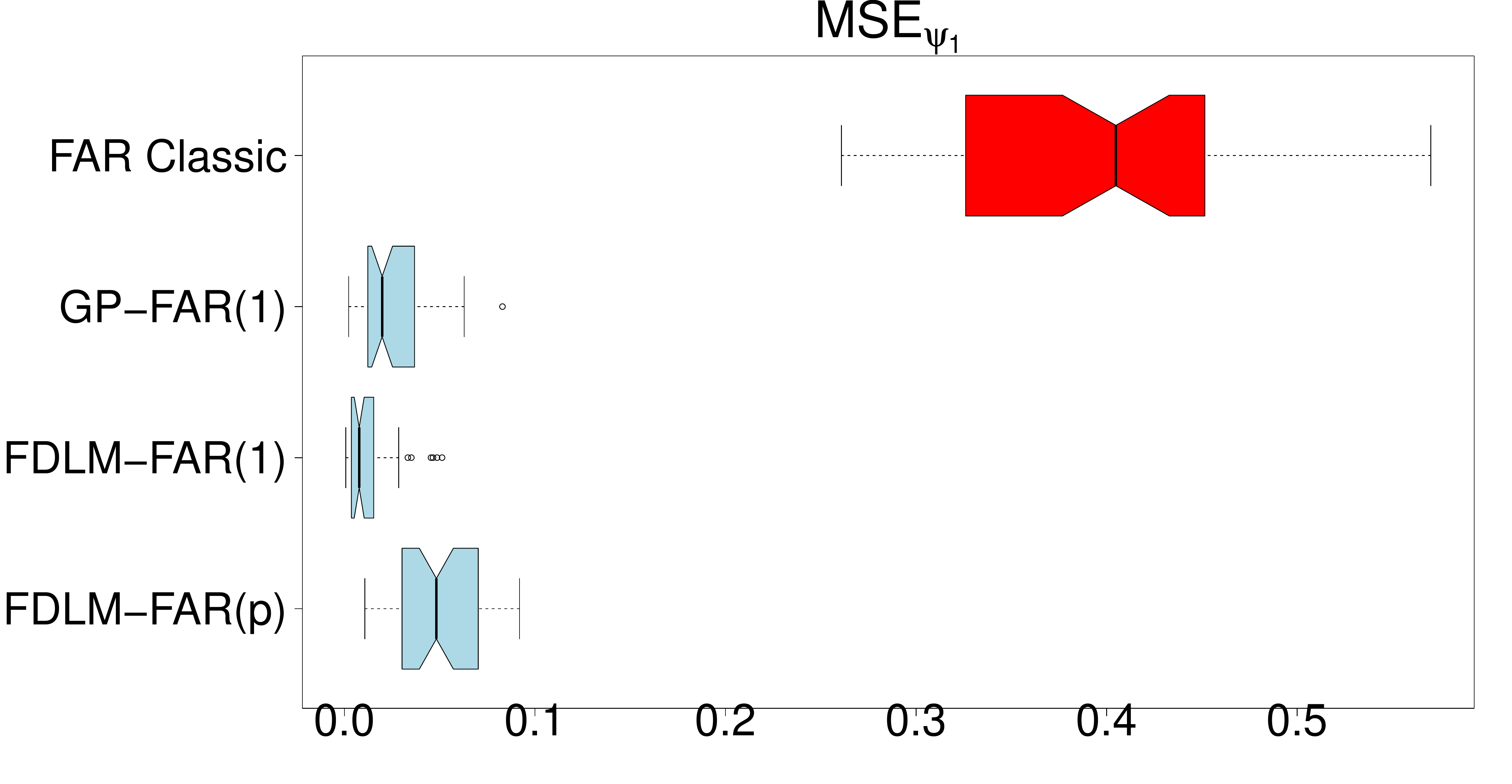}
\includegraphics[width=.47\textwidth]{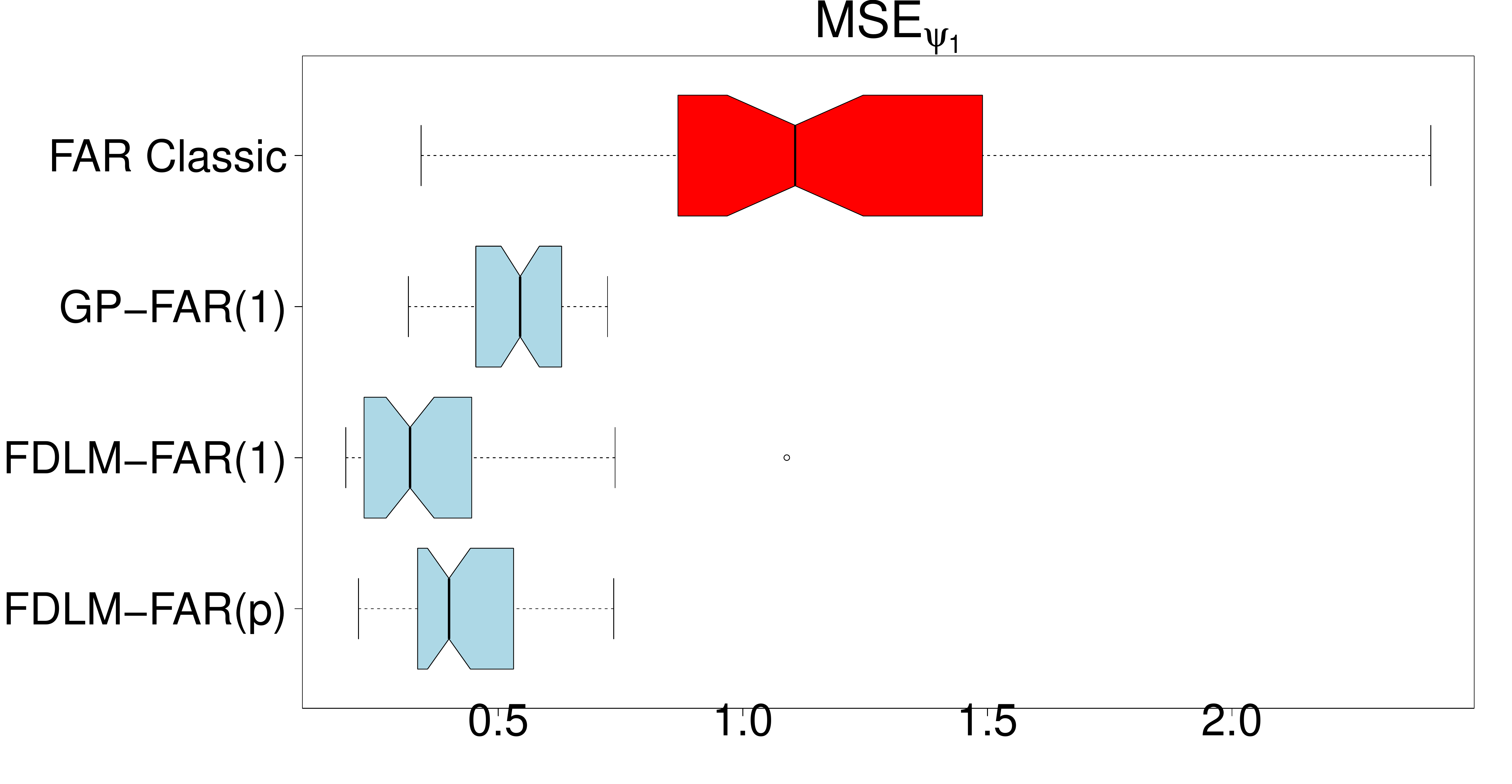}
\includegraphics[width=.47\textwidth]{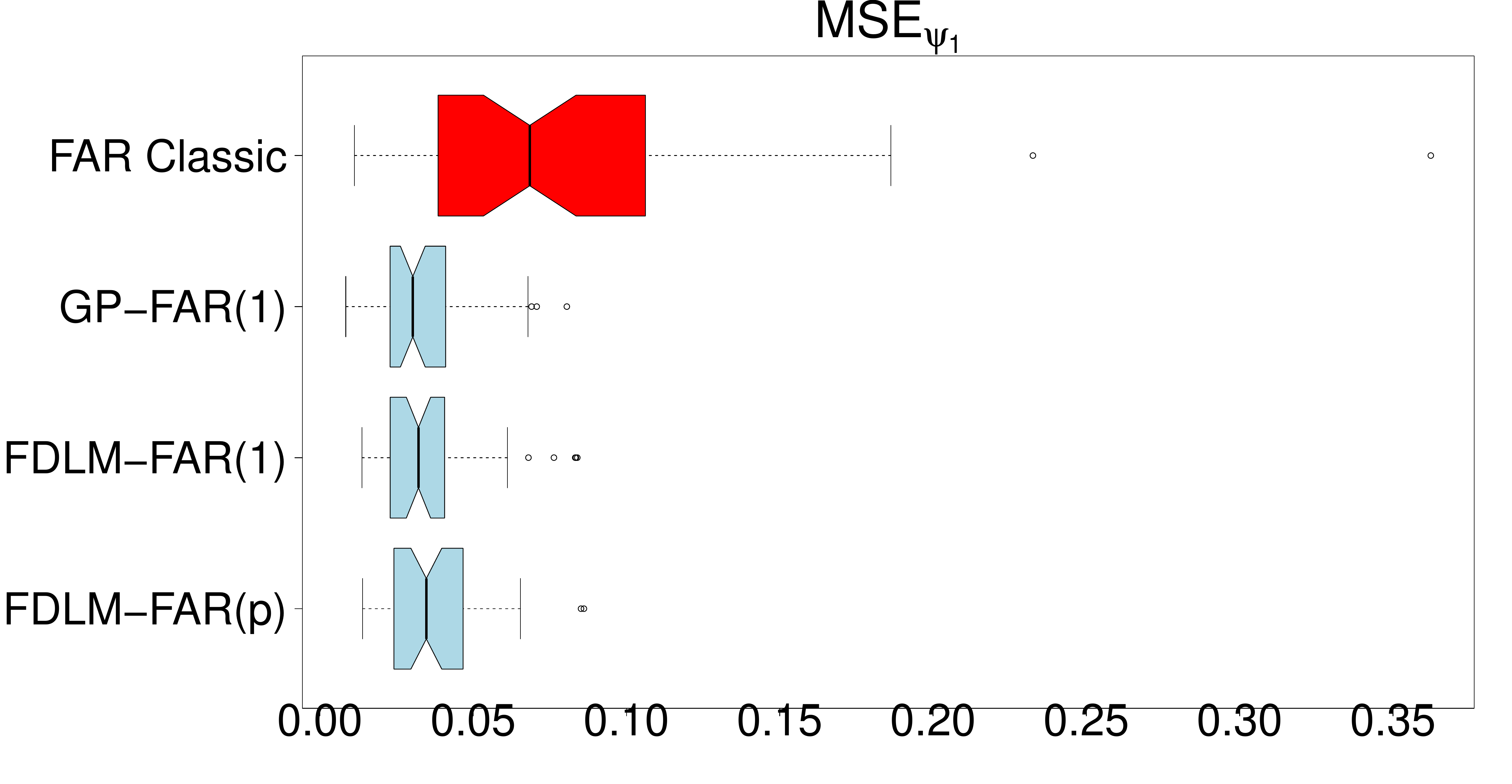}
\includegraphics[width=.47\textwidth]{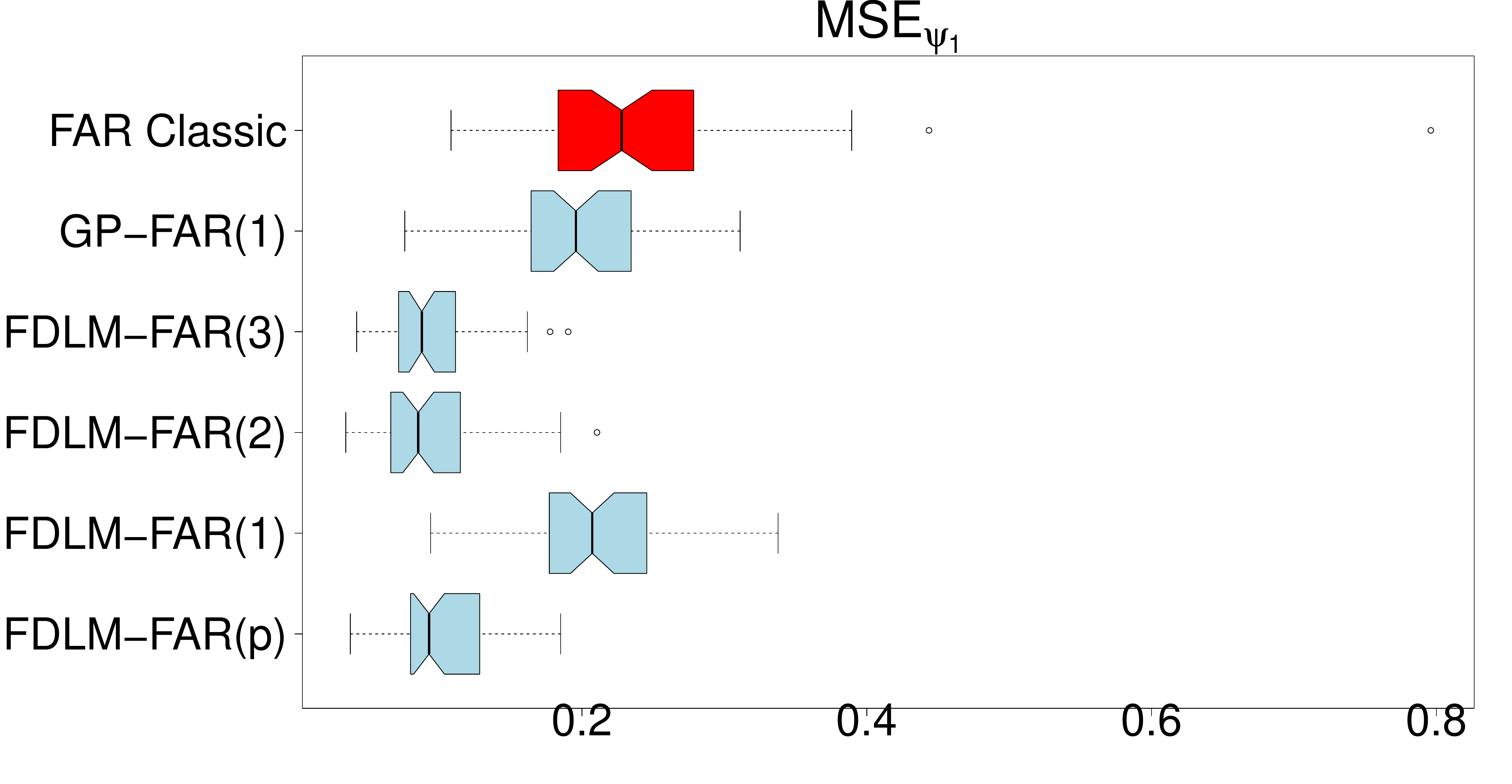}
\caption{$MSE_{\psi_1}$ under various designs. {\bf Top left:} FAR(1), $T=350$, sparse-random design with the Linear-$u$ kernel and smooth GP innovations. {\bf Top right:}  FAR(1), $T=50$, sparse-random design with the Bimodal-Gaussian kernel and non-smooth GP innovations. {\bf Bottom left:}  FAR(1), $T=350$, sparse-fixed design with the Bimodal-Gaussian kernel and smooth GP innovations. {\bf Bottom right:}  FAR(2), $T=125$, sparse-fixed design with Bimodal-Gaussian and Linear$-\tau$ kernels and smooth GP innovations. Estimates of $\psi_1$ are far superior for the proposed methods, including the FAR($p$) with model averaging. 
\label{fig:mseFAR}}
\end{center}
\end{figure}

We computed $MSFE_e$ and $MSE_{\psi_1}$ under a variety of sampling designs, each for $N=50$ simulations, and present the results for a few important cases in Figures \ref{fig:figMSFE}
  and \ref{fig:mseFAR}, respectively. The figures are color-coded: multivariate methods are green, existing functional data methods are red, the proposed methods are blue, and the oracle is gold.

  
For the sparse designs in Figure \ref{fig:figMSFE}, the proposed methods are all superior to the competitors, and in some cases nearly achieve the oracle performance, even though the oracle is unaffected by sparsity.   Figure \ref{fig:mseFAR} shows that the  proposed methods also offer a substantial improvement in $\psi_1$ estimation. Importantly, the  proposed model with model averaging is competitive with the known $p$ model for both forecasting and estimation of $\psi_1$. The model averaging procedure of Section \ref{psiEst} typically identifies the true $p$ with high probability, with a mild tendency to overestimate $p$. However, this behavior is encouraging:  the bottom right panel of Figure \ref{fig:mseFAR}, in which $p=2$, suggests that overestimating the lag (FDLM-FAR($3$)) is preferable to underestimating  the lag (FDLM-FAR($1$), GP-FAR($1$)) for $\psi_1$ estimation. {\color{blue} FDLM-FAR($1$) is competitive with GP-FAR($1$), even when the parametric Gaussian process model assumes the correct (smooth) innovation distribution, which suggests that the FDLM implementation of Section \ref{gaussSection} provides an adequate approximation. 
}
Under the dense design (see Section C of the web supplement), the improvements of the proposed methods over  existing functional data methods are less substantial, and for $T=350$ the functional data methods all nearly achieve the oracle performance. The proposed methods, however, again provide superior recovery of $\psi_1$.  In general, we find that the functional data methods, in particular the proposed approaches,  outperform the multivariate methods, especially in the dense design. 
   We conclude that the proposed methods provide highly competitive forecasts and superior FAR kernel recovery in a wide variety of important settings. 

 

\begin{figure}[h!]
\begin{center}
\includegraphics[width=.95\textwidth]{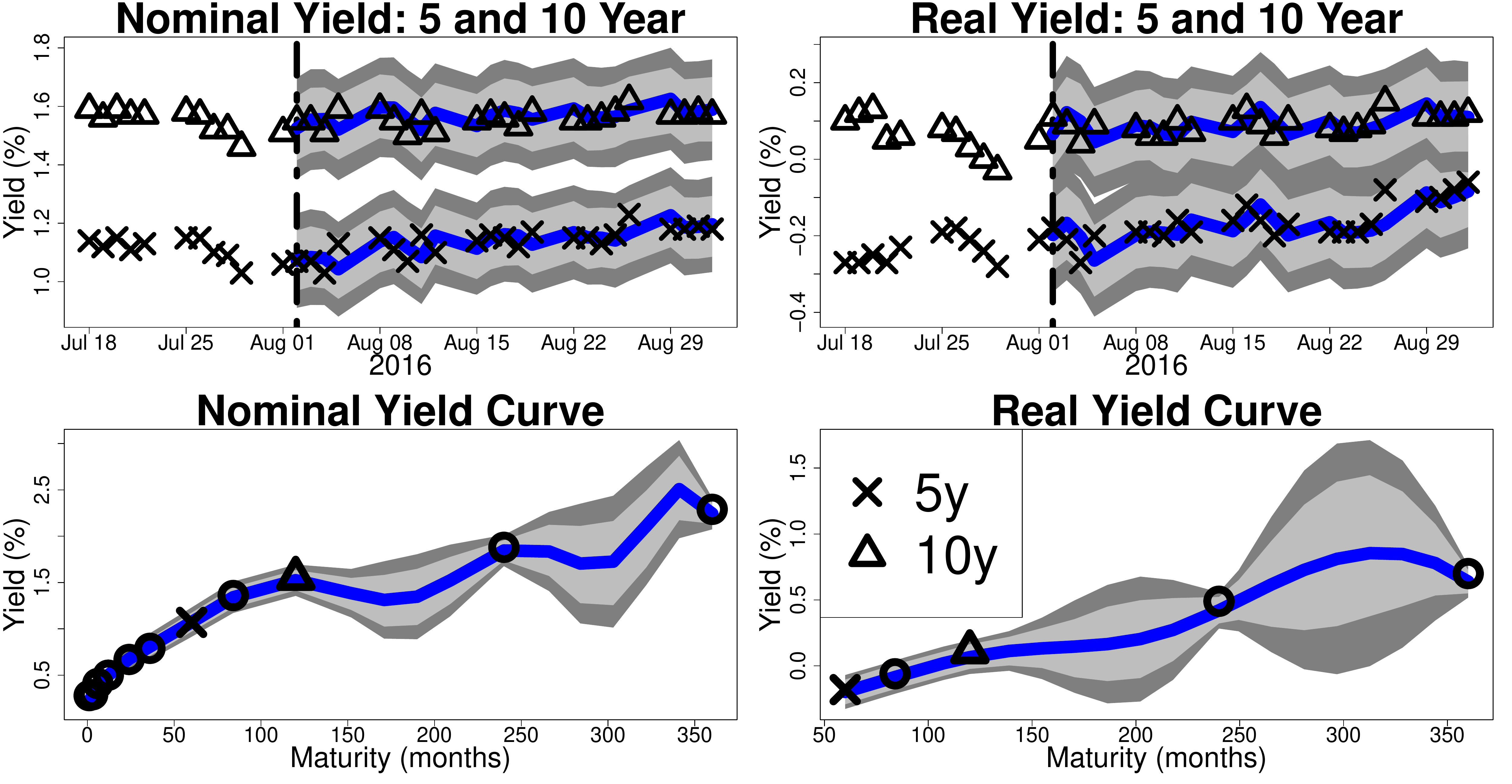}
\caption{One-step nominal ({\bf left}) and real ({\bf right}) yield curve forecasts during 2016. {\bf Top:} Time series of five  ($\times$) and ten ($\triangle$) year observed maturities with one-step forecasts. {\bf Bottom: } Observed (points) and forecast (line) curves on 8/2/16, corresponding to the dotted vertical line in the top panels. Posterior means (blue) and 95\% pointwise and simultaneous prediction bands (light gray and dark gray, respectively) estimated using 10,000 MCMC simulations after a burn-in of 5,000. 
\label{fig:nomReal}}
\end{center}
\end{figure}

{\color{blue}
\section{Forecasting Nominal and Real Yield Curves}\label{yields}
We apply the proposed methods to model and forecast nominal and real yield curves. Yield curves are important in a variety of economic and  financial applications, such as evaluating economic and monetary conditions, pricing  fixed-income securities, generating forward curves, computing inflation premiums, and monitoring business cycles \citep{bolder2004empirical}. In practice, the U.S. real yield curve is estimated using Treasury Inflation-Protected Securities (TIPS), for which payments are adjusted according to the Consumer Price Index for All Urban Consumers (CPI-U) to provide investors with protection against inflation. U.S. nominal and TIPS yield curve data are published daily by the Federal Reserve, which uses actively-traded securities to fit a quasi-cubic spline for each curve. Estimates of the real and nominal yield curves are provided for maturities $\mathcal{T}_t^R = \{60, 84, 120, 240, 360\}$ and $\mathcal{T}_t^N =  \{1,3,6,12,24,36\}\cup \mathcal{T}_t^R$ months, respectively. Notably, the real yield is observed sparsely, and only at longer maturities. The small number of available maturities for real yields presents a challenge for existing functional time series models, and provides an interesting comparison with the nominal yield, for which there are more observed maturities.

To assess the performance of the proposed model, we conducted an extensive forecasting study using daily nominal and real yield curve data. Beginning in 2003, we construct nine consecutive yet non-overlapping 18-month subperiods for estimation ($T \approx 375$); the corresponding starting dates are given in Table \ref{table:rmsfeNominal}. For the month following each estimation period, we compute both one- and five-step (i.e., one business week) forecasts  ($\approx 20$ and $\approx 15$ time points, respectively) for \emph{both} the nominal and real yields. In all cases, the nominal and real yields are modeled separately in order to provide additional comparisons. 

We compute forecasts for the proposed methods by simulating from the forecasting distribution in the DLM \eqref{dlmEval}. For computational convenience, we update only the DLM state parameters $\{\utwi{\mu}_t\}$ during the forecast periods, and fix the remaining parameters based on the estimation periods. We also rescale the observation points $\mathcal{T}_t^R$ and $\mathcal{T}_t^N$ such that $\mathcal{T}_t^R, \mathcal{T}_t^N \subset \mathcal{T}= [0,1]$. We compute forecasts using the competing methods described in Section \ref{sims}, which use all available data for each forecast. For further comparisons, we include two popular parametric yield curve models based on the Nelson-Siegel parametrization \citep{nelson1987parsimonious}:  \citet[DL]{diebold2006forecasting}, which extends the Nelson-Siegel model to the dynamic setting via a two-step estimation procedure, and \citet[DRA]{diebold2006macroeconomy} which is similar to DL, but instead estimates parameters jointly using maximum likelihood within a state space model; see the web supplement for implementation details.

The one- and five-step root mean squared forecasting errors (RMSFEs) for the nominal yields and real yields are in Tables \ref{table:rmsfeNominal} and  \ref{table:rmsfeReal}, respectively. We omit unstable DRA forecasts, as well as multi-step forecasts for FAR Classic, which are unavailable. For both data sets, the proposed methods|denoted FAR(1) and FAR($p$), using the lag selection procedure with $p_{max}=3$|are consistently among the best forecasters for all time periods, and outperform the existing functional data forecasts by a wide margin. For the nominal yields, the FAR(1) provides the best one-step forecasts aggregated across all time periods. For the real yields, the proposed methods are again among the most competitive, particularly in the periods since the financial crisis. Echoing the results in \cite{diebold2006forecasting}, the RW forecast is a difficult benchmark to clear, and the existing functional data models typically fail to do so. By comparison, the proposed FAR forecasts are highly competitive across all time periods and for both the nominal and (sparsely-observed) real yields.

An important feature of the proposed FAR model is the ability to compute exact (up to MCMC error) credible bands for parameters of interest, including forecasts. Such uncertainty quantification is unavailable for the RW forecast, which is our primary competitor in this application. For illustration, we compute pointwise and simultaneous credible bands for  one-step forecasts during August 2016 in Figure \ref{fig:nomReal}. For both nominal and real yields, the credible bands are tighter for shorter maturities and widen in regions of unobserved points, which is appropriate behavior for a nonparametric method. 
}

\section{Concluding Remarks}
The proposed hierarchical FAR($p$) model provides a useful framework for estimation, inference, and forecasting functional time series data. Our model  is especially suited for sparsely or irregularly sampled curves and for curves sampled with non-negligible measurement error, and produces best linear predictors in a general FAR($p$) setting, thereby dominating many competing functional time series models. The  FDLM provides a more flexible, computationally efficient, and stable approach for modeling (innovation) covariance functions. Our model averaging procedure provides an effective solution to the problem of specifying $p$, and produces highly competitive forecasts. The simulation analysis and yield curve application suggest that the proposed FAR($p$) model may improve forecasting and estimation in a wide range of settings, and the efficient MCMC sampling algorithm allows us to perform exact (up to MCMC error and prior misspecification) inference for important parameters.

While we assumed independent factors (and therefore independent innovations) in Section \ref{gaussSection}, we can relax this assumption and allow $\utwi{\Sigma}_e$ to be a stochastic process evolving over time. In this more general framework, the FDLM \eqref{fdlm} can accommodate  stochastic volatility or heavier-tailed distributions for the factors, yet retains the computational simplifications of \eqref{covMatInv} and Theorem \ref{fdlmKriging}.  Letting $\utwi{\Sigma}_t = \mbox{diag}\left(\{\sigma_{j,t}^2\}_{j=1}^{J_\epsilon}\right)$, the (time-dependent) innovation covariance function is $K_{\epsilon_t}(\tau,u)\equiv \mbox{Cov}\left(\epsilon_t(\tau), \epsilon_t(u)\right) = \sum_{j=1}^{J_\epsilon} \sigma_{j,t}^2 \phi_j(\tau)\phi_j(u) + \sigma_\eta^2 \mathbf{1}\{\tau = u\}$.  By modeling each $\{\sigma_{j,t}^2\}_{t=1}^T$ for $j=1,\ldots,J_\epsilon$ with an independent stochastic volatility model (e.g., \citealp{kim1998stochastic}), the time-dependence of $\{\sigma_{j,t}^2\}$ will propagate to the innovation covariance functions, $K_{\epsilon_t}$. Similar modifications can accommodate scale-mixtures of Gaussian distributions for the factors \citep{fernandez2000bayesian} to induce more general distributions for the innovation process, $\{\epsilon_t\}$. These generalizations are particularly important for financial applications, for which stochastic volatility models and heavy-tailed distributions are commonly appropriate.

Future work will investigate more adaptive FAR($p$) models for longer, possibly nonstationary functional time series through stochastic volatility, time-varying $\psi_\ell$, and regime shifts. {\color{blue} Important extensions also include modeling multiple functional responses $Y_t(\tau) \in \mathbb{R}^d$ for $d > 1$, which requires a model for both the auto- and cross-correlations, and incorporating exogenous predictors. In both cases, the DLM framework of \eqref{dlmEval} offers a promising platform for pursuing these extensions.}





\bibliographystyle{apalike}
\bibliography{BFDLMbib}

\begin{thebibliography}{}

\bibitem[Aue et~al., 2015]{aue2015prediction}
Aue, A., Norinho, D.~D., and H{\"o}rmann, S. (2015).
\newblock On the prediction of stationary functional time series.
\newblock {\em Journal of the American Statistical Association},
  110(509):378--392.

\bibitem[Behseta et~al., 2005]{behseta2005hierarchical}
Behseta, S., Kass, R.~E., and Wallstrom, G.~L. (2005).
\newblock Hierarchical models for assessing variability among functions.
\newblock {\em Biometrika}, 92(2):419--434.

\bibitem[Besse et~al., 2000]{besse2000autoregressive}
Besse, P.~C., Cardot, H., and Stephenson, D.~B. (2000).
\newblock Autoregressive forecasting of some functional climatic variations.
\newblock {\em Scandinavian Journal of Statistics}, pages 673--687.

\bibitem[Bolder et~al., 2004]{bolder2004empirical}
Bolder, D., Johnson, G., and Metzler, A. (2004).
\newblock {\em An empirical analysis of the {{C}}anadian term structure of
  zero-coupon interest rates}.
\newblock Bank of Canada.

\bibitem[Bosq, 2000]{bosq2000linear}
Bosq, D. (2000).
\newblock {\em Linear processes in function spaces: theory and applications},
  volume 149.
\newblock Springer Science \& Business Media.

\bibitem[Bosq and Blanke, 2008]{bosq2008inference}
Bosq, D. and Blanke, D. (2008).
\newblock {\em Inference and prediction in large dimensions}, volume 754.
\newblock John Wiley \& Sons.

\bibitem[Cardot et~al., 1999]{cardot1999functional}
Cardot, H., Ferraty, F., and Sarda, P. (1999).
\newblock Functional linear model.
\newblock {\em Statistics \& Probability Letters}, 45(1):11--22.

\bibitem[Chen and Li, 2015]{chen2015adaptive}
Chen, Y. and Li, B. (2015).
\newblock An adaptive functional autoregressive forecast model to predict
  electricity price curves.
\newblock {\em Journal of Business \& Economic Statistics},
  (just-accepted):1--56.

\bibitem[Crainiceanu et~al., 2005]{crainiceanu2005bayesian}
Crainiceanu, C., Ruppert, D., and Wand, M.~P. (2005).
\newblock Bayesian analysis for penalized spline regression using {{WinBUGS}}.
\newblock {\em Journal of Statistical Software}, 14(14):1--24.

\bibitem[Cressie and Wikle, 2011]{cressie2011statistics}
Cressie, N. and Wikle, C.~K. (2011).
\newblock {\em Statistics for spatio-temporal data}.
\newblock John Wiley \& Sons.

\bibitem[Damon and Guillas, 2002]{damon1z2002inclusion}
Damon, J. and Guillas, S. (2002).
\newblock The inclusion of exogenous variables in functional autoregressive
  ozone forecasting.
\newblock {\em Environmetrics}, 13:759--774.

\bibitem[Damon and Guillas, 2005]{damon2005estimation}
Damon, J. and Guillas, S. (2005).
\newblock Estimation and simulation of autoregressive hilbertian processes with
  exogenous variables.
\newblock {\em Statistical Inference for Stochastic Processes}, 8(2):185--204.

\bibitem[Didericksen et~al., 2012]{didericksen2012empirical}
Didericksen, D., Kokoszka, P., and Zhang, X. (2012).
\newblock Empirical properties of forecasts with the functional autoregressive
  model.
\newblock {\em Computational Statistics}, 27(2):285--298.

\bibitem[Diebold and Li, 2006]{diebold2006forecasting}
Diebold, F.~X. and Li, C. (2006).
\newblock Forecasting the term structure of government bond yields.
\newblock {\em Journal of Econometrics}, 130(2):337--364.

\bibitem[Diebold et~al., 2006]{diebold2006macroeconomy}
Diebold, F.~X., Rudebusch, G.~D., and Aruoba, B.~S. (2006).
\newblock The macroeconomy and the yield curve: a dynamic latent factor
  approach.
\newblock {\em Journal of Econometrics}, 131(1):309--338.

\bibitem[Durbin and Koopman, 2002]{durbin2002simple}
Durbin, J. and Koopman, S.~J. (2002).
\newblock A simple and efficient simulation smoother for state space time
  series analysis.
\newblock {\em Biometrika}, 89(3):603--616.

\bibitem[Earls and Hooker, 2014]{earls2014bayesian}
Earls, C. and Hooker, G. (2014).
\newblock Bayesian covariance estimation and inference in latent {{G}}aussian
  process models.
\newblock {\em Statistical Methodology}, 18:79--100.

\bibitem[Fernandez and Steel, 2000]{fernandez2000bayesian}
Fernandez, C. and Steel, M.~F. (2000).
\newblock Bayesian regression analysis with scale mixtures of normals.
\newblock {\em Econometric Theory}, 16(01):80--101.

\bibitem[Gelman, 2006]{gelman2006prior}
Gelman, A. (2006).
\newblock Prior distributions for variance parameters in hierarchical models
  (comment on article by {{B}}rowne and {{D}}raper).
\newblock {\em Bayesian Analysis}, 1(3):515--534.

\bibitem[Hays et~al., 2012]{FDFM}
Hays, S., Shen, H., and Huang, J.~Z. (2012).
\newblock Functional dynamic factor models with application to yield curve
  forecasting.
\newblock {\em The Annals of Applied Statistics}, 6(3):870--894.

\bibitem[Horv{\'a}th and Kokoszka, 2012]{horvath2012inference}
Horv{\'a}th, L. and Kokoszka, P. (2012).
\newblock {\em Inference for functional data with applications}, volume 200.
\newblock Springer Science \& Business Media.

\bibitem[Hyndman and Khandakar, 2008]{hyndman2008automatic}
Hyndman, R. and Khandakar, Y. (2008).
\newblock Automatic time series forecasting: The forecast package for {{R}}.
\newblock {\em Journal of Statistical Software}, 27(1):1--22.

\bibitem[Hyndman and Ullah, 2007]{hyndman2007robust}
Hyndman, R.~J. and Ullah, M.~S. (2007).
\newblock Robust forecasting of mortality and fertility rates: a functional
  data approach.
\newblock {\em Computational Statistics \& Data Analysis}, 51(10):4942--4956.

\bibitem[Kargin and Onatski, 2008]{kargin2008curve}
Kargin, V. and Onatski, A. (2008).
\newblock Curve forecasting by functional autoregression.
\newblock {\em Journal of Multivariate Analysis}, 99(10):2508--2526.

\bibitem[Kaufman et~al., 2010]{kaufman2010bayesian}
Kaufman, C.~G., Sain, S.~R., et~al. (2010).
\newblock Bayesian functional {{ANOVA}} modeling using gaussian process prior
  distributions.
\newblock {\em Bayesian Analysis}, 5(1):123--149.

\bibitem[Kim et~al., 1998]{kim1998stochastic}
Kim, S., Shephard, N., and Chib, S. (1998).
\newblock Stochastic volatility: likelihood inference and comparison with
  {{ARCH}} models.
\newblock {\em The Review of Economic Studies}, 65(3):361--393.

\bibitem[Kokoszka, 2012]{kokoszka2012dependent}
Kokoszka, P. (2012).
\newblock Dependent functional data.
\newblock {\em ISRN Probability and Statistics}, 2012.

\bibitem[Kokoszka and Reimherr, 2013]{kokoszka2013determining}
Kokoszka, P. and Reimherr, M. (2013).
\newblock Determining the order of the functional autoregressive model.
\newblock {\em Journal of Time Series Analysis}, 34(1):116--129.

\bibitem[Korobilis, 2013]{korobilis2013var}
Korobilis, D. (2013).
\newblock {{VAR}} forecasting using {{B}}ayesian variable selection.
\newblock {\em Journal of Applied Econometrics}, 28(2):204--230.

\bibitem[Kowal et~al., 2016]{kowal2014bayesian}
Kowal, D.~R., Matteson, D.~S., and Ruppert, D. (2016).
\newblock A {{B}}ayesian multivariate functional dynamic linear model.
\newblock {\em Journal of the American Statistical Association}.
\newblock (in press).

\bibitem[Kuo and Mallick, 1998]{kuo1998variable}
Kuo, L. and Mallick, B. (1998).
\newblock Variable selection for regression models.
\newblock {\em Sankhy{\=a}: The Indian Journal of Statistics, Series B}, pages
  65--81.

\bibitem[Laurini, 2014]{laurini2014dynamic}
Laurini, M.~P. (2014).
\newblock Dynamic functional data analysis with non-parametric state space
  models.
\newblock {\em Journal of Applied Statistics}, 41(1):142--163.

\bibitem[Mat{\'e}rn, 2013]{matern2013spatial}
Mat{\'e}rn, B. (2013).
\newblock {\em Spatial variation}, volume~36.
\newblock Springer Science \& Business Media.

\bibitem[Neal, 1999]{neal1999regression}
Neal, R.~M. (1999).
\newblock Regression and classification using {{G}}aussian process priors.
\newblock {\em Bayesian Statistics}, 6:475--501.

\bibitem[Neal, 2003]{neal2003slice}
Neal, R.~M. (2003).
\newblock Slice sampling.
\newblock {\em Annals of Statistics}, pages 705--741.

\bibitem[Nelson and Siegel, 1987]{nelson1987parsimonious}
Nelson, C.~R. and Siegel, A.~F. (1987).
\newblock Parsimonious modeling of yield curves.
\newblock {\em Journal of Business}, 60(4):473.

\bibitem[Plummer et~al., 2006]{coda}
Plummer, M., Best, N., Cowles, K., and Vines, K. (2006).
\newblock {{CODA}}: Convergence diagnosis and output analysis for {{MCMC}}.
\newblock {\em R News}, 6(1):7--11.

\bibitem[Ramsay, 2006]{ramsay2006functional}
Ramsay, J.~O. (2006).
\newblock {\em Functional data analysis}.
\newblock Wiley Online Library.

\bibitem[Ramsay et~al., 2014]{fdaPackage}
Ramsay, J.~O., Wickham, H., Graves, S., and Hooker, G. (2014).
\newblock {\em fda: Functional Data Analysis}.
\newblock R package version 2.4.4.

\bibitem[Rasmussen and Williams, 2006]{rasmussen2006gauss}
Rasmussen, C.~E. and Williams, C.~K. (2006).
\newblock Gaussian processes for machine learning.
\newblock {\em The MIT Press}.

\bibitem[Shi and Choi, 2011]{shi2011gaussian}
Shi, J.~Q. and Choi, T. (2011).
\newblock {\em Gaussian process regression analysis for functional data}.
\newblock CRC Press.

\bibitem[Wand and Ormerod, 2008]{wand2008semiparametric}
Wand, M. and Ormerod, J. (2008).
\newblock On semiparametric regression with {{O'Sullivan}} penalized splines.
\newblock {\em Australian \& New Zealand Journal of Statistics},
  50(2):179--198.

\bibitem[West and Harrison, 1997]{westDLM}
West, M. and Harrison, J. (1997).
\newblock {\em Bayesian Forecasting and Dynamic Models}.
\newblock Springer.

\bibitem[Wood, 2003]{wood2003thin}
Wood, S.~N. (2003).
\newblock Thin plate regression splines.
\newblock {\em Journal of the Royal Statistical Society: Series B (Statistical
  Methodology)}, 65(1):95--114.

\bibitem[Yao et~al., 2005]{yao2005functional}
Yao, F., M{\"u}ller, H.-G., and Wang, J.-L. (2005).
\newblock Functional data analysis for sparse longitudinal data.
\newblock {\em Journal of the American Statistical Association},
  100(470):577--590.

\end{thebibliography}

\section*{Appendix}

\subsection*{Priors}
The prior for $\{\utwi{\mu}_t\}_{t=1}^T$ is determined by \eqref{dlmEval}. Let $\utwi{b}_{\psi}$ be a $J_{\psi}$-dimensional vector of cubic B-spline basis functions with $ \min\{|\mathcal{T}_o|/2, 35\} = (J_{\psi} - 4)$ equally-spaced interior knots. The tensor product expansion $\psi_\ell(\tau, u) = \utwi{b}_{\psi}'(\tau) \utwi{\Theta}_{\psi_\ell} \utwi{b}_{\psi}(u) = \left( \utwi{b}_{\psi}'(u) \otimes \utwi{b}_{\psi}'(\tau)\right) \utwi{\theta}_{\psi_\ell}$, where $\utwi{\Theta}_{\psi_\ell}$ is a $J_{\psi} \times J_{\psi}$ matrix of unknown coefficients and  $\utwi{\theta}_{\psi_\ell} =  \mbox{vec}\left(\utwi{\Theta}_{\psi_\ell}\right)$, is computationally convenient for the FAR surfaces $\{\psi_\ell\}_{\ell=1}^p$. 
The Gaussian prior  $\left[\utwi{\theta}_{\psi_\ell}| \lambda_{\psi_\ell}\right] \sim N\left(\utwi{0}, \lambda_{\psi_\ell}^{-1} \utwi{\Omega}_{\psi_\ell}^{-1}\right)$ induces a Gaussian process prior on $\psi_\ell$,  where $\utwi{\Omega}_{\psi_\ell}$ is a penalty matrix and $\lambda_{\psi_\ell}$ is a smoothing parameter. The  standard roughness penalty $\int \int \left\{ \frac{\partial^2}{\partial u_1}\psi_\ell(u_1, u_2) + 2\frac{\partial^2}{\partial u_1\partial u_2}\psi_\ell(u_1, u_2)  +\frac{\partial^2}{\partial u_2}\psi_\ell(u_1, u_2) \right\}  \, d u_1 \, du_2$ can be expressed as  $\utwi{\theta}_{\psi_\ell}' \utwi{\Omega}_{2} \utwi{\theta}_{\psi_\ell}$ for a known singular matrix $\utwi{\Omega}_{2}$. To obtain a proper prior, which is necessary for our model averaging procedure, we  combine the roughness penalty with a nonstationarity penalty: a sufficient condition for stationarity of $ Y_t $ in model \eqref{farpInt} is $\sum_{\ell=1}^p \int\int \psi_\ell^2(\tau,u) \, d\tau \, d u < 1$, which can be expressed as $\sum_{\ell=1}^p \utwi{\theta}_{\psi_\ell}'\utwi{\Omega}_0 \utwi{\theta}_{\psi_\ell} < 1$ where $\utwi{\Omega}_0$ is a known invertible matrix. We use the prior precision matrix $\utwi{\Omega}_{\psi_\ell} = \utwi{\Omega}_2 + \kappa_\ell \utwi{\Omega}_0$, which penalizes roughness of $\psi_\ell$ and provides shrinkage toward stationarity, where the trade-off is determined by $\kappa_\ell$. Simulations suggest that the posterior distribution is not sensitive to the choice of $\kappa_\ell$; we fix $\kappa_\ell = 1$ for the simulations and assume $\log\left(\kappa_\ell\right) \sim N\left(0, 4\right)$ for the application. 
For the smoothing parameter $\lambda_{\psi_\ell}$, we use the half-Cauchy prior of \cite{gelman2006prior}, which provides excellent mixing of the states $\{s_\ell\}$ in the model averaging procedure.  The prior may be expressed hierarchically via the auxiliary variables $\tilde{\lambda}_{\psi_\ell} \sim \mbox{Gamma}\left(\frac{1}{2}, \frac{1}{2}\right)$, $\tilde{\xi}_{\psi_\ell} \sim N\left(0, 10^6\right)$, and $\utwi{\tilde{\theta}}_{\psi_\ell} \sim N\left(\utwi{0}, \tilde{\lambda}_{\psi_\ell}^{-1} \utwi{\Omega}_{\psi_\ell}^{-1}\right)$, with the identification $\utwi{\theta}_{\psi_\ell} = \tilde{\xi}_{\psi_\ell} \utwi{\tilde{\theta}}_{\psi_\ell}$ and $\lambda_{\psi_\ell} = \tilde{\xi}_{\psi_\ell}^{-2}\tilde{\lambda}_{\psi_\ell}$.

 We use the conditionally conjugate inverse-Gamma priors  $\sigma_\nu^{-2},   \sigma_\eta^{-2} \sim \mbox{Gamma}(10^{-3}, 10^{-3})$ for the measurement error precision and the FDLM approximation error precision, respectively. In some cases, we may prefer smoother sample paths of $\mu_t$, but the  paths will not be smooth when $\sigma_\eta^2$ is large. If increasing $J_\epsilon$ is infeasible or undesirable, fixing  $\sigma_\eta^2$ at some small value, such as $\sigma_\eta^2 = 10^{-6}$, often works well, and can be interpreted as a jitter term for computing a valid inverse of $\utwi{K}_\epsilon$ \citep{neal1999regression}. Assuming  the FDLM \eqref{fdlm} for the innovation covariance $\utwi{K}_\epsilon$, the factors are distributed $\utwi{e}_t \stackrel{iid}{\sim}N(\utwi{0}, \utwi{\Sigma}_e)$ with $\utwi{\Sigma}_e = \mbox{diag}\left(\{\sigma_j^2\}_{j=1}^{J_\epsilon}\right)$, although many generalizations are available \citep{kowal2014bayesian}. To enforce the ordering constraints $\sigma_1^2 > \sigma_2^2 > \cdots > \sigma_{J_\epsilon}^2 > 0$, recall that the joint distribution (of the precisions) may be written
$\left[\sigma_1^{-2}, \ldots, \sigma_{J_\epsilon}^{-2}\right] = \left[\sigma_{J_\epsilon}^{-2}\right]  \prod_{j=1} ^{J_\epsilon - 1} \left[\sigma_j^{-2} | \sigma_{j+1}^{-2}, \ldots, \sigma_{J_\epsilon}^{-2}\right]$. A noninformative joint prior that respects the constraints is  fully specified by $\sigma_{J_\epsilon}^{-2} \sim \mbox{Gamma}\left(10^{-3}, 10^{-3}\right)$   and $\left[\sigma_j^{-2} | \sigma_{j+1}^{-2}, \ldots, \sigma_{J_\epsilon}^{-2}\right] = \left[\sigma_j^{-2} | \sigma_{j+1}^{-2}\right] \sim \mbox{Uniform}\left(0, \sigma_{j+1}^{-2} \right)$ for $j=1,\ldots, J_\epsilon-1$. The FLCs are $\phi_j(\tau) = \utwi{b}_\phi'(\tau) \utwi{\xi}_j$, where $\utwi{b}_\phi$ is a low-rank thin plate spline basis with knot locations determined by the quantiles of the observation points, $\mathcal{T}_o$, $\utwi{\xi}_j \sim N(\utwi{0}, \utwi{\Lambda}_{\phi_j})$, and $\utwi{\Lambda}_{\phi_j}^{-1}$ is the low-rank thin plate spline penalty matrix. We follow \cite{wand2008semiparametric} in the singular value decomposition-based diagonalization of the  penalty matrix, so that $\utwi{\Lambda}_{\phi} = \mbox{diag}\left(10^8, 10^8, \lambda_{\phi_j}^{-1},\ldots, \lambda_{\phi_j}^{-1}\right)$, which places a noninformative prior on the constant and linear components of the thin plate spline basis, which are  unpenalized. The prior precision $\lambda_{\phi_j}$ is common among the nonlinear components, and corresponds to the smoothing parameter for the regression function $\phi_j$. Following \cite{gelman2006prior}, we place uniform priors on the standard deviations $\lambda_{\phi_j}^{-1/2} \sim \mbox{Uniform}\left(0, 10^4\right)$, which implies the prior for the precision  $[\lambda_{\phi_j}] \propto \lambda_{\phi_j}^{-3/2} \mathbf{1}\{\lambda_{\phi_j} > 10^{-8}\}$. The  upper bound for the prior standard deviation is selected to match the noninformative components of $\utwi{\Lambda}_{\phi_j}$. The orthonormality constraint is enforced during sampling, which we discuss in Section A of the web supplement. We assume the same parametrization and prior distribution for the mean function, $\mu(\tau) = \utwi{b}_\phi'(\tau) \utwi{\theta}_\mu$.


\subsection*{Proof of Theorem \ref{linearOpt}}

To prove Theorem \ref{linearOpt}, we use the following well-known results:

\begin{proposition}\label{propGauss}
For random vectors $\utwi{\delta}$ and $\utwi{Y}$ with known mean and covariance, the \emph{unique best linear predictor} of $\utwi{\delta}$ given $\utwi{Y}$ is $\mathbb{E}_{\mathbb{G}}[\utwi{\delta}|\utwi{Y}]$, where $\mathbb{E}_{\mathbb{G}}$ is the expectation computed under the assumption that $(\utwi{\delta}',\utwi{Y}')'$ is jointly Gaussian. 
\end{proposition}

\begin{proposition}[\citealp{westDLM}]\label{propDLM}
Under a DLM such as model \eqref{dlmEval},  the random vectors $\utwi{y}_{1:T} = (\utwi{y}_1',\ldots,\utwi{y}_T')'$ and  $\utwi{\mu}_{1:T} = (\utwi{\mu}_1',\ldots,\utwi{\mu}_T')'$ are jointly Gaussian, conditional on the remaining parameters. In addition, all conditionals and marginals of the joint distribution of $(\utwi{y}_{1:T}', \utwi{\mu}_{1:T}')'$ are Gaussian. 
\end{proposition}
Note that we could extend $\utwi{\mu}_{1:T}$ to include $\utwi{\mu}$, which is also a Gaussian random vector. Following Propositions \ref{propGauss} and \ref{propDLM}, the proof of Theorem \ref{linearOpt} is straightforward:
\begin{proof}(Theorem \ref{linearOpt})
Let $\mathcal{T}_e$ be fixed and finite such that $\mathcal{T}_e \subset  \mathcal{T}$. Given this choice of $\mathcal{T}_e$, we can form the DLM \eqref{dlmEvalnon} with the appropriately modified terms. Similarly, we can form the Gaussian DLM \eqref{dlmEval}.  Proposition \ref{propDLM} implies that $(\utwi{y}_{1:T}', \utwi{\mu}_{1:T}')'$ under model \eqref{dlmEval} and conditional on $\utwi{\Theta}$ is jointly Gaussian. Therefore, for any $\utwi{\delta}, \utwi{Y} \subseteq \mathcal{D}_T \cup \{\mu_t(\tau): \tau \in \mathcal{T}_e, t=1,\ldots,T\}$, i.e., any subvectors of $(\utwi{y}_{1:T}', \utwi{\mu}_{1:T}')'$, the distribution of $[\utwi{\delta} | \utwi{Y}, \utwi{\Theta}]$ is Gaussian. Proposition \eqref{propGauss} implies that $\utwi{\hat{\delta}}(\utwi{Y}|\utwi{\Theta}) \equiv \mathbb{E}[\utwi{\delta} | \utwi{Y}, \utwi{\Theta}]$, computed under the Gaussian DLM \eqref{dlmEval}, is the unique best linear predictor of $[\utwi{\delta} | \utwi{Y}, \utwi{\Theta}]$ under the DLM \eqref{dlmEvalnon}. 
\end{proof}

 \begin{table}[ht!]
\centering
\begin{tabular}{rc|ccc|cc|cc|cc}
\multicolumn{11}{c}{Nominal Yields: $h$-Step Root Mean Squared Forecast Errors (RMSFEs)} \\ \hline 
&$h$ & RW &  Mean &  VAR-Y & DL &  DRA  & FAR Classic &  VAR-FPC &  FAR(1) & FAR($p$)    \\
\hline  
\multirow{2}{*}{2/03} & $1$ & 0.0488 & 0.4554 & \emph{0.0487} & 0.1218 & 0.1440 & 0.1641 & 0.1631 & 0.0498 & 0.0516 \\ 
  & $5$ & 0.0966 & 0.4369 & 0.0904 & 0.1409 & 0.8221 & - & 0.1941 & \emph{0.0879} & 0.1002 \\ 
  \hline
\multirow{2}{*}{8/04} & $1$ & 0.0253 & 1.1079 & \emph{0.0252} & 0.0877 & - & 0.1113 & 0.1127 & 0.0281 & 0.0281 \\ 
 & $5$ & 0.0525 & 1.1279 & \emph{0.0383} & 0.0953 & - & - & 0.1435 & 0.0412 & 0.0505 \\ 
    \hline
\multirow{2}{*}{2/06} & $1$ & 0.1710 & 0.5408 & 0.1809 & 0.2206 & - & 0.3349 & 0.3334 & 0.1682 & \emph{0.1673} \\ 
  & $5$ & \emph{0.4534} & 0.5971 & 0.5885 & 0.4927 & - & - & 0.5928 & 0.4680 & 0.4627 \\ 
\hline
\multirow{2}{*}{8/07} & $1$& 0.0833 & 1.3125 & 0.0860 & 0.1817 & 0.1854 & 0.1168 & 0.1173 & 0.0806 & \emph{0.0793} \\ 
& $5$ & 0.1345 & 1.3146 & 0.1402 & 0.2099 & 0.2998 & - & 0.1292 & 0.1537 & \emph{0.1233} \\ 
  \hline
\multirow{2}{*}{2/09} & $1$  & \emph{0.0487} & 0.5268 & 0.0517 & 0.1376 & 0.0917 & 0.1406 & 0.1398 & 0.0488 & 0.0760 \\ 
 & $5$ & \emph{0.0894} & 0.5560 & 0.1227 & 0.1872 & 0.1451 & - & 0.1990 & 0.1323 & 0.2608 \\ 
 \hline
\multirow{2}{*}{8/10} & $1$ & 0.0344 & 0.5063 & 0.0333 & 0.1920 & 0.0878 & 0.0551 & 0.0554 & \emph{0.0291} & 0.0292 \\ 
 & $5$  & 0.0583 & 0.4999 & 0.0603 & 0.1950 & 0.1356 & - & 0.0724 & \emph{0.0452} & 0.0495 \\ 
\hline
\multirow{2}{*}{2/12} & $1$   & 0.0383 & 0.5329 & 0.0384 & 0.0953 & 0.1915 & 0.0464 & 0.0463 & 0.0312 & \emph{0.0311} \\ 
& $5$& 0.0951 & 0.5522 & 0.0915 & 0.1240 & 0.2476 & - & 0.0989 & 0.0760 & \emph{0.0734} \\ 
 \hline
\multirow{2}{*}{8/13} & $1$  & 0.0463 & 0.4169 & \emph{0.0443} & 0.0621 & 0.0692 & 0.0634 & 0.0644 & 0.0547 & 0.0676 \\ 
& $5$  & 0.1210 & 0.3842 & 0.1104 & 0.1423 & 0.1448 & - & \emph{0.1100} & 0.1208 & \emph{0.1100}\\ 
\hline
\multirow{2}{*}{2/15} & $1$ & 0.0329 & 0.3085 & 0.0320 & 0.1125 & 0.1001 & 0.0594 & 0.0606 & \emph{0.0305} & 0.0321 \\ 
    & $5$& 0.0420 & 0.3080 & 0.0403 & 0.1149 & 0.1202 & - & 0.0697 & \emph{0.0393} & 0.0441 \\ 
     \hline
\end{tabular}
\caption{$h$-step RMSFEs for nominal yields, grouped (left to right) by multivariate methods, parametric yield curve models, existing functional data methods, and proposed hierarchical FAR methods. The minimum RMSFE in each row is italicized. 
\label{table:rmsfeNominal}}
\end{table}
\begin{table}[h!]
\centering
\begin{tabular}{rc|ccc|cc|cc|cc}
\multicolumn{11}{c}{Real Yields: $h$-Step Root Mean Squared Forecast Errors (RMSFEs)} \\ \hline 
& $h$ & RW &  Mean &  VAR-Y & DL &  DRA  & FAR Classic &  VAR-FPC &  FAR(1) & FAR($p$)    \\
\hline  
\multirow{2}{*}{2/03} & $1$ & \emph{0.0490} & 0.1629 & 0.0504 & 0.0499 & 0.0492 & 0.1366 & 0.1329 & 0.0509 & 0.0572 \\ 
& $5$ & 0.1001 & 0.1585 & 0.1040 & 0.1017 & 0.1128 & - & 0.1525 & \emph{0.0967} & 0.1110 \\ 
 \hline
\multirow{2}{*}{8/04} & $1$ & 0.0331 & 0.3827 & 0.0337 & 0.0353 & 0.0528 & 0.0431 & 0.0440 & 0.0331 & \emph{0.0326} \\ 
& $5$  & 0.0724 & 0.3924 & 0.0707 & 0.0792 & 0.1690 & - & 0.0721 & 0.0679 &  \emph{0.0651} \\ 
  \hline
\multirow{2}{*}{2/06} & $1$ & 0.0429 & 0.1089 & 0.0428 & 0.0448 & 0.0453 & 0.0529 & 0.0533 & \emph{0.0424} & \emph{0.0424} \\ 
  & $5$  & 0.0934 & 0.1082 & 0.0858 & 0.0957 & 0.1362 & - & 0.0920 & 0.0852 &  \emph{0.0835} \\ 
\hline
\multirow{2}{*}{8/07} & $1$ & \emph{0.0802} & 0.2150 & 0.0896 & 0.0944 & 0.1979 & 0.1212 & 0.1202 & 0.0898 & 0.0880 \\ 
& $5$& \emph{0.1866} & 0.2309 & 0.2268 & 0.2504 & 1.1843 & - & 0.1916 & 0.2051 & 0.1980 \\ 
 \hline
\multirow{2}{*}{2/09} & $1$  & \emph{0.0519} & 0.5162 & 0.0544 & 0.0643 & 0.1229 & 0.0736 & 0.0749 & 0.0526 & 0.0541 \\ 
& $5$ & \emph{0.0798} & 0.5262 & 0.1100 & 0.1092 & 0.3606 & - & 0.1092 & 0.0992 & 0.1046 \\ 
\hline
\multirow{2}{*}{8/10} & $1$ & 0.0490 & 0.7836 & 0.0492 & 0.0591 & 0.0663 & 0.0800 & 0.0762 & 0.0488 & \emph{0.0486} \\ 
& $5$ & 0.0735 & 0.7845 & 0.0787 & 0.0794 & 0.1815 & - & 0.0959 &  \emph{0.0727} & 0.0744 \\ 
\hline
\multirow{2}{*}{2/12} & $ 1$  & \emph{0.0602} & 0.8838 & 0.0612 & 0.0675 & 0.1492 & 0.0906 & 0.0853 & 0.0610 & 0.0608 \\ 
& $5$& 0.1845 & 0.9250 & 0.1958 & 0.1897 & 1.7442 & - & 0.2034 &  \emph{0.1840} & 0.1846 \\ 
\hline
\multirow{2}{*}{8/13} & $1$  & 0.0526 & 0.3242 & 0.0506 & 0.0736 & - & 0.0613 & 0.0610 & 0.0500 & \emph{0.0492} \\ 
& $5$ & 0.1551 & 0.2981 & 0.1278 & 0.1380 & - & - & 0.1246 & 0.1407 &  \emph{0.1239} \\ 
\hline
\multirow{2}{*}{2/15} & $1$ & 0.0328 & 0.3088 & 0.0327 & 0.0439 & 0.1529 & 0.0776 & 0.0779 & \emph{0.0325} & 0.0336 \\ 
 & $5$& 0.0489 & 0.3104 & 0.0521 & 0.0562 & - & - & 0.0816 &  \emph{0.0466} & 0.0543 \\ 
     \hline
\end{tabular}
\caption{$h$-step RMSFEs for real yields, grouped (left to right) by multivariate methods, parametric yield curve models, existing functional data methods, and proposed hierarchical FAR methods. The minimum RMSFE in each row is italicized. 
\label{table:rmsfeReal}}
\end{table}

 \clearpage\LARGE\noindent {\bf Web Supplement}\large
 
 \appendix

\section{Initialization and MCMC Sampling Algorithm}

\subsection{Initialization}\label{initSection}
We initialize the unknown functions using splines and the remaining parameters using conditional maximum likelihood estimators. We first estimate  $\utwi{\mu}$ as a smooth mean of $\{\utwi{y}_t\}_{t=1}^T$, evaluated at $\mathcal{T}_e$. Next, we estimate  each $\utwi{\mu}_t$ by fitting a spline to $\utwi{y}_t - \utwi{Z}_t\utwi{\mu}$ for $t=1,\ldots,T$ using  the \texttt{R} function \texttt{smooth.spline}. Since sparse observation points may lead to unstable initializations of $\utwi{\mu}_t$, we compute the median degrees of freedom implied by the spline fits for $t=1,\ldots,T$, and then recompute the splines for $t=1,\ldots,T$ using this common degrees of freedom parameter. 
Conditional on these estimates, we estimate $\sigma_\nu^2$, $\{\utwi{\theta}_{\psi_1},\ldots, \utwi{\theta}_{\psi_p}\}$, and $\{{\lambda}_{\psi_1},\ldots, {\lambda}_{\psi_p}\}$ using the maximum likelihood estimators, and initialize $\utwi{\tilde{\theta}}_{\psi_\ell} = \utwi{\theta}_{\psi_\ell}$, $\tilde{\lambda}_{\psi_\ell} = 1$, and $\tilde{\xi}_{\psi_\ell} = \lambda_{\psi_\ell}^{-1/2}$. From these estimators, we compute the innovations  $\utwi{\epsilon}_t$ for $t=1,\ldots, T$. We initialize the FDLM parameters using the initialization algorithm of \cite{kowal2014bayesian} based on the singular value decomposition (SVD) of 
$(\utwi{\epsilon}_1, \ldots, \utwi{\epsilon}_T)' = \utwi{U}_e\utwi{D}_e\utwi{V}_e'$. For the FLCs, we let $\utwi{\Phi}$ equal the first $J_\epsilon$ columns of $\utwi{V}_e$ and then estimate $\utwi{\Xi}$ to minimize   $||\utwi{\Phi} - \utwi{B}_\phi \utwi{\Xi}||^2$. For the factors, we let $(\utwi{e}_1, \ldots, \utwi{e}_T)'$ be the first $J_\epsilon$ columns of $(\utwi{U}_e\utwi{D}_e)$, and then estimate $\{\sigma_j^2\}$ and $\sigma_\eta^2$ using the conditional maximum likelihood estimators. Since $\sum_{k=1}^j \sigma_k^2/\sum_{k} \sigma_k^2$ estimates the proportion of variance of $\utwi{\epsilon}_t$ explained by the first $j$ factors, we set $J_\epsilon$ to be the smallest number of factors that explain at least 95\% of the  variance of $\utwi{\epsilon}_t$.  While more sophisticated procedures are available for selecting $J_\epsilon$, such as DIC and marginal likelihood, we find that this simple approach performs  well in simulations. 


\subsection{Gibbs Sampling Algorithm} 
We propose to sample from the joint posterior distribution using a Gibbs sampler with the following steps:
\begin{enumerate}
\item FAR process, $Y_t$:
\begin{enumerate}
\item Centered FAR process, $\mu_t$: form the DLM (6) and sample $\left[\{\utwi{\mu}_t\}_{t=1}^T | \cdots \right]$  {\it jointly} using the state space sample of \cite{durbin2002simple} implemented in the \texttt{R} package \texttt{KFAS}. 
\item Mean function, $\mu(\tau) = \utwi{b}_\phi'(\tau) \utwi{\theta}_\mu$: sample $\left[\utwi{\theta}_\mu | \cdots \right] \sim N(\utwi{A}_\mu \utwi{a}_\mu, \utwi{A}_\mu)$ where  
\begin{align*}
\utwi{A}_\mu^{-1} &= \utwi{\Lambda}_\mu^{-1} + \sigma_\nu^{-2} \sum_{t=1}^T\utwi{B}_\phi' \utwi{Z}_t'\utwi{Z}_t \utwi{B}_\phi,\\
\utwi{a}_\mu &=  \sigma_\nu^{-2} \sum_{t=1}^T  \utwi{B}_\phi'\utwi{Z}_t' (\utwi{y}_t - \utwi{Z}_t\utwi{\mu}_t),
\end{align*}
and
  $\utwi{\Lambda}_{\mu} = \mbox{diag}\left(10^8, 10^8, \lambda_{\mu}^{-1},\ldots, \lambda_{\mu}^{-1}\right)$.  We sample the smoothing parameter  $[\lambda_{\mu} | \cdots ] \sim \mbox{Gamma}\left(\frac{1}{2}(J_{\mu} - 3), \frac{1}{2}\sum_{j=3}^{J_{\mu}} \theta_{\mu, j}^2\right)$ restricted to $\lambda_{\mu} > 10^{-8}$ (see the $\sigma_j^{-2}$ sampler below), where $J_{\mu}$ $( = J_\phi)$ is the dimension of $\utwi{\theta}_{\mu}$ and $\theta_{\mu, j}$ is the $j$th component of $\utwi{\theta}_{\mu}$. 
\end{enumerate}
Set $Y_t = \mu_t + \mu$ or, in vector form,  $\utwi{Y}_t = \utwi{\mu}_t  + \utwi{\mu}$.
\item Measurement error precision, $\sigma_\nu^{-2}$: sample $$[\sigma_\nu^{-2} | \cdots] \sim \mbox{Gamma}\left( 10^{-3} + \frac{1}{2}\sum_{t=1}^T m_t, 10^{-3} +   \frac{1}{2} \sum_{t=1}^T \sum_{i=1}^{m_t} (y_{i,t} - \mu(\tau_{i,t}) - \mu_t(\tau_{i,t}))^2\right).$$

\item The FAR kernels, $\psi_1,\ldots, \psi_p$: using the \cite{gelman2006prior} prior and parametrization of  $\utwi{\theta}_{\psi_\ell} = \tilde{\xi}_{\psi_\ell} \utwi{\tilde{\theta}}_{\psi_\ell}$,   where $\psi_\ell(\tau, u) = \utwi{b}_\psi'(\tau, u)\utwi{\theta}_{\psi_\ell}$ and $\utwi{B}_{\psi} = (\utwi{b}_{\psi}(\tau_1), \ldots, \utwi{b}_{\psi}(\tau_M))'$, we sample
\begin{enumerate}
\item $ \utwi{\tilde{\theta}}_{\psi} = (\utwi{\tilde{\theta}}_{\psi_1}', \ldots, \utwi{\tilde{\theta}}_{\psi_p}')'$ {\it jointly} from $[ \utwi{\tilde{\theta}}_{\psi}| \cdots ] \sim  N(\utwi{A}_{\psi} \utwi{a}_{\psi}, \utwi{A}_{\psi})$, where 
\begin{align*}
\utwi{A}_{\psi}^{-1}[\ell, k] &=
\begin{cases}
\lambda_{\psi_\ell} \utwi{\Omega}_{\psi_\ell} +  s_\ell \tilde{\xi}_{\psi_\ell}^2\left[(\utwi{B}_{\psi}'\utwi{Q}) \left\{ \sum_{t=p+1}^T  \utwi{\mu}_{t-\ell} \utwi{\mu}_{t-\ell}'\right\} (\utwi{B}_{\psi}'\utwi{Q})' \right] \otimes \left[ \utwi{B}_{\psi}' \utwi{K}_\epsilon^{-1} \utwi{B}_{\psi}\right],  &  \ell = k \\
s_\ell s_k   \tilde{\xi}_{\psi_\ell} \tilde{\xi}_{\psi_k}\left[(\utwi{B}_{\psi}'\utwi{Q}) \left\{ \sum_{t=p+1}^T \utwi{\mu}_{t-\ell} \utwi{\mu}_{t-k}'\right\} (\utwi{B}_{\psi}'\utwi{Q})' \right] \otimes \left[ \utwi{B}_{\psi}' \utwi{K}_\epsilon^{-1} \utwi{B}_{\psi}\right] & \ell \ne k
\end{cases}\\
\utwi{a}_{\psi}[\ell] &= s_\ell \tilde{\xi}_{\psi_\ell}\mbox{vec}\left( \utwi{B}_{\psi}'\utwi{K}_\epsilon^{-1} \left\{\sum_{t=p+1}^T  \utwi{\mu}_t \utwi{\mu}_{t-\ell}' \right\} (\utwi{B}_{\psi}' \utwi{Q})'\right),
\end{align*}
$\utwi{A}_{\psi}^{-1}[\ell,k]$ is the $ (\ell,k)$th block of $\utwi{A}_{\psi}^{-1}$ of dimension $J_\psi^2 \times J_\psi^2$ and $\utwi{a}_{\psi}[\ell]$ is the $\ell$th subvector of $\utwi{a}_\psi$  of length $J_\psi^2$;

\item For $\ell=1,\ldots,p$, sample  $\left[\tilde{\xi}_{\psi_\ell} | \cdots \right] \sim N\left(A_{\tilde{\xi}_{\psi_\ell}} a_{\tilde{\xi}_{\psi_\ell}}, A_{\tilde{\xi}_{\psi_\ell}}\right)$, where 
\begin{align*}
A_{\tilde{\xi}_{\psi_\ell}} ^{-1} &= 10^{-6} + \utwi{\tilde{\theta}}_{\psi}'\left( \left[(\utwi{B}_{\psi}'\utwi{Q}) \left\{ \sum_{t=p+1}^T  \utwi{\mu}_{t-\ell} \utwi{\mu}_{t-\ell}'\right\} (\utwi{B}_{\psi}'\utwi{Q})' \right] \otimes \left[ \utwi{B}_{\psi}' \utwi{K}_\epsilon^{-1} \utwi{B}_{\psi}\right]\right) \utwi{\tilde{\theta}}_{\psi}
,\\
a_{\tilde{\xi}_{\psi_\ell}} &=  \utwi{\tilde{\theta}}_{\psi}'\mbox{vec}\left( \utwi{B}_{\psi}'\utwi{K}_\epsilon^{-1} \left\{\sum_{t=p+1}^T  \left[\utwi{\mu}_t - \sum_{k\ne \ell} s_k \utwi{G}(\psi_k)\utwi{\mu}_{t-k}\right] \utwi{\mu}_{t-\ell}' \right\} (\utwi{B}_{\psi}' \utwi{Q})'\right),
\end{align*}
sample $\left[\tilde{\lambda}_{\psi_\ell} | \cdots \right] \sim  \mbox{Gamma}\left(\frac{1}{2} + J_\psi^2/2,  \frac{1}{2} +  \utwi{\theta}_{\psi_\ell}'\utwi{\Omega}_{\psi_\ell}\utwi{\theta}_{\psi_\ell} /2 \right)$, and, if $\kappa_\ell$ is unknown, sample $\kappa_\ell$ using the slice sampler \citep{neal2003slice}. Set $\utwi{\theta}_{\psi_\ell} = \tilde{\xi}_{\psi_\ell} \utwi{\tilde{\theta}}_{\psi_\ell}$ and update $\utwi{\Omega}_{\psi_\ell}$. 
\item For the model averaging procedure, sample  $[s_\ell | \cdots] $ (in random order), i.e., set $s_\ell = 1$ if $\log O_{10}^{post} > \log(1/U - 1)$ and $s_\ell=0$ otherwise, where   $U \sim \mbox{Uniform}(0,1)$,
$\log O_{10}^{post}$ is the log-posterior odds 
$$
\log O_{10}^{post} =  -\frac{1}{2}
\sum_{t=p+1}^T \left[\utwi{\mu}_{t-\ell}' \utwi{K}_\epsilon^{-1} \utwi{\mu}_{t-\ell} - 2\left(\utwi{\mu}_t - \sum_{k\ne \ell} s_k \utwi{G}(\psi_k)\utwi{\mu}_{t-k} \right)'\utwi{K}_\epsilon^{-1} \utwi{\mu}_{t-\ell}\right]+ \log O_{10}^{prior}, 
$$
and 
$ \log O_{10}^{prior} = \log \mathbb{P}(s_\ell=1 | s_k, k \ne \ell) - \log \mathbb{P}(s_\ell=0 | s_k, k\ne \ell)$ is the log-prior odds. 
\end{enumerate}

\item The innovation covariance, $\utwi{K}_\epsilon$, under the FDLM:
\begin{enumerate}
\item The factors, $\{\utwi{e}_t\}_{t=1}^T$: using the prior $\utwi{e}_t \stackrel{iid}{\sim} N(\utwi{0}, \utwi{\Sigma}_e)$ and  the conditional likelihood $\utwi{\epsilon}_t = \utwi{\mu}_t - \sum_{\ell=1}^p \utwi{G}(\psi_\ell) \utwi{\mu}_{t-\ell} = \utwi{\Phi} \utwi{e}_{t}   + \utwi{\eta}_t$, sample $[\utwi{e}_t | \cdots ] \sim N(\utwi{A}_e \utwi{a}_{e_t}, \utwi{A}_e)$, where 
\begin{align*}
\utwi{A}_e^{-1} &= \sigma_\eta^{-2} \utwi{\Phi}'\utwi{\Phi}  + \utwi{\Sigma}_e^{-1} =  \mbox{diag}\left(\{\sigma_{\eta}^{-2} + \sigma_j^{-2}\}_{j=1}^{J_\epsilon} \right)\\
\utwi{a}_{e_t} &= \sigma_\eta^{-2} \utwi{\Phi}'\utwi{\epsilon}_t.
\end{align*}
Note that $\utwi{A}_e$ is time-invariant and diagonal, so we can sample $\{\utwi{e}_t\}_{t=1}^T$ jointly and efficiently. 
\item The factor precisions, $\sigma_j^{-2}$: sample $[\sigma_{J_\epsilon}^{-2} | \cdots ] \sim \mbox{Gamma}\left(10^{-3} + \frac{T}{2}, 10^{-3} + \frac{1}{2} \sum_{t=1}^T e_{J_\epsilon,t}^2  \right)$; then, for $j = J_\epsilon - 1, \ldots, 1$, set $\sigma_j^{-2} = F_\phi^{-1}(U; s_\phi, r_{\phi_j})$, where $F_\phi$ is the distribution function for a Gamma random variable with shape parameter $s_\phi = (T-1)/2$ and rate parameter  $r_{\phi_j} = \sum_{t=1}^T e_{j,t}^2/2$, and 
$U \sim \mbox{Uniform}\left(a_{\phi_j} ,b_{\phi_j}\right)$ where $a_{\phi_j} = F_\phi(0;  s_\phi, r_{\phi_j})$ and  $b_{\phi_j} = F_\phi(\sigma_{j+1}^{-2};  s_\phi, r_{\phi_j})$. 
\item The approximation error precision, $\sigma_\eta^{-2}$: sample $$[\sigma_\eta^{-2} | \cdots ] \sim \mbox{Gamma}\left( 10^{-3} + \frac{TM}{2}, 10^{-3} + \frac{1}{2} \sum_{t=1}^T ||\utwi{\epsilon}_t - \utwi{\Phi}\utwi{e}_t ||^2\right)$$
where $||\cdot||^2$ denotes the Euclidean distance. 
\item The factor loading curves: for $j=1,\ldots,J_\epsilon$ (in random order), sample $\utwi{\xi}_j \sim N(\utwi{A}_{\xi_j}\utwi{a}_{\xi_j}, \utwi{A}_{\xi_j})$, where 
\begin{align*}
\utwi{A}_{\xi_j}^{-1} &=  \utwi{\Lambda}_{\phi_j}^{-1} + \sigma_\eta^{-2} \left( \sum_{t=1}^T e_{j,t}^2\right)\utwi{B}_\phi'\utwi{B}_\phi, \\
\utwi{a}_{\xi_j} &= \sigma_\eta^{-2} \utwi{B}_\phi' \sum_{t=1}^T e_{j,t} \left(\utwi{\epsilon}_t -  \utwi{B}_\phi\sum_{k \ne j}\utwi{\xi}_{k} e_{k,t}\right).
\end{align*}
To enforce the orthogonality constraint, we condition on the linear constraints $\left(\utwi{B}_\phi \utwi{\xi}_k\right)' \utwi{B}_\phi \utwi{\xi}_j = 0 $ for $k \ne j$; since $\utwi{\xi}_j$ is Gaussian and $\utwi{\xi}_k$ is conditioned upon, the resulting distribution is Gaussian with easily computable moments, which  is also convenient for efficient sampling; see \cite{kowal2014bayesian} for more details. After sampling from the conditional distribution, we normalize  the sampled vector $\utwi{\xi}_j$, so that the orthonormality constraint is enforced at every MCMC iteration. We sample the corresponding smoothing parameters $[\lambda_{\phi_j} | \cdots ] \sim \mbox{Gamma}\left(\frac{1}{2}\left(J_\phi - 3\right),  \frac{1}{2}\sum_{k=3}^{J_\phi} \xi_{j, k}^2 \right)$
restricted to $\lambda_{\phi_j} > 10^{-8}$, 
where $\xi_{j,k}$ is the $k$th component of $\utwi{\xi}_j$. 
\end{enumerate}
Finally, we form the covariance and precision matrices $\utwi{K}_\epsilon$ and $\utwi{K}_\epsilon^{-1}$, respectively, using the sampled components. Since the orthonormality constraint $\utwi{\Phi}'\utwi{\Phi} = \utwi{I}_{J_\epsilon}$ is enforced at every MCMC iteration, we can compute $\utwi{K}_\epsilon^{-1}$ directly and efficiently using (8).
\end{enumerate}

When the sample size $T$ or the number of evaluation points $M$ is large (i.e., $T > 10,000$ or $M > 50$), the \cite{durbin2002simple} joint sampler is computationally inefficient. Instead, we may use a single-move sampler for $\{\utwi{\mu}_t\}_{t=1}^T$, in which we sample from the full conditional distribution of each $[\utwi{\mu}_t | \utwi{\mu}_s, s\ne t]$ separately for $t=1,\ldots,T$ (in random order). The single-move sampler is more computationally efficient, but is typically less MCMC efficient. The   FDLM  provides a closed form for $\utwi{K}_\epsilon^{-1}$, which  substantially reduces computation time when $M$ is large. 

The tensor product basis for $\psi_\ell$ provides a computational simplification for jointly sampling the FAR kernel basis coefficients, $\utwi{\theta}_{\psi}$. Importantly, the dimension of the Kronecker product for computing $\utwi{A}_{\psi}^{-1}$ is determined by the number of basis functions, $J_{\psi}$, which is bounded by $35$ in our specification, and may be smaller for some applications. For other bivariate bases, such as the thin plate spline basis, such simplifications are not readily available, and the Kronecker product scales with the number of evaluation points, $M$.

In the model averaging procedure, there is a nontrivial concern about the ability of the MCMC sampler to move between states. When $s_\ell = 0$, $\psi_\ell$ does not appear in the likelihood (9), so the Gibbs sampler will draw $\psi_\ell$ from its prior. Therefore, the prior for $\psi_\ell$ must be proper; if it is nonetheless noninformative, then the draws of $\psi_\ell$ from the prior distribution may not be reasonable for (9), so the next MCMC sample of $s_\ell$ will  be zero with high probability. To alleviate this problem, we fix $s_\ell= 1$ for all $\ell$ during a short burn-in period, so that each $\psi_\ell$ is well-estimated and therefore more likely to be included in the model if it is relevant. In both simulations and the yield curve application, the \cite{gelman2006prior} parametrization for $\psi_\ell$ sampling discussed in the Appendix provides excellent mixing among the states $\{s_\ell\}_{\ell=1}^{p_{max}}$.

 \section{Additional Theoretical Results}
 \subsection{Proof of Proposition 1}
 Let ${\Psi}(B)$ be a polynomial in the backshift operator $B$ of order $p$, so that $\Psi(B)Y_t  = (1 - \Psi_1B - \Psi_2B^2 - \cdots - \Psi_pB^p)Y_t = Y_t - \sum_{\ell=1}^p \Psi_\ell(Y_{t-\ell})$, where $\{\Psi_\ell\}_{\ell=1}^p$ are bounded linear operators on $L^2(\mathcal{T})$. Similarly, let $\Theta(B)$ be a polynomial in the backshift operator $B$ of order $q$, where $\{\Theta\}_{\ell=1}^q$ are bounded linear operators on $L^2(\mathcal{T})$. A   \emph{functional autoregressive moving average process} of order $(p,q)$, written FARMA($p,q$), is defined by 
 $\Psi(B)(Y_t - \mu) = \Theta(B)\epsilon_t $, where $\{\epsilon_t\}$ is a white noise process in $L^2(\mathcal{T})$ and $\mu$ is the unconditional mean of $Y_t$. The FAR($p$) model may be written compactly as $\Psi(B)(Y_t - \mu) = \epsilon_t$. By assumption, we observe the process $\{y_t\}$, where $y_t = Y_t + \nu_t$ and $\{\nu_t\}$ is a white noise process in $L^2(\mathcal{T})$ independent of $\{\epsilon_t\}$. Rewriting the observation equation $y_t - \mu = Y_t  - \mu + \nu_t$ and applying $\Psi(B)$, we have 
 $\Psi(B)(y_t - \mu) = \Psi(B)(Y_t  - \mu) + \Psi(B)\nu_t = \epsilon_t + \Psi(B)\nu_t.$ It remains to show that $Z_t \equiv  \epsilon_t + \Psi(B)\nu_t$ is a \emph{functional moving average process} of order $p$, or equivalently, FARMA($0, p$). Clearly, $X_t \equiv \Psi(B)\nu_t$ is FARMA($0, p$). By Proposition 10.2 in \cite{bosq2008inference}, $C_p^{X} \ne 0$ and $C_\ell^{X} =0$ for $\ell > p$, where $C_\ell^X$ is the covariance operator of $X_t$ defined by $C_\ell^X(x) \equiv \mathbb{E}\left[ \langle X_t, x\rangle X_{t+\ell}\right]$ for $x \in L^2(\mathcal{T})$. Let $C_\ell^Z$ and $C_\ell^\epsilon$ denote the covariance operators for $Z_t$ and $\epsilon_t$, respectively. Then $C_\ell^Z(x) = \mathbb{E}\left[ \langle Z_t, x\rangle Z_{t+\ell}\right]= \mathbb{E}\left[ \langle  \epsilon_t + X_t, x\rangle  \left(\epsilon_{t+\ell} +X_{t+\ell}\right)\right] = C_\ell^\epsilon(x) + C_\ell^X(x) $, using independence of $\{\epsilon_t\}$ and $\{\nu_t\}$. Since $\epsilon_t$ is white noise, $C_\ell^\epsilon = 0$ for $\ell>0$, from which it follows that $C_p^Z \ne 0$ and $C_\ell^Z =0$ for $\ell>p$. Proposition 10.2 in \cite{bosq2008inference} implies that $Z_t$ is FARMA($0,p$), so we conclude that $y_t$ is FARMA($p,p$).

 
\subsection{DLM Recursions and Special Cases of Theorem 1} 
For completeness, we provide the standard DLM recursion formulas for model (6). 
Let $\mathcal{D}_t = \{\utwi{y}_t, \utwi{y}_{t-1},\ldots,\utwi{y}_1\} \cup \mathcal{D}_0$ be the information available at time $t$, where $\mathcal{D}_0$ represents the information prior to $t=1$. For our purposes|in particular, for the Gibbs sampling algorithm|we let $\mathcal{D}_0 = \{\mu, \sigma_\nu^2, \psi, K_\epsilon\}$ (denoted by $\utwi{\Theta}$ in Theorem 1). We may compute full conditional posterior distributions from model (6) using standard DLM recursions (e.g., \citealp{westDLM}). For simplicity, let $\utwi{G} = \utwi{G}(\psi)$. 
Suppose that $[\utwi{\mu}_{t-1} | \mathcal{D}_{t-1}] \sim N(\utwi{m}_{t-1}, \utwi{C}_{t-1})$. The {\it prior at time $t$} is 
  $[\utwi{\mu}_t|\mathcal{D}_{t-1}] \sim N(\utwi{a}_t, \utwi{R}_t)$, where $\utwi{a}_t = \utwi{G}\utwi{m}_{t-1}$ and $\utwi{R}_t   = \utwi{G}\utwi{C}_{t-1}\utwi{G}' + \utwi{K}_\epsilon$. The {\it one-step forecast at time $t$} is $[\utwi{y}_t | \mathcal{D}_{t-1}] \sim N(\utwi{f}_t, \utwi{Q}_t)$, where $\utwi{f}_t = \utwi{Z}_t \utwi{\mu} + \utwi{Z}_t \utwi{a}_t = \utwi{Z}_t(\utwi{\mu} + \utwi{G}\utwi{m}_{t-1})$ and $\utwi{Q}_t = \utwi{Z}_t\utwi{R}_t\utwi{Z}_t' + \sigma_\nu^2\utwi{I}_{m_t}$. The {\it posterior at time $t$} is $[\utwi{\mu}_{t} | \mathcal{D}_{t}] \sim N(\utwi{m}_{t}, \utwi{C}_{t})$, where $\utwi{m}_t = \utwi{C}_t^{-1} \left(\utwi{R}_t^{-1}\utwi{a}_t + \sigma_{\nu}^{-2}\utwi{Z}_t'(\utwi{y}_t - \utwi{Z}_t\utwi{\mu})\right)$ and $\utwi{C}_t^{-1} = \utwi{R}_t^{-1} + \sigma_\nu^{-2}\utwi{Z}_t'\utwi{Z}_t$, or, more commonly, $\utwi{m}_t = \utwi{a}_t + \utwi{A}_t\utwi{r}_t$, $\utwi{A}_t = \utwi{R}_t\utwi{Z}_t'\utwi{Q}_t^{-1}$, $\utwi{r}_t = \utwi{y}_t - \utwi{f}_t$, and $\utwi{C}_t = \utwi{R}_t - \utwi{A}_t\utwi{Q}_t\utwi{A}_t'$. The $h$-step forecast of the functional observations is $\mathbb{E}[\utwi{y}_{t+h}|\mathcal{D}_t]  = \mathbb{E}[\utwi{Z}_{t+h}\utwi{\mu} + \utwi{Z}_{t+h}\utwi{\mu}_{t+h} + \utwi{\nu}_{t+h}|\mathcal{D}_t] = \utwi{Z}_{t+h}\utwi{\mu} + \utwi{Z}_{t+h}\mathbb{E}[\utwi{\mu}_{t+h}|\mathcal{D}_t]$, where $\mathbb{E}[\utwi{\mu}_{t+h}|\mathcal{D}_t] = \utwi{G}^h \utwi{m}_t$, which is the $h$-step forecast of $\utwi{\mu}_t$. 

 Some special cases of Theorem 1 are proved in \cite{westDLM}:
 
\noindent {\bf Corollary B.2.1} (Theorem 4.10, \citealp{westDLM}). \emph{The unique best linear predictor of the filtering random variable $[\utwi{\mu}_t|\mathcal{D}_t]$ is $\utwi{m}_t$}
 
 \noindent {\bf Corollary B.2.2} (Corollary 4.7, \citealp{westDLM}). \emph{The unique best linear predictor of the one-step forecast $[\utwi{\mu}_t|\mathcal{D}_{t-1}]$ is $\utwi{a}_t$. The unique best linear predictor of the one-step forecast $[\utwi{y}_t|\mathcal{D}_{t-1}]$ is $\utwi{f}_t$.}
 


\subsection{Proof of Theorem 2} 
Suppose $\tau^* \in \mathcal{T}$ such that $\tau^* \not \in  \mathcal{T}_e$. The full conditional distribution of $\mu_t(\tau^*)$ is 
\begin{align*}
 \left[\mu_t(\tau^*) | \{\utwi{\mu}_r\}_{r=1}^T,  \utwi{\Theta}, \mathcal{D}_s\right] &\propto  \left[\utwi{y}_1,\ldots,\utwi{y}_s | \mu_t(\tau^*) ,\{\utwi{\mu}_r\}_{r=1}^T,\utwi{\Theta}\right]\times  \left[\mu_t(\tau^*) | \{\utwi{\mu}_r\}_{r=1}^T,  \utwi{\Theta} \right]\\ 
 &\propto  \left[\mu_t(\tau^*) | \{\utwi{\mu}_r\}_{r=1}^T,  \utwi{\Theta}\right],
\end{align*}
since the likelihood term is constant with respect to $\mu_t(\tau^*)$: $\mathcal{T}_o \subseteq \mathcal{T}_e$, so $\tau^*\not\in\mathcal{T}_e$ implies   $\tau^*\not\in\mathcal{T}_o$, and therefore $\mu_t(\tau^*)$ does not appear in the likelihood of model (4). For $p=1$, the conditional Gaussian process prior for $\mu_t$ implied by model (4) under the approximation (5) is  $[\mu_t | \mu_{t-1}, \psi, K_\epsilon] \sim \mathcal{GP}\left(\utwi{\psi}'(\cdot)\utwi{Q}\utwi{\mu}_{t-1}, K_\epsilon\right)$, where   $\utwi{\psi}'(\tau) = \left(\psi(\tau, \tau_1), \ldots, \psi(\tau, \tau_M)\right)$, $\utwi{Q}$ is a known quadrature weight matrix, and $\utwi{\mu}_{t-1} = (\mu_{t-1}(\tau_1),\ldots,\mu_{t-1}(\tau_M))'$ is the function $\mu_{t-1}$ evaluated at each $\tau \in \mathcal{T}_e$. Notably, $\tau^*\not\in\mathcal{T}_e$ implies that $\mu_t(\tau^*)$ does not appear in the conditional mean function for $\mu_{t+1}$, so we may further simplify the distribution of $\mu_t(\tau^*)$:
$$
 \left[\mu_t(\tau^*) | \{\utwi{\mu}_r\}_{r=1}^T,  \utwi{\Theta}, \mathcal{D}_s\right] \propto  \left[\mu_t(\tau^*) | \utwi{\mu}_t, \utwi{\mu}_{t-1},  \utwi{\Theta}\right].
$$
To compute this distribution, we use the definition of a Gaussian process, which implies the following joint distribution of $\mu_t(\tau^*)$ and $\utwi{\mu}_t$, conditional on $\utwi{\mu}_{t-1}$, $\psi$, and $K_\epsilon$:
$$
\begin{pmatrix} \mu_t(\tau^*) \\ \utwi{\mu}_t \end{pmatrix} \sim N\left(
\begin{pmatrix}\utwi{\psi}'(\tau^*)\utwi{Q}\utwi{\mu}_{t-1} \\ \utwi{\Psi}\utwi{Q}\utwi{\mu}_{t-1} \end{pmatrix},
\begin{pmatrix} K_\epsilon(\tau^*, \tau^*) &  \utwi{K}_\epsilon(\tau^*) \\  \utwi{K}_\epsilon'(\tau^*) &  \utwi{K}_\epsilon \end{pmatrix}\right),
$$
where $\utwi{\Psi} = \{\psi(\tau_i, \tau_k)\}_{i,k = 1}^M$ and $\utwi{K}_\epsilon(\tau^*) = (K_\epsilon(\tau^*, \tau_1),\ldots,K_\epsilon(\tau^*, \tau_M))$. Conditioning on $\utwi{\mu}_t$ induces the desired distribution 
$[\mu_t(\tau^*) | \utwi{\mu}_t, \utwi{\mu}_{t-1}, \psi, {K}_\epsilon ] \sim N\left(m_t(\tau^*), K_t(\tau^*)\right)$, where $m_t(\tau^*) = \utwi{\psi}'(\tau^*)\utwi{Q}\utwi{\mu}_{t-1}  + \utwi{K}_\epsilon(\tau^*) \utwi{K}_\epsilon^{-1}\left(\utwi{\mu}_t - \utwi{\Psi}\utwi{Q}\utwi{\mu}_{t-1} \right)$ and $K_t(\tau^*) = K_\epsilon(\tau^*, \tau^*) - \utwi{K}_\epsilon(\tau^*)\utwi{K}_\epsilon^{-1} \utwi{K}_\epsilon'(\tau^*)$. Under the FDLM, the following useful simplifications are available: $
K_\epsilon(\tau^*, \tau^*) =  \sigma_\eta^2 + \utwi{\phi}'(\tau^*)\utwi{\Sigma}_e\utwi{\phi}(\tau^*)$, $\utwi{K}_\epsilon(\tau^*) = \utwi{\phi}'(\tau^*)\utwi{\Sigma}_e\utwi{\Phi}'$, and using (8),  $\utwi{K}_\epsilon^{-1}  =  \sigma_\eta^{-2} \utwi{I}_M - \sigma_\eta^{-2}\utwi{\Phi}\utwi{\tilde{\Sigma}}_e\utwi{\Phi}'$, where  $\utwi{\phi}'(\tau^*) = (\phi_1(\tau^*),\ldots,\phi_{J_\epsilon}(\tau^*))$, $\utwi{\Sigma}_e = \mbox{diag}\left(\{\sigma_j^2\}_{j=1}^{J_\epsilon}\right)$, $\utwi{\Phi} = (\utwi{\phi}(\tau_1), \ldots, \utwi{\phi}(\tau_M))' $, and $\utwi{\tilde{\Sigma}}_e =\mbox{diag}\left(\{\sigma_j^2/(\sigma_\eta^2+\sigma_j^2)\}_{j=1}^{J_\epsilon}\right)$. 
By substitution, we derive
\begin{align*}
m_t(\tau^*) &= \utwi{\psi}'(\tau^*)\utwi{Q}\utwi{\mu}_{t-1}  + \utwi{K}_\epsilon(\tau^*) \utwi{K}_\epsilon^{-1}\left(\utwi{\mu}_t - \utwi{\Psi}\utwi{Q}\utwi{\mu}_{t-1} \right)\\
&= \utwi{\psi}'(\tau^*)\utwi{Q}\utwi{\mu}_{t-1} +  \utwi{\phi}'(\tau^*)\utwi{\Sigma}_e\utwi{\Phi}'\left(\sigma_\eta^{-2} \utwi{I}_M - \sigma_\eta^{-2}\utwi{\Phi}\utwi{\tilde{\Sigma}}_e\utwi{\Phi}'\right)\left(\utwi{\mu}_t - \utwi{\Psi}\utwi{Q}\utwi{\mu}_{t-1}\right)\\
&= \utwi{\psi}'(\tau^*)\utwi{Q}\utwi{\mu}_{t-1} +   \utwi{\phi}'(\tau^*)\utwi{\tilde\Sigma}_e\utwi{\Phi}' \left(\utwi{\mu}_t - \utwi{\Psi}\utwi{Q}\utwi{\mu}_{t-1}\right),
\end{align*}
using the constraint $\utwi{\Phi}'\utwi{\Phi} = \utwi{I}_{J_\epsilon}$ and the simplification  $\sigma_\eta^{-2}\utwi{\Sigma}_e - \sigma_\eta^{-2} \utwi{\Sigma}_e\utwi{\tilde{\Sigma}}_e = \utwi{\tilde\Sigma}_e$. 
Similarly, 
\begin{align*}
K_t(\tau^*) &=  K_\epsilon(\tau^*, \tau^*) - \utwi{K}_\epsilon(\tau^*)\utwi{K}_\epsilon^{-1} \utwi{K}_\epsilon'(\tau^*)\\
&= \sigma_\eta^2 + \utwi{\phi}'(\tau^*)\utwi{\Sigma}_e\utwi{\phi}(\tau^*) - \utwi{\phi}'(\tau^*)\utwi{\Sigma}_e\utwi{\Phi}'\left(\sigma_\eta^{-2} \utwi{I}_M - \sigma_\eta^{-2}\utwi{\Phi}\utwi{\tilde{\Sigma}}_e\utwi{\Phi}'\right)\utwi{\Phi}\utwi{\Sigma}_e\utwi{\phi}(\tau^*)\\
 &=  \sigma_\eta^2 +\sigma_\eta^2 \utwi{\phi}'(\tau^*)\utwi{\tilde \Sigma}_e\utwi{\phi}(\tau^*),
\end{align*}
which is time-invariant. 
Extensions for $p>1$ only require modification of the mean function:  $m_t(\tau^*) =\sum_{\ell=1}^p \utwi{\psi}_\ell'(\tau^*)\utwi{Q}\utwi{\mu}_{t-\ell} +  \utwi{\phi}'(\tau^*)\utwi{\tilde\Sigma}_e\utwi{\Phi}' \left(\utwi{\mu}_t - \sum_{\ell=1}^p \utwi{\Psi}_\ell \utwi{Q}\utwi{\mu}_{t-\ell} \right)$, where $\utwi{\psi}_\ell'(\tau) = \left(\psi_\ell(\tau, \tau_1), \ldots, \psi_\ell(\tau, \tau_M)\right)$ and $\utwi{\Psi}_\ell = \{\psi_\ell(\tau_i, \tau_k)\}_{i,k = 1}^M$.


\section{Additional Simulation Results}
\renewcommand{\thefigure}{C\arabic{figure}} \setcounter{figure}{0}

\begin{figure}[h]
\begin{center}
\includegraphics[width=.48\textwidth]{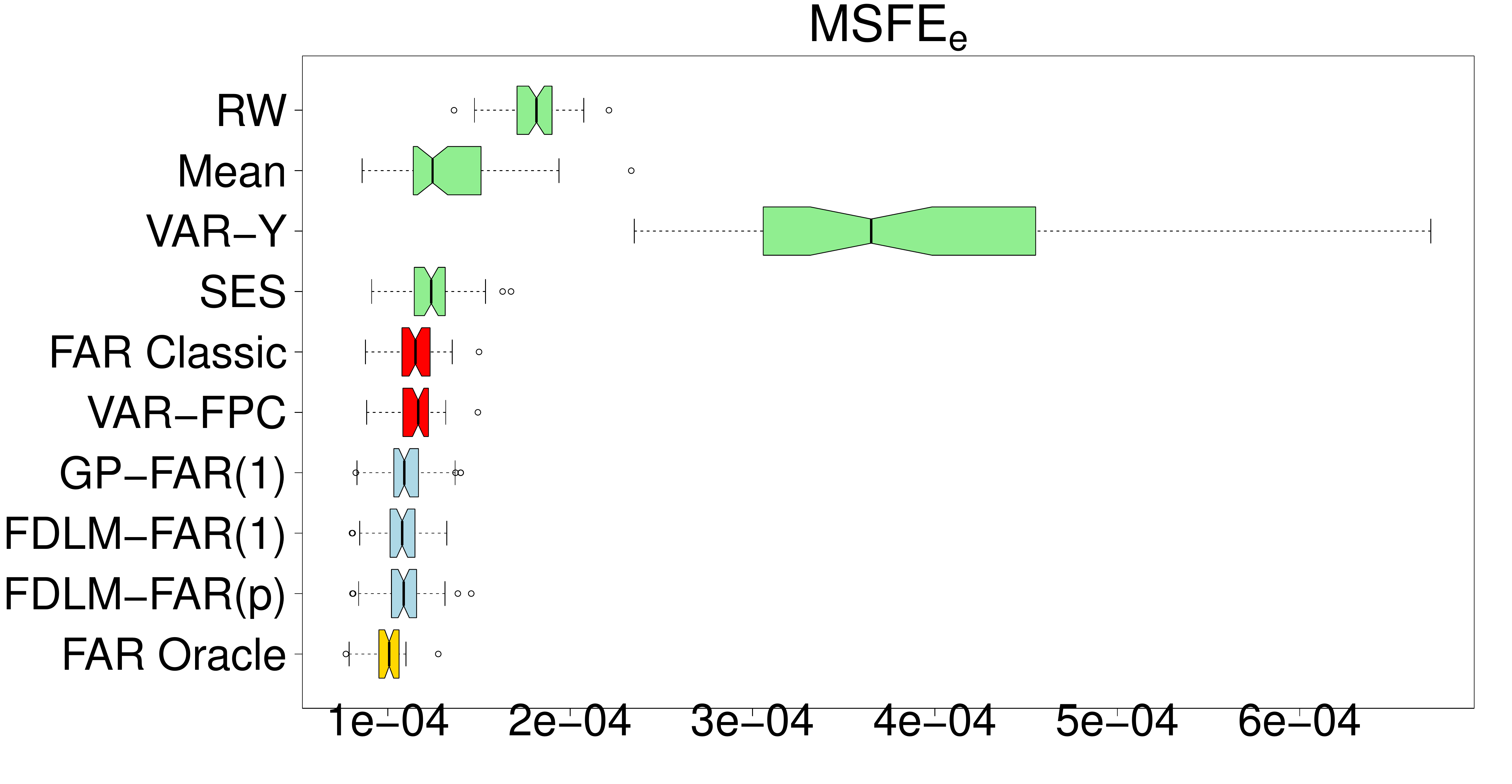}
\includegraphics[width=.48\textwidth]{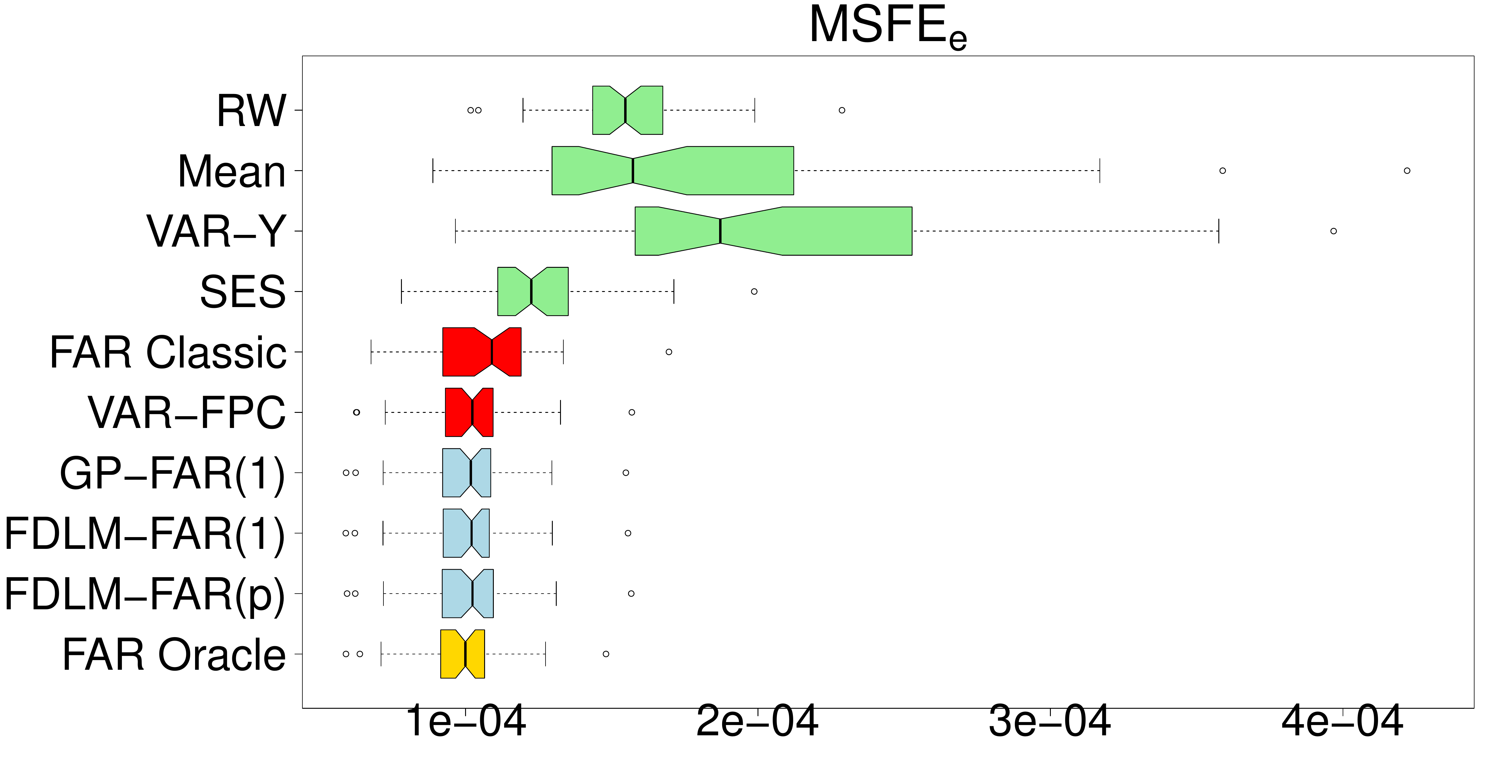}
\includegraphics[width=.48\textwidth]{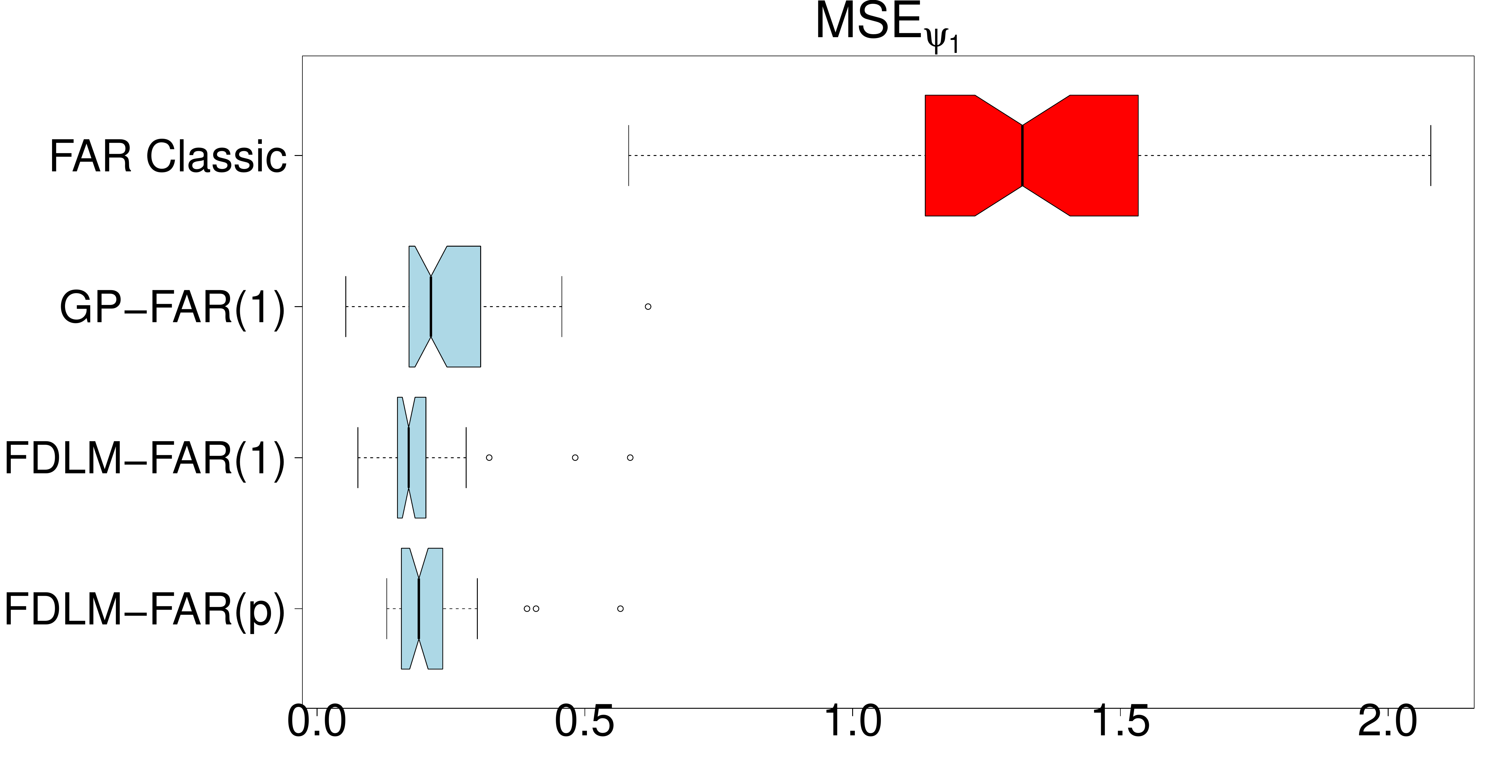}
\includegraphics[width=.48\textwidth]{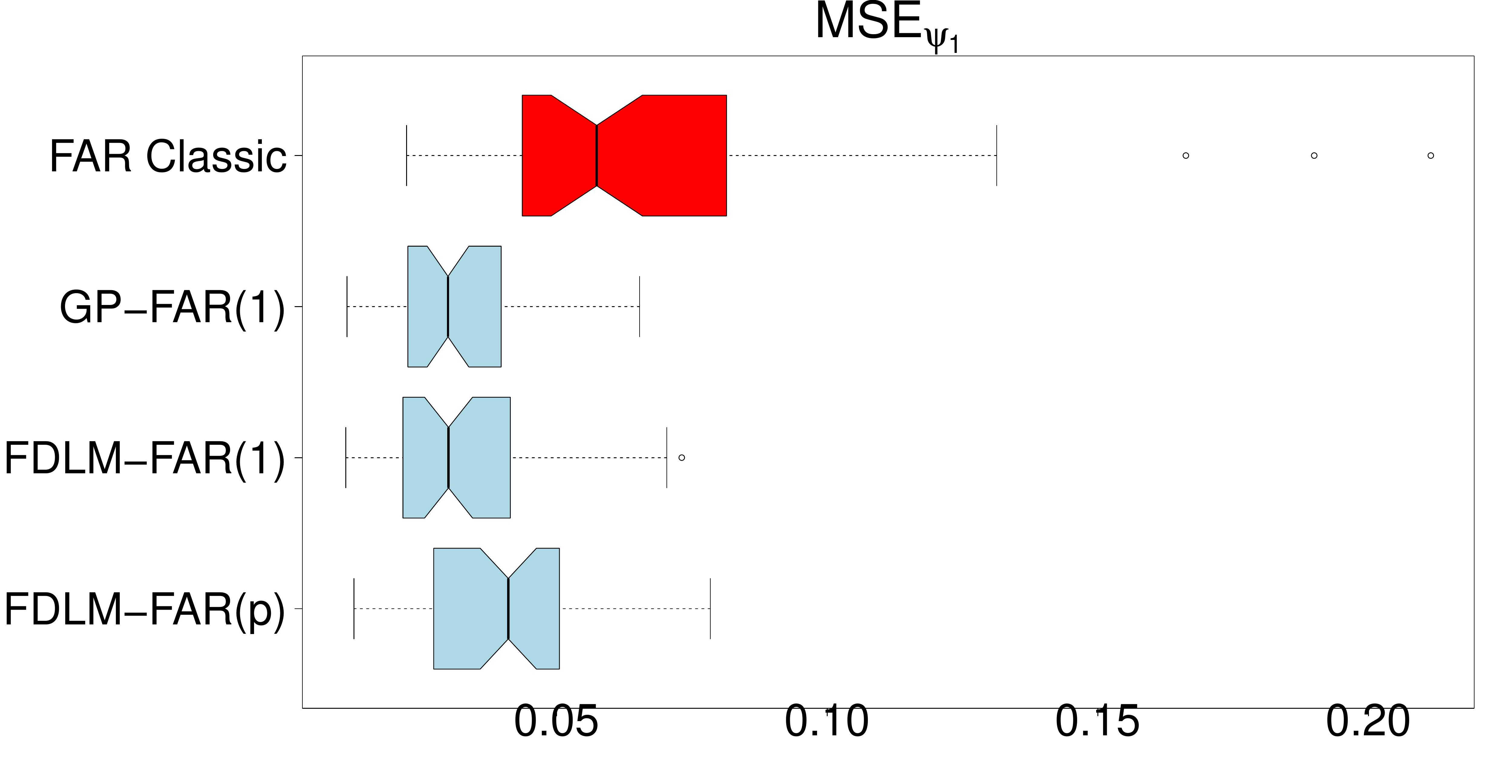}
\caption{$MSFE_e$ ({\bf top}) and corresponding $MSE_{\psi_1}$ ({\bf bottom}) under various designs. {\bf Left:} FAR(1), $T=50$, dense design with the Bimodal-Gaussian kernel and non-smooth GP innovations. {\bf Right:}  FAR(1), $T=350$, dense design with the Bimodal-Gaussian kernel and smooth GP innovations. The proposed methods provide superior forecasts and nearly achieve the oracle performance, despite the presence of sparsity.  \label{fig:figMSEapp}}
\end{center}
\end{figure}

In Figure \ref{fig:figMSEapp}, we display the results from FAR(1) simulations under the dense design, while varying both smoothness of $\epsilon_t$ and the sample size, $T$. The functional data methods all nearly achieve the oracle performance, and are superior to the multivariate methods. 
These results confirm the findings of \cite{didericksen2012empirical}: when $T$ is large and the observation points are dense in $\mathcal{T}$, existing functional data methods can nearly achieve the oracle performance, even when $\psi_1$ is estimated poorly. The proposed methods, particularly with the FDLM (FDLM-FAR($1$) and FDLM-FAR($p$)), outperform existing functional data methods for non-smooth GP innovations, and again are far superior for $\psi_1$ estimation. The uncertainty of $p$ incorporated into the lag selection procedure (FDLM-FAR($p$)) does not appear to inhibit forecasting or estimation of $\psi_1$ substantially. 

\begin{figure}[h]
\begin{center}
\includegraphics[width=.75\textwidth]{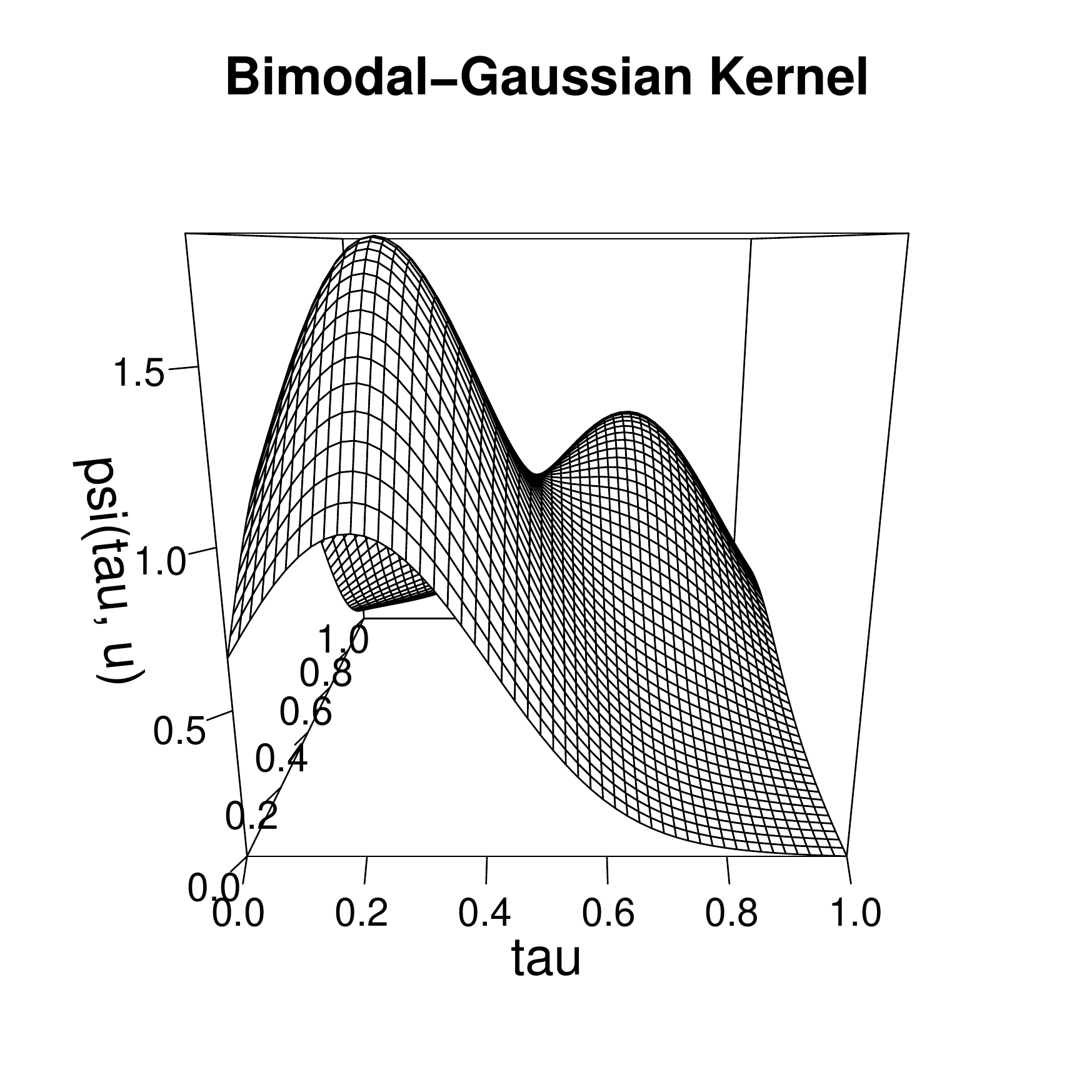}
\caption{The  {\it Bimodal-Gaussian} kernel, $\psi(\tau, u) \propto \frac{0.75}{\pi (0.3)(0.4)}\exp\{-(\tau-0.2)^2/(0.3)^2 - (u-0.3)^2/(0.4)^2\}  + \frac{0.45}{\pi (0.3)(0.4)} \exp\{-(\tau-0.7)^2/(0.3)^2 - (u-0.8)^2/(0.4)^2\}$, normalized so that  $\int\int \psi_\ell^2(\tau,u) \, d \tau \, d u = 0.8$. \label{fig:figBGkrnl}}
\end{center}
\end{figure}

For further clarity, we plot the {\it Bimodal-Gaussian} kernel in Figure \ref{fig:figBGkrnl}, which is featured prominently in our simulation study. 

\section{Additional Details for the Yield Curve Application}
\renewcommand{\thefigure}{D\arabic{figure}} \setcounter{figure}{0}
\renewcommand{\thetable}{D\arabic{table}}\setcounter{table}{0}

{\color{blue}

We include MCMC diagnostics for the yield curve application. All diagnostics were computed using the \texttt{R} package \texttt{coda} \citep{coda}. In Figures \ref{fig:tracePlotNominal} and \ref{fig:tracePlotReal}, we provide trace plots for the one-step forecast distributions for the nominal and real yield curves, respectively, on a single day in 2016 across selected maturities. The mixing is very efficient, which is confirmed by  effective sample sizes which exceed 5,000 in all cases.

In our yield curve forecasting study of Section 7, we included two popular parametric yield curve models based on the Nelson-Siegel parametrization \citep{nelson1987parsimonious}:  \citet[DL]{diebold2006forecasting} and \citet[DRA]{diebold2006macroeconomy}. The Nelson-Siegel basis is  defined by $f_1(\tau) =1$, $f_2(\tau | \lambda_{NS}) =  \frac{1 - \exp(-\tau\lambda_{NS})}{\tau\lambda_{NS}}$, and $f_3(\tau | \lambda_{NS}) =  \frac{1 - \exp(-\tau\lambda_{NS})}{\tau\lambda_{NS}} - \exp(-\tau\lambda_{NS})$, where $\lambda_{NS}$ is an unknown parameter. For both DL and DRA, the yield curve $Y_t(\tau)$ for time $t$ and time to maturity $\tau$ is written as a linear combination of the Nelson-Siegel basis function, for which the corresponding weights are dynamic:
\begin{align}\label{draObs}
Y_t(\tau) &= \bm{f}'(\tau | \lambda_{NS})\bm{\beta}_t + \epsilon_t(\tau), \\
\label{draEv}
\left(\bm{\beta}_t - \bm \mu_\beta\right) &= \bm{A} \left(\bm{\beta}_{t-1} - \bm \mu_\beta\right)  + \bm{\eta}_t
\end{align}
where $\bm{f}'(\tau | \lambda_{NS}) = (f_1(\tau), f_2(\tau| \lambda_{NS}), f_3(\tau| \lambda_{NS}))$, $\bm \beta_t$ is the corresponding 3-dimensional vector of dynamic weights with unconditional mean $\bm \mu_\beta$, and $\bm{A}$ is the $3\times 3$ evolution matrix. For implementation purposes, assume that the yield curve is observed at a fixed set of maturities $\tau_1,\ldots,\tau_M$, so that \eqref{draObs} becomes
\begin{equation}\label{draObs2} 
\bm{y}_t = \bm{F}_{NS}\bm \beta_t + \bm \epsilon_t
\end{equation}
where $\bm y_t = (Y_t(\tau_1), \ldots, Y_t(\tau_M))'$, $\bm{F}_{NS} = \left(\bm{f}(\tau_1 | \lambda_{NS}), \ldots, \bm{f}(\tau_M | \lambda_{NS})\right)'$, and $\bm \epsilon_t = (\epsilon_t(\tau_1), \ldots, \epsilon_t(\tau_M))'$.

The DL approach fixes $\lambda_{NS} = 0.0609$ and then estimates the parameters using a multi-step procedure. First, the weights $\{\bm \beta_t\}$ are estimated using ordinary least squares from \eqref{draObs2}. Next, the evolution matrix $\bm{A}$ in \eqref{draEv} is estimated as a VAR coefficient matrix, conditional on $\{\bm \beta_t\}$. \cite{diebold2006forecasting} note that constraining $\bm A$ to be diagonal may improve forecasting in some cases. Finally, $h$-step forecasts $\bm{\hat y}_{T+h}$ are computed via $\bm{\hat y}_{T+h} = \bm{F}_{NS} \bm{\hat \beta}_{T+h}$, where $\bm{\hat \beta}_{T+h}$ is the $h$-step forecast computed from the VAR in \eqref{draEv}. 

Alternatively, the DRA approach combines \eqref{draObs2} and \eqref{draEv} into a state space model, with error distributions  $\bm \epsilon_t \stackrel{iid}{\sim} N(\bm 0, \bm H)$ independent of $\bm \eta_t \stackrel{iid}{\sim} N(\bm 0, \bm Q)$. DRA assume that $\bm{H}$ is diagonal; we further assume that $\bm{Q}$ is diagonal, which helps stabilize  computations. The unknown parameters $\{\lambda_{NS}, \bm{A}, \bm{H}, \bm{Q}\}$ are then estimated jointly using maximum likelihood based on the Kalman filter. Following DRA, we model $\lambda_{NS}$ and the diagonal elements of $\bm H$ and $\bm Q$ on the log-scale to ensure positivity in the optimization routine. Conditional on the maximum likelihood estimates for these parameters, DRA use standard state space computations to construct forecasts for the response vector, $\bm y_t$.

 \begin{figure}[h]
\begin{center}
\includegraphics[width=1\textwidth]{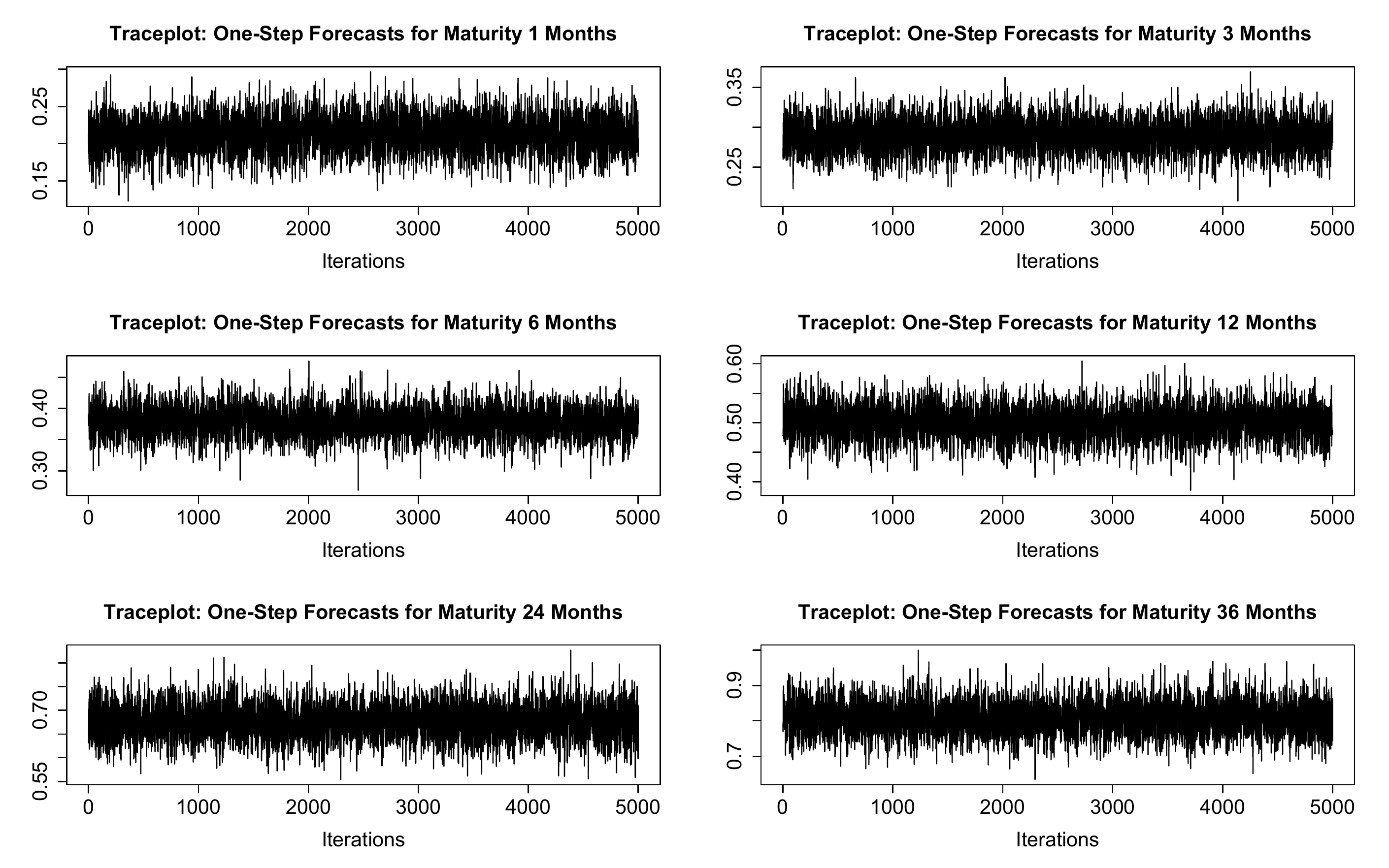}
\caption{Traceplot for one-step forecasts for nominal yield curves at selected maturities during 2016.  \label{fig:tracePlotNominal}}
\end{center}
\end{figure}

 \begin{figure}[h]
\begin{center}
\includegraphics[width=1\textwidth]{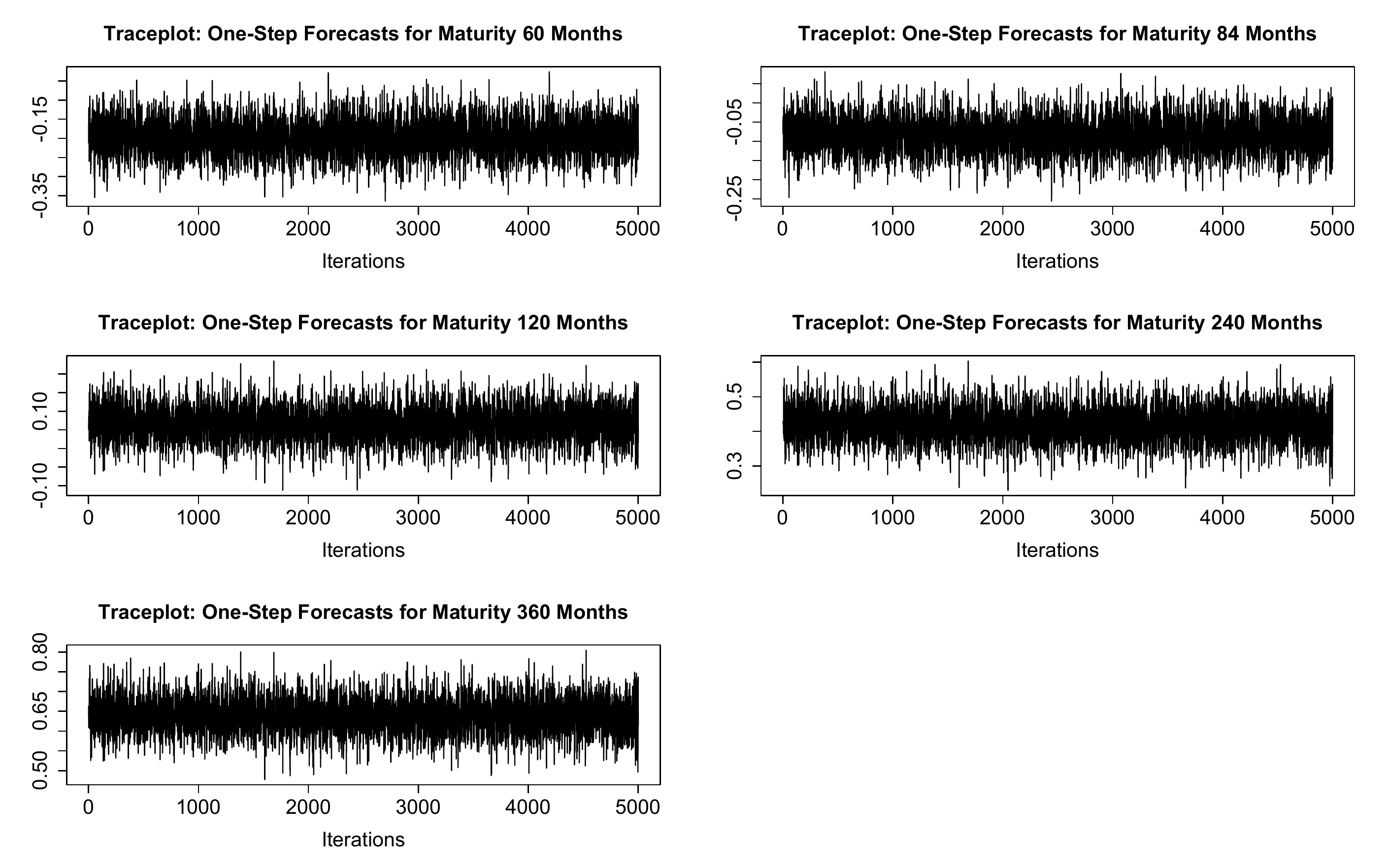}
\caption{Traceplot for one-step forecasts for real yield curves at selected maturities during 2016.  \label{fig:tracePlotReal}}
\end{center}
\end{figure}
}


\section{Additional Details on the Quadrature Approximation}
\renewcommand{\thefigure}{E\arabic{figure}} \setcounter{figure}{0}
\renewcommand{\thetable}{E\arabic{table}}\setcounter{table}{0}
 {\color{blue}  
Consider the integral in the FAR(1) evolution equation, $\mathcal{I}(\tau) \equiv \int \psi(\tau, u) \mu_{t-1}(u) \, du$, where we omit dependence of $\mathcal{I}$ on $t$ for notational simplicity. In the proposed methodology, we approximate this integral using quadrature: $\mathcal{I}(\tau) \approx \mathcal{I}_M(\tau) \equiv   \left(\psi(\tau, \tau_1), \ldots, \psi(\tau, \tau_M)\right) \utwi{Q} \utwi{\mu}_{t-1}$, where $\{\tau_1,\ldots,\tau_M\} = \mathcal{T}_e \subset \mathcal{T}$ is the set of unique evaluation points, $\bm{Q}$ is a known $M\times M$ quadrature matrix, and $\bm\mu_{t-1} = (\mu_{t-1}(\tau_1),\ldots,\mu_{t-1}(\tau_M))'$ is the function $\mu_{t-1}$ evaluated at the evaluation points. It is important to assess how the accuracy of the approximation of $\mathcal{I}$ by $\mathcal{I}_M$ depends in $M$, and in particular to determine a value of $M$ sufficiently large to produce reasonable approximations in practice. However, there is a tradeoff: the state vector in the dynamic linear model is $M$-dimensional, so increasing $M$ indiscriminately may unnecessarily increase computation time. 

We conducted a sensitivity analysis based on the simulations from Section 6 of the main paper. In particular, we use the Bimodal-Gaussian kernel, $\psi(\tau, u) \propto \frac{0.75}{\pi (0.3)(0.4)}\exp\{-(\tau-0.2)^2/(0.3)^2 - (u-0.3)^2/(0.4)^2\}  + \frac{0.45}{\pi (0.3)(0.4)} \exp\{-(\tau-0.7)^2/(0.3)^2 - (u-0.8)^2/(0.4)^2\}$, normalized so that  $\int\int \psi_\ell^2(\tau,u) \, d \tau \, d u = 0.8$. The Bimodal-Gaussian kernel is nonlinear, and therefore is inherently more difficult to approximate using linear quadrature methods, such as the trapezoidal rule. For the other component of the integrand, $\mu_{t-1}$, we simulate $\mu_{t-1} \sim \mathcal{GP}(0, K_\epsilon)$ using the covariance function parameterization $K_\epsilon= \sigma^2 R_\rho$,  where $R_\rho$ is the Mat\'{e}rn correlation function $
R_\rho(\tau, u) = \left\{2^{\rho_1-1}\Gamma(\rho_1)\right\}^{-1} \left( ||\tau - u||/\rho_2\right)^{\rho_1} K_{\rho_1}\!\left( ||\tau-u||/\rho_2\right)
$,  $\Gamma(\cdot)$ is the gamma function, $K_{\rho_1}$ is the modified Bessel function of order $\rho_1$, and $\utwi{\rho} = (\rho_1, \rho_2)$ are parameters \citep{matern2013spatial}. We let $\sigma = 0.01$ and $\utwi{\rho} = (\rho_1, 0.1)$, with $\rho_1 = 2.5$ for smooth (twice-differentiable) sample paths and $\rho_1 = 0.5$ for non-smooth (continuous, non-differentiable) sample paths. Comparisons between these cases are important: the non-smooth setting is substantially more challenging for approximations.

For each simulated value of $\mu_{t-1} \sim \mathcal{GP}(0, K_\epsilon)$, we compute $\mathcal{I}_{200}(\tau)$, which we use as a proxy for the true (but unknown) integral value $\mathcal{I}(\tau)$, and compare it to $\mathcal{I}_M(\tau)$ for $M \in \{5, 10, 15, 20, 25, 30, 40, 50, 60, 70, 80, 90, 100 \}$. Note that the approximation induced by $\mathcal{I}_{200}(\tau)$ is also used to generate the simulations of Section 6 in the main paper. We measure accuracy using the \emph{relative absolute error} (RAE) and the \emph{standardized squared error} (SSE), defined respectively by
\begin{equation}\label{quadErrors}
R_M =   \int \left| \frac{ \mathcal{I}_{200}(\tau) - \mathcal{I}_M(\tau)}{\mathcal{I}_{200}(\tau)} \right| \, d\tau, \quad S_M=   \int \frac{ \left(\mathcal{I}_{200}(\tau) - \mathcal{I}_M(\tau)\right)^2}{\sigma^2}  \, d\tau,
\end{equation}
which we compute for each simulation. We report the pointwise medians for each $R_M$ and $S_M$ as a function of $M$ in Figure \ref{fig:quadApprox}. As expected, for fixed $M$, the integral approximation is more accurate when $\mu_{t-1}$|and therefore the integrand|is smooth. Nonetheless, the relative gains of increasing $M$ decline quickly for $M > 20$ in both cases. 

}

 \begin{figure}[h]
\begin{center}
\includegraphics[width=1\textwidth]{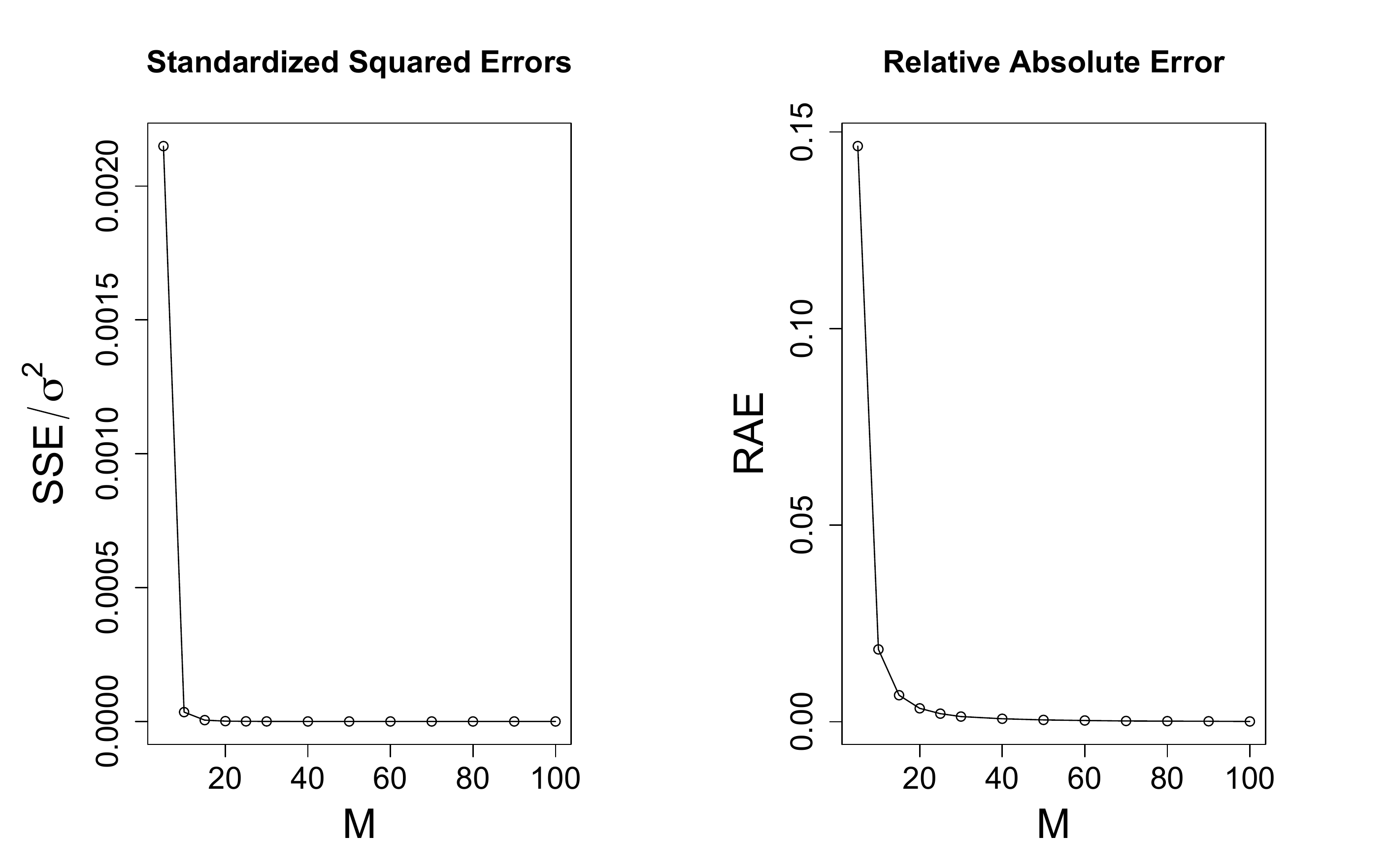}
\includegraphics[width=1\textwidth]{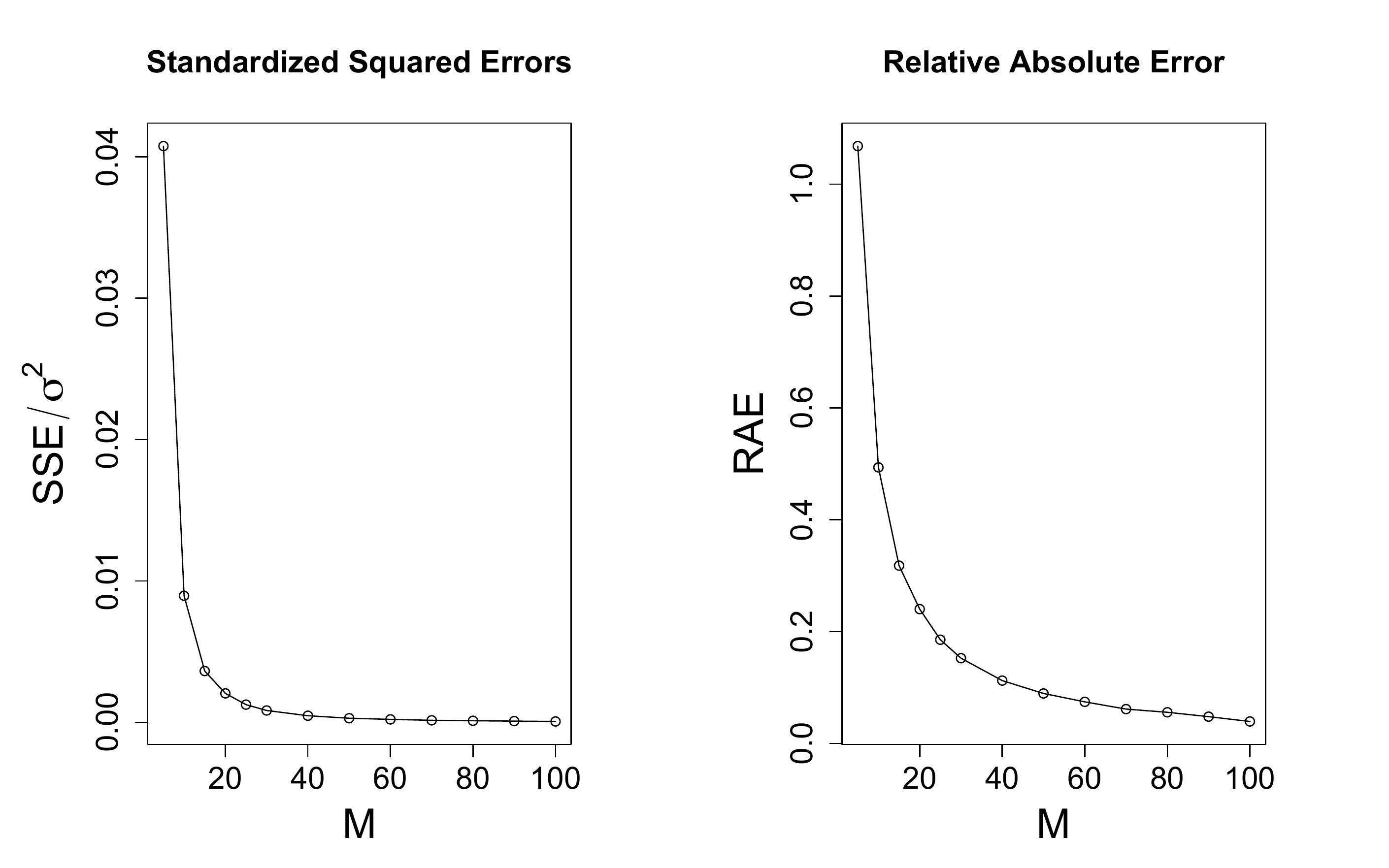}
\caption{Standardized squared errors and relative absolute errors for smooth (\textbf{top}) and non-smooth (\textbf{bottom}) integrands. The errors are small in magnitude, particularly in the smooth case, and decay quickly for $M > 20$. \label{fig:quadApprox}}
\end{center}
\end{figure}


\end{document}